%% file: popident1.tex
\documentclass{article}
\usepackage[latin1]{inputenc}

\usepackage{amsmath,amssymb,amsbsy}
\usepackage[english]{babel}
\usepackage{graphicx}
\usepackage{subfigure}
\usepackage{natbib}

\newtheorem{definition}{Definition}

\newcommand{\Ent}{\mathrm{\mathop{Ent}}}
\newcommand{\st}{\mathrm{\mathop{s.t.}}}
\newcommand{\T}{^\mathrm{T}}
\newcommand{\Real}{\mathbb{R}}

\newcommand{\one}{\mathbf{1}}

\usepackage{tikz}
\usetikzlibrary{arrows}
\usetikzlibrary{shapes}
\usetikzlibrary{backgrounds}
\usetikzlibrary{patterns}
\tikzstyle{graph}=[line width=1.2pt,join=round]

\title{Estimation of biochemical network parameter distributions in cell populations}
\author{Steffen Waldherr \and
Jan Hasenauer \and Frank Allg\"ower}

\date{Institute for Systems Theory and Automatic Control,\\
Universit\"at Stuttgart,
Pfaffenwaldring 9,
Stuttgart, Germany
}

\begin{document}
\maketitle

\begin{abstract}
Populations of heterogeneous cells play an important role in many
biological systems.
In this paper we consider systems where each cell can be modelled
by an ordinary differential equation.
To account for heterogeneity, parameter values are different among individual cells,
subject to a distribution function which is part of the model specification.

Experimental data for heterogeneous cell populations can be obtained
from flow cytometric fluorescence microscopy.
We present a heuristic approach to use such data for estimation of the
parameter distribution in the population.
The approach is based on generating simulation data for samples in
parameter space.
By convex optimisation, a suitable
probability density function for these samples is computed.

To evaluate the proposed approach, we consider artificial
data from a simple model of the tumor necrosis factor (TNF) signalling pathway.
Its main characteristic is a bimodality in the TNF response:
a certain percentage of cells undergoes apoptosis upon stimulation, while the
remaining part stays alive.
We show how our modelling approach allows to identify the reasons that underly
the differential response.

Keywords: estimation of probability distribution, population models, TNF signalling
\end{abstract}


\section{Introduction}
\label{sec:intro}


Modelling in systems biology typically aims at achieving a quantitative 
description of intracellular signal transduction or
differentiation processes at the cellular level.
Most models describe a ``typical single
cell'' on the basis of experimental data obtained from cell populations.
However, to understand dynamical behaviour
within heterogeneous cell populations, a consideration of many cells within
the whole population is mandatory.

Phenotypic heterogeneity in genetically identical cells arises mainly 
from stochasticity in biochemical reactions,
unequal partitioning of cellular material at
cell division \citep{Mantzaris2007}, or epigenetic differences \citep{Avery2006}.
When considering cells with high mutation rate, such as cancer cells, also
genotypic heterogeneity plays a major role.
For this paper, we choose to model heterogeneity by differences in parameter
values of the model describing the process of interest.
The model structure is on the contrary assumed to be identical in
all cells, as it usually represents the physical interactions among molecules,
which should be independent of the cell's state.
The parametric approach is well suited for genetic and epigenetic differences.
We assume that interactions among cells in the population can be neglected
for the process to be studied.
This is indeed the case for many relevant signalling pathways, and
is also an implicit assumption in many single-cell models.
The distribution of parameter values within the considered cell population
is described by a suitable multivariate probability distribution function,
which needs to be part of the model specification.
Mathematical modelling of such a process will typically result in
a non-linear partial differential
equation for the probability distributions of the state variables \citep{Mantzaris2007}.
Since this is very hard to deal with, we propose to use a sample
based approach, consisting of a large collection of ordinary differential equation
systems of identical structure, but with differing parameter values which
are subject to a specified parameter distribution function.

In this paper, we explore the possibility to estimate the parameter
distribution function using experimental large-scale measurements of the distributions
of system variables within the cell population.
Such data is available on a suitable scale from a newly developed measurement technology,
the flow cytometric fluorescence microscopy \citep{OrtynPer2007}, which
is a combination of classical flow cytometry and fluorescence microscopy.

Classical flow cytometry is a long-established tool to obtain
distributions of system variables in heterogeneous cell populations \citep{PerezNol2006}.
To measure the activity of signalling proteins, suitable fluorescence
markers are introduced into the cells.
A stream of several thousand cells per second
is then injected into the measurement device, and
the fluorescence intensity of each individual cell can be
measured.
While static flow cytometry measurements are very common in experimental setups,
corresponding time course data are rarely collected and are typically quite
sparse \citep{GardnerCan2000}.

Fluorescence microscopy is another established experimental tool, where microscopic
images from a population of fluorescently labelled cells are collected
and evaluated by image analysis.
With the classical technical implementation, fluorescence microscopy 
is limited to small sample numbers.
Yet percentages of cells showing a particular feature, such as an apoptotic
phenotype, are commonly measured at several time instances.
Also distributions of relevant variables over a time course
have been measured \citep{MettetalMuz2006}, but the technology is
not widely used in dynamical modelling studies.

Flow cytometric fluorescence microscopy now combines flow cytometry 
with single cell fluorescence microscopy by taking microscopic images
of individual cells while they pass the flow cytometer.
This allows to collect and analyse microscopic images of several thousand fluorescently
labelled cells per minute \citep{OrtynPer2007}, with
technological requirements similar to classical flow cytometry.
In this way, distributions of signalling protein activities can be measured
efficiently in large populations of heterogeneous cells \citep{GeorgeFan2006}.
Although the technology has not been used so far to obtain distributions of
relevant variables at several time instances, such measurements are now 
becoming experimentally feasible.

Estimation of parameter distributions in model collections that represent a heterogeneous
population is a long-standing topic in pharmacodynamics \citep{AI-BannaKel1990}.
However, a crucial difference between pharmacological experiments and
cell population measurements is that in pharmacodynamics, samples are taken from
the same individuals at all time points, measurements are linked to individuals, 
and as a consequence individual trajectories are known.
This is not the case in fluorescence microscopy, where each individual cell is measured
only once, and for each time point only the distribution of the measured variable
within the population is recorded.

Other established approaches to parameter estimation of probabilistic systems
usually consider a problem setup where the output of a single cell is directly considered
as available measurement data at all time instances.
This is quite different to our setup, where each individual cell can only be
taken for measurement once, and thus only the distribution of output variables
within the population is reliably known for all sampling times.
As a consequence, established approaches of parameter estimation seem not
to be well suited to deal with this problem.

In this paper, we present a heuristic approach to estimate the parameter distribution
from the distributions of measured variables.
In a first step, simulation data is generated for a suitable choice of
parameter samples.
As such, the approach is related to classical particle filters \citep{Doucet2001}.
However, instead of an iterative updating, we construct a convex optimisation problem
that produces a suitable weighting for the considered parameter samples.
This weighting can directly be transformed into a probability distribution
for the parameter values.

The paper is structured as follows.
In Section~\ref{sec:population-model}, the population
modelling framework that we are using is introduced.
In Section~\ref{sec:estimation}, we present the proposed method
to estimate the parameter distribution function of the model based
on population measurements.
Section~\ref{sec:tnf-model} describes the application of the proposed
method to simulated data for a model of the TNF signalling pathway, and
discusses how to use population modelling in order to evaluate
differences in cellular behavior within a heterogeneous cell population.

\textit{Notation:}
Denote by $[\check z,\hat z] \subset \Real^k$ the hyperrectangle
$\lbrace z \in \Real^k: \check z_i \leq z_i \leq \hat z_i, i=1,\ldots,k \rbrace$.

\section{Parameter-distributed population models}
\label{sec:population-model}

For the purpose of this paper, a model of a biochemical reaction network in
a population of $N$ cells is given by the collection of differential equations
\begin{equation}
\begin{aligned}
\dot x^{(i)}(t) &= f(x^{(i)}(t),\pi^{(i)}),& x^{(i)}(0) = x_0^{(i)}, \\
y^{(i)}(t) &= h(x^{(i)}(t)), &\quad i=1,\ldots,N
\end{aligned}
\label{eq:system}
\end{equation}
with state variables $x(t) \in \Real^n$, measured variables $y(t)\in\Real^q$,
and parameters $\pi\in\Real^r$.
The index $i$ specifies the individual cells within the population.
We collect the parameters and initial condition in the extended parameter
vector $p^{(i)} = (\pi^{(i)}, x_0^{(i)})\in\Real^{m}$, where $m = n+r$.

We assume that the population is heterogeneous, where heterogeneity is accounted
for by differences in parameter values among individual cells.
The distribution of parameters and initial conditions is given by a cumulative 
probability distribution function $\Phi:\Real^m \rightarrow [0,1]$
which is part of the model specification, i.e.\ parameter values
and initial conditions for the
cell with index $i$ are subject to the probability distribution
\begin{equation}
\begin{aligned}
\mathrm{Prob}(p_1^{(i)}\leq p_1,\ldots,p_m^{(i)}\leq p_m) = \Phi(p_1,\ldots,p_m).
\end{aligned}
\end{equation}

Due to the measurement technology, the output of every individual cell can only
be measured once during the course of an experiment,
because the cell is removed from the population for the measurement.
Thus, instead of considering the measured output $y^{(i)}$ directly,
it makes more sense to consider the distribution of $y^{(i)}$
at sampling times $t_k$, $k=1,\ldots,K$ as an output.
At each sampling time, $M$ cells are selected arbitrarily from the
population and subjected to measurement.
We assume that $M$ is large enough such that a reliable approximation of
the output distribution within the whole population can be obtained.
The measurement taken from \eqref{eq:system} is thus given by functions
\begin{equation}
\begin{aligned}
\Psi_k(y) &= \mathrm{Prob}(y^{(i)}(t_k) \leq y), \quad k=1,\ldots,K, \; i=1,\ldots,N.
\end{aligned}
\end{equation}
The goal of parameter distribution estimation is to compute the function
$\Phi(p_1,\ldots,p_m)$ from knowledge of the functions $\Psi_k(y)$ and
the model structure \eqref{eq:system}.
Typically the measurement of $\Psi_k(y)$ is discretized over suitable hyperrectangles
in the variable $y$.
Since the number of cells being measured is finite, the values of $\Psi_k(y)$
are discrete as well, although for most considerations we can assume the
number of cells large enough to neglect this.

\section{Parameter estimation method for population models}
\label{sec:estimation}

\subsection{Maximum Entropy approach to probability density estimation}

The proposed estimation method is based on the simulation of \eqref{eq:system}
for all parameter values and initial conditions contained in a finite sample 
\begin{equation*}
\begin{aligned}
\mathcal P = \lbrace p^{(i)}\in\Real^m: i=1,\ldots,M\rbrace.
\end{aligned}
\end{equation*}
A good choice for a sampling set is the so called Latin hypercube, which
ensures that the total range of the relevant parameter set is captured \citep{Stein1987}.

\begin{definition}
A finite set $\mathcal P \subset [\check p,\hat p] \subset \Real^m$ 
is called a \emph{latin hypercube in $\Real^m$
with sampling density $d$}, if it contains exactly one point $p\in\Real^m$
such that 
\begin{equation}
\begin{aligned}
(\alpha-1)\frac{\hat p_i-\check p_i}{d} < p_i - \check p_i < \alpha\frac{\hat p_i-\check p_i}{d}
\end{aligned}
\end{equation}
for each $i=1,\ldots,m$ and $\alpha=1,\ldots,d$.
\end{definition}

Having chosen the sampling set $\mathcal P$ as a Latin hypercube, the goal
is to estimate the fraction of cells, $\varphi(p^{(i)})$,
which have an extended parameter vector close to $p^{(i)}$, 
such that the weighted simulated trajectories approximate the measured
population dynamics reasonably well. This fraction
is a measure for the relative contribution of the
neighbourhood of $p^{(i)}$ to the cell population response.
Additionally, the fractions have to sum up to one,
\begin{equation}
\begin{aligned}
\sum_{i=1}^{M} \varphi(p^{(i)}) = 1.
\end{aligned}
\end{equation}
Hence, $\varphi(p^{(i)})$ can be interpreted as
an approximation of the probability density function at $p^{(i)}$.

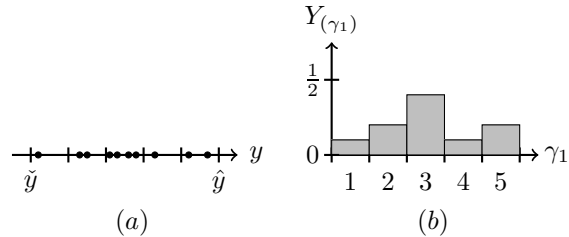
\begin{figure}[b]
\begin{center}
\begin{tikzpicture}
\draw[->,thick] (0.00,0.00) -- (3.00,0.00);
\foreach \x in {0.25,0.75,1.25,1.75,2.25,2.75}
	\draw[-,thick] (\x,-.1) -- (\x,0.1);
\foreach \x in {0.35,0.90,1.00,1.30,1.4,1.55,1.65,1.9,2.35,2.60}{
	\filldraw (\x,0) circle (0.04cm);
	}
\path[thick] (3.25,0.00) node {$y$};
\path[thick] (0.25,-.35) node {$\check y$};
\path[thick] (2.75,-.35) node {$\hat y$};
\path[thick] (1.6,-.9) node {$(a)$};

\draw[->,thick] (4.25,0.00) -- (4.25,1.50);
\draw[->,thick] (4.25,0.00) -- (7.00,0.00);
\path[thick] (7.25,0.00) node {$\gamma_1$};
\path[thick] (4.25,1.80) node {$Y_{(\gamma_1)}$};
\foreach \x in {4.25,4.75,5.25,5.75,6.25,6.75}
	\draw[-,thick] (\x,-.1) -- (\x,0.1);
\path[thick] (4.5,-.35) node {$1$};
\path[thick] (5.0,-.35) node {$2$};
\path[thick] (5.5,-.35) node {$3$};
\path[thick] (6.0,-.35) node {$4$};
\path[thick] (6.5,-.35) node {$5$};
\filldraw[fill=lightgray,draw=black] (4.25,0.00) rectangle (4.75,0.2);
\filldraw[fill=lightgray,draw=black] (4.75,0.00) rectangle (5.25,0.4);
\filldraw[fill=lightgray,draw=black] (5.25,0.00) rectangle (5.75,0.8);
\filldraw[fill=lightgray,draw=black] (5.75,0.00) rectangle (6.25,0.2);
\filldraw[fill=lightgray,draw=black] (6.25,0.00) rectangle (6.75,0.4);
\path[thick] (5.6,-.9) node {$(b)$};
\draw[-,thick] (4.15,0.00) -- (4.35,0.00);
\draw[-,thick] (4.15,1.00) -- (4.35,1.00);
\path[thick] (4.00,0.00) node {$0$};
\path[thick] (4.00,1.00) node {$\frac{1}{2}$};

\end{tikzpicture}
\end{center}
\caption{Illustration of the discretized popula\-tion distribution. 
$(a)$ shows as dots the measured outputs at time $t_k$ and $(b)$ 
depicts the corresponding discretized population distribution for $\beta = 5$.}
\label{fig:descretization of output space}
\end{figure}

In order to calculate $\varphi(p^{(i)})$,
the output space is divided into hyperrectangles.
For each sampling time, the fraction of cells of the population
which is contained in each hyperrectangle is computed
according to the following definition.
An illustration is shown in Figure \ref{fig:descretization of output space}.

\begin{definition}
A $q$-dimensional array $Y(t_k) \in \Real^{\beta_1 \times \ldots \times \beta_q}$
is called a \emph{discretized population distribution} 
at time $t_k$ with discretization vector $\beta = \left[\beta_1,\ldots,\beta_q\right]$, if
\begin{equation}
\begin{aligned}
Y_{(\gamma_1,\ldots,\gamma_q)}(t_k) & = \mathrm{Prob}\bigl(y^{(i)}(t_k) \in [\check y^{\gamma},\hat y^{\gamma}]\bigr),
\end{aligned}
\end{equation}
where
$\gamma_1 = 1,\ldots, \beta_1, \; \ldots, \; \gamma_q = 1,\ldots, \beta_q$,
and $\check y_i^{\gamma} = \check y_i + (\gamma_i - 1)\dfrac{\hat y_i - \check y_i}{\beta_i}$ and
$\hat y_i^{\gamma} = \check y_i + \gamma_i \dfrac{\hat y_i - \check y_i}{\beta_i}$, $i = 1,\ldots,m$.
Hereby $\check y_i$ and $\hat y_i$ are the minimal respectively 
maximal values of output $i$ which are measured.
\end{definition}

The array $Y(t_k)$ can be interpreted as a discrete approximation of
the probability density function of the outputs.
To compute the $\varphi(p^{(i)})$, $i=1,\ldots,M$,
the system \eqref{eq:system} is simulated for every $p^{(i)} \in \mathcal P$
and the obtained outputs are discretized.

\begin{definition}
A $q$-dimensional array $\tilde{Y}^{(i)}(t_k) \in \Real^{\beta_1 \times \ldots \times \beta_q}$
is called a \emph{discretized trajectory of $\tilde{y}^{(i)}$} at time points $t_k$, $k=1,\ldots,K$,
with discretization vector $\beta = \left[\beta_1,\ldots,\beta_q\right]$, if
\begin{equation}
\begin{aligned}
\tilde{Y}^{(i)}_{(\gamma_1,\ldots,\gamma_q)}(t_k) & = 
\begin{cases}
1, \quad \textnormal{if } \tilde{y}^{(i)}(t_k) \in [\check y^{\gamma},\hat y^{\gamma}]\\
0, \quad \textnormal{otherwise},
\end{cases}\\
& \gamma_1 = 1,\ldots, \beta_1, \; \ldots, \; \gamma_q = 1,\ldots, \beta_q.
\end{aligned}
\end{equation}
Hereby $\tilde{y}^{(i)}$ is the output of the system obtained
by simulation with $p^{(i)}$.
$\check y^{\gamma}$, $\hat y^{\gamma}$ are defined as before.
\end{definition}

The array $\tilde{Y}^{(i)}(t_k)$ can also be interpreted as an approximation
of the probability density function for cells with the parameter vector $p^{(i)}$.
Given the discretized measured population dynamics and
the discretized simulated trajectories,
we intend to compute the fractions, $\varphi(p^{(i)})$, 
such that the difference between the weighted sum of simulated trajectories
and the measured population dynamics,
\begin{equation}
\begin{aligned}
\Delta Y(\varphi,t_k) &= \sum_{i=1}^{M} \varphi(p^{(i)}) \, \tilde{Y}^{(i)}(t_k) - Y(t_k),
\end{aligned}
\label{eq: difference of measurement and simulation}
\end{equation}
is zero, for $k = 1,\ldots,K$, where
\begin{equation}
\begin{aligned}
\varphi = \left[\varphi(p^{(1)}),\ldots,\varphi(p^{(M)})\right]^T.
\end{aligned}
\end{equation}
This problem could be solved using standard
least square techniques, but this leads in many cases
to dramatical overfitting. Especially, if the 
measurement data does not contain enough information to fit 
the parameter distribution, a spiky probability density function is obtained.
Least square techniques generally select a minimum norm solution for
underdetermined systems, so it can happen that parameters which are not identifiable
show a peak in the probability density function at a single point.

This should be avoided, because it is desirable that the resulting distribution
indicates whether a parameter is identifiable given the measured data or not.
Therefore, we choose an entropy based approach
to determine $\varphi(p^{(i)})$.

\begin{definition}
The function $\Ent: \Real^M \rightarrow \Real$ given by
\begin{equation}
\begin{aligned}
\Ent(\varphi) = - \sum_{i=1}^{M} \varphi(p^{(i)}) \, \mathrm{ln}(\varphi(p^{(i)})).
\end{aligned}
\end{equation}
is called the \emph{entropy} of $\varphi$.
\end{definition}

Given an underdetermined system of equations, the maximum entropy
approach selects the solution which contains least information, and thus
avoids adding artificial information to the measurement data \citep{MacKay2003}.
In our case this implies that the ``flattest'' probability distribution
which fits all the constraints is selected in the optimisation problem. 
Thus, if a parameter is not identifiable,
we obtain a very flat distribution and no information is added.

The entropy approach yields the optimisation problem
\begin{equation}
	\begin{array}{lrll}
	\max	& \Ent(\varphi)\\
	\st		& \Delta Y(\varphi,t_k)&= 0, & k = 1,\ldots,K\\
			& \one\T \varphi &= 1 \\
			& \varphi &\geq 0,
	\end{array}
	\label{eq: optimization problem without errors}
\end{equation}
where $\one = [1,\ldots,1]^T \in \Real^{M}$.
The solution of \eqref{eq: optimization problem without errors} is the weighting vector $\varphi$,
with the highest entropy which exactly reproduces the discretized measured
population dynamics.
Unfortunately, \eqref{eq: optimization problem without errors} is very likely to be infeasible
because even if the equation
\begin{equation}
\begin{aligned}
\Delta Y(\varphi,t_k) &= 0, \quad k = 1,\ldots,K,
\end{aligned}
\end{equation}
is underdetermined, it cannot be ensured that a solution exists.
Reasons are measurement errors and small cell numbers in measurements,
but primarily an insufficient discretization of the parameter space.
To improve the feasibility, small discrepancies between the measured
and the weighted simulated population are allowed.
This leads to the relaxed problem
\begin{equation}
	\begin{array}{lrll}
	\max	& \Ent(\varphi)\\
	\st		& \Delta Y(\varphi,t_k)&\in [- \hat e,\hat e], & k = 1,\ldots,K\\
			& \one^T \varphi &= 1 \\
			& \varphi &\geq 0, \\
	\end{array}
	\label{eq: optimization problem with errors}
\end{equation}
where $\Delta Y(\varphi,t_k) \in [-\hat e,\hat e]$ denotes the constraint that 
each element of $\Delta Y$ is bounded between $- \hat e$ and $+ \hat e$.
As before, the other constraints are that all weights
sum up to one, and that all weights are greater than
or equal to zero.

We are left with the problem to define the error bound $\hat e$.
A known constraint is that $\hat e \in [0,1]$.
To obtain estimation results that fit the measurements as good as possible, $\hat e$ 
is decreased to the minimal value for which 
\eqref{eq: optimization problem with errors} is still feasible.
This is done via a bisection algorithm.

The relaxed optimisation problem \eqref{eq: optimization problem with errors} is convex.
The entropy is concave and the constraints are linear.
For the class of convex optimisation problems efficient solvers
exist, for instance the primal-dual-interior point method
\citep{BoydVan2004}. 
Convergence to the global maximum in polynomial time can be guaranteed.

Based on the solution of \eqref{eq: optimization problem with errors},
an estimate $\hat\Phi$ for the parameter distribution function $\Phi$ is
computed as
\begin{equation}
\begin{aligned}
\hat \Phi(p) = \sum_{i: p \geq p^{(i)} \in \mathcal P} \varphi(p^{(i)}).
\end{aligned}
\label{eq:phi-estimate}
\end{equation}

\subsection{Distribution estimation for independent parameters}

A simplifying yet convenient assumption is that parameter values
and initial conditions are independently distributed.
Although not strictly true in most cases, it is reasonable to
make this simplification also if parameters are only weakly correlated.
In this case, the probability distribution
function can be decomposed as
\begin{equation}
\label{eq:density-decomposition}
\begin{aligned}
\Phi(p) = \Phi_1(p_1) \Phi_2(p_2) \cdots \Phi_m(p_m),
\end{aligned}
\end{equation}
where $\Phi_i(p_i)$ denotes the distribution function for the $i$-th parameter.

Based on the estimate $\hat\Phi$ \eqref{eq:phi-estimate}, estimates for the
individual distribution functions can be computed by marginalising the other
parameters, i.e.\ taking
$\Phi_i(p_i) = \lim_{p_j \rightarrow \infty,\,j\neq i} \Phi(p)$.
Thus an estimate $\hat\Phi_i$ for the individual distributions
is obtained as
\begin{equation}
\begin{aligned}
\hat\Phi_i(p_i) = \sum_{j: p_i \geq p_i^{(j)}\in\mathcal P} \varphi(p^{(j)})
\end{aligned}
\end{equation}
for $i=1,\ldots,m$.

\section{Application to a TNF signal transduction model}
\label{sec:tnf-model}

\subsection{Motivation for population modelling}

TNF is a signalling hormone involved in the inflammatory response of
mammalian cells.
It can induce programmed cell death (apoptosis) via the caspase cascade, 
but has also anti-apoptotic effects via the NF-$\kappa$B pathway \citep{WajantPfi2003}.
For many cell types, stimulation with TNF will incude apoptosis in a certain
percentage of the population, while the remaining cells stay alive.
The reasons for this heterogeneous behaviour are unclear, but
of great interest for biological research in TNF signalling.
However, a major obstacle to the direct experimental study of the process
is that the behaviour of individual cells cannot be monitored
on a population scale over the time scale of interest.
To overcome this problem, we propose the use of population modelling
and estimation of parameter distributions from experimental population data.
With a suitable model, a collection of single cell trajectories 
can be clustered according
to the individual cell's fate and compared for characteristic
differences in parameters or early-stage cell behaviour.

In this paper, we use artificial measurement data, generated from simulations, for two reasons.
First, suitable experimental data is not yet available.
Second, the purpose of the paper is more an evaluation of the estimation method
itself than its application in biological research.
Also, since no general results on parameter identifiability in the considered problem
are available, such a study should be done in each application of the method to evaluate
identifiability properties.

\subsection{Presentation of the TNF signal transduction model}

%

The model is based on earlier work from \cite{ChavesEis2008} and is built
from known activating and inhibitory interactions among key signalling proteins.
It includes as state variables activities of the caspases 8 and 3 (C8, C3), the transcription
factor NF-$\kappa$B and its inhibitor I-$\kappa$B.
The model is given by the ODE system
\begin{equation}
\label{eq:tnf-model}
\begin{aligned}
\dot x_1 &= -x_1 + \frac{1}{2}(\beta_4(x_3)\alpha_1(u)+\alpha_3(x_2)) \\
\dot x_2 &= -x_2 + \alpha_2(x_1)\beta_3(x_3) \\
\dot x_3 &= -x_3 + \beta_2(x_2)\beta_5(x_4) \\
\dot x_4 &= -x_4 + \frac{1}{2}(\beta_1(u)+\alpha_4(x_3)).
\end{aligned}
\end{equation}
The state variables $x_i$, $i=1,\ldots,4$ are bounded between $0$ and $1$ and
denote the relative activities of the signalling proteins C8, C3, NF-$\kappa$B 
and I-$\kappa$B, respectively.
The functions $\alpha_j(x_i)$, $j=1,\ldots,4$ represent activating connections
and are given by $\alpha_j(x_i) = \frac{x_i^2}{a_j^2 + x_i^2}$.
Correspondingly, $\beta_j(x_i) = \frac{b_j^2}{b_j^2 + x_i^2}$, $j=1,\ldots,5$ 
represent inhibiting connections.
$a_j$ and $b_j$ are parameters with values between $0$ and $1$,
representing activation and inhibition thresholds, respectively.
The input $u$ denotes the external TNF stimulus.
Nominal parameter values are given in Table~\ref{tab:tnf-parameters}.

\begin{table}
\begin{center}
\begin{tabular}{c|ccccc}
$i$ & 1 & 2 & 3 & 4 & 5 \\ \hline
$a_i$ & $0.6$ & $0.2$ & $0.2$ & $0.5$ &  \\
$b_i$& $0.4$ & $0.7$ & $0.3$ & $0.5$ & $0.4$
\end{tabular}\end{center}
\caption{Nominal parameter values for the TNF signalling model \eqref{eq:tnf-model}.}
\label{tab:tnf-parameters}
\end{table} 

\subsection{Results of parameter distribution estimation}
\label{ssec:tnf-estimation}

To evaluate the proposed approach we
consider a virtual experimental setup
in which the caspase 3 and NF-$\kappa$B
activity is measured at the time points $t \in \left\{0, 0.5, 1, 2, 4, 6, 8, 10, 15, 20\right\}$
by flow cytometric microscopy.
For each time point, the outputs of 10000 simulated cells are obtained, resulting
in an output density distribution for each time point.
Measurement errors are neglected in this example
to make the interpretion of the results as
simple as possible.

\begin{figure}[t]
\centering
\begin{tikzpicture}
\input{matlab_plots/measdist_t=0.pgf}
\end{tikzpicture}
\begin{tikzpicture}
\input{matlab_plots/measdist_t=2.pgf}
\end{tikzpicture}
\begin{tikzpicture}
\input{matlab_plots/measdist_t=4.pgf}
\end{tikzpicture} \\[3ex]
\begin{tikzpicture}
\input{matlab_plots/measdist_t=6.pgf}
\end{tikzpicture}
\begin{tikzpicture}
\input{matlab_plots/measdist_t=10.pgf}
\end{tikzpicture}
\begin{tikzpicture}
\input{matlab_plots/measdist_t=20.pgf}
\end{tikzpicture}
\caption{Measured output distributions for caspase 3 activity at the considered
sampling times. The horizontal axis gives the caspase 3 activity, the size of
the bars indicates the relative frequency in the population.}
\label{fig:outputs}
\end{figure}
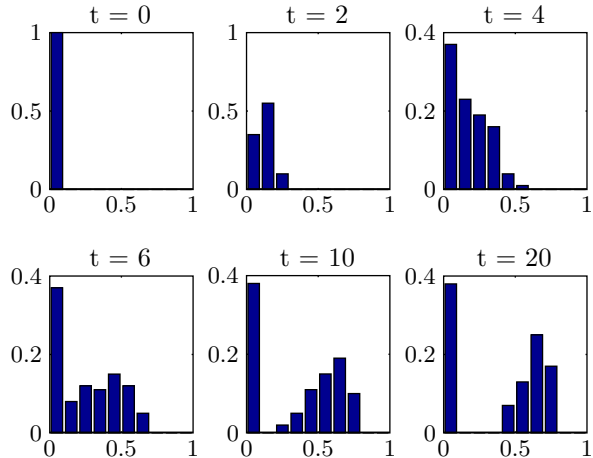

\begin{figure}[tb]
\centering
\begin{tikzpicture}
\input{matlab_plots/pdf_a_1.pgf}
\end{tikzpicture} \quad
\begin{tikzpicture}
\input{matlab_plots/pdf_a_4.pgf}
\end{tikzpicture} \\[3ex]
\begin{tikzpicture}
\input{matlab_plots/pdf_b_2.pgf}
\end{tikzpicture} \quad
\begin{tikzpicture}
\input{matlab_plots/pdf_b_3.pgf}
\end{tikzpicture}
\caption{Comparison of actual (dashed) and estimated (full) parameter probability function.}
\label{fig:pdf}
\end{figure}
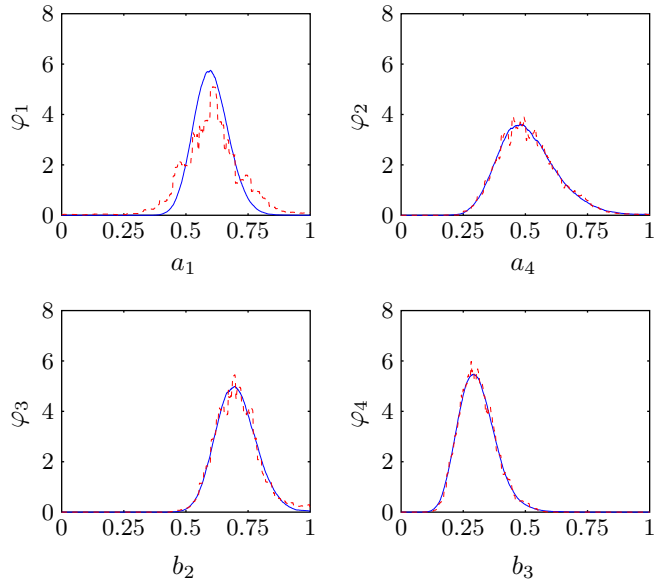

For this example, we assume heterogeneity in the 
parameters $a_1$, $a_4$, $b_2$, and $b_3$.
For the generation of measurement data,
each of the heterogeneous parameters is assumed to be distributed according to a log-normal
distribution around the nominal values given in table \ref{tab:tnf-parameters}.
Each cell is assumed to have an initial condition which corresponds to the steady state
with $x_1 = x_2 = 0$
for $u = 0$, where $x_3$ and $x_4$ depend on the individual parameter values.
The considered heterogeneity is interesting, because it results in a bimodal
response of the population to a TNF pulse applied during the time interval $0 < t < 2$.
The output distributions in caspase 3 activity that would be measured in
this case are shown in Figure~\ref{fig:outputs}.
About 35 \% of the population returns to zero caspase 3 activity, while the remaining
cells show sustained caspase activity.
The probability density functions of the parameter values are shown
in Figure~\ref{fig:pdf}.

For the identification of the parameter distribution,
the lower and upper bounds of all parameters
are set to be 0 respectively 1. The sampling density $d$
of the Latin hypercube is set to $2000$ and $\beta = [10, 10]^T$ is
selected as discretization vector for the outputs.
The probability density functions that are estimated for the
different parameters by our method are depicted in
figure \ref{fig:pdf}, in comparison to the real density
functions.
As can be seen in the figure, the probability density functions
of $a_4$, $b_2$, and $b_3$ are approximated very well.
Also for $a_1$ we can see good agreement.
The distributions peak at approximately the same point
and the shape is roughly the same.

Although all parameters are identifiable, there are huge differences
in the identifiability of the single parameters.
The cumulative probability functions of $a_4$ and $b_3$
can be estimated very well with a sampling density
$d$ of only $250$ (results not shown). For the approximation of $b_2$ and especially 
$a_1$ more samples in the parameter space have to be taken.
This can be related to the observation from the analysis done in 
Section~\ref{ssec:model-analysis}, that $a_4$ and $b_3$ are of high relevance for the 
bimodal response of the cells, while the other two parameters do not
have a high influence on this property.

For the considered example, the computation is quite efficient.
Computation time on a standard desktop computer was on the order of a few minutes,
and most of the time was spent computing trajectories for individual
parameter values.
In the proposed algorithm, this task can easily be parallelised for more complex models. 

\subsection{Analysis of the population model}
\label{ssec:model-analysis}

In this section, we discuss the biological conclusions that can
be drawn from a computational analysis of the population model \eqref{eq:tnf-model}
with parameter distributions as used in Section~\ref{ssec:tnf-estimation}.
For the considered system, it is of particular interest to distinguish
between cells that undergo apoptosis and cells that stay alive.
In apoptotic cells, the state variable $x_2$
(caspase 3 activity) tends to a positive value larger than a threshold $\theta$.
For non-apoptotic cells, $x_2$ returns to zero after a small transient
rise.

In order to investigate the underlying differences that lead to such
a differential behaviour, 
we consider a sample of parameter values $p^{(i)}$ taken from the distributions specified
in Section \ref{ssec:tnf-estimation},
giving rise to trajectories $x^{(i)}(t)$ of \eqref{eq:tnf-model}.
The parameter samples are clustered into the apoptotic set $\mathcal A$ and
the non-apoptotic set $\mathcal L$ by the criterium
\begin{equation}
\begin{aligned}
\mathcal A &= \lbrace p^{(i)} \mid x^{(i)}_2(T_{end}) \geq \theta \rbrace,
&\ \mathcal L &= \lbrace p^{(i)} \mid p^{(i)} \notin \mathcal A \rbrace,
\end{aligned}
\end{equation}
where $\theta = 0.3$ for this study.

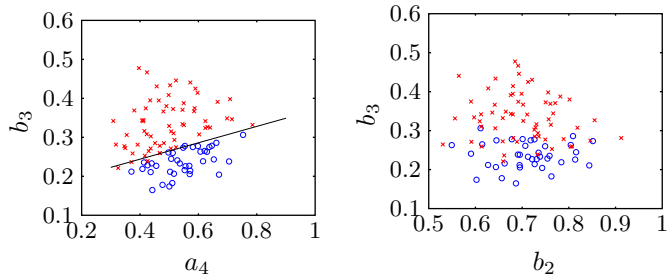
\begin{figure}[t]
\centering
\begin{tikzpicture}
\input{matlab_plots/a4_b3.pgf}
\end{tikzpicture}
\quad
\begin{tikzpicture}
\input{matlab_plots/b2_b3.pgf}
\end{tikzpicture}
\caption{Comparison of parameter values for apoptotic ({\color{red}x}) 
and non-apoptotic ({\color{blue}o}) cells and approximate separation \eqref{eq:param-separation} in the
$a_4$--$b_3$ plane.}
\label{fig:param-compare}
\end{figure}

First let us compare the sets $\mathcal A$ and $\mathcal L$ by directly
examining the respective parameter values.
As seen from Figure~\ref{fig:param-compare}, the differences between the cells
can mainly be explained from differences in the value of the parameter
$b_3$, which is the threshold for NF-$\kappa$B to inhibit caspase 3 activation, and
the parameter $a_4$, which is the threshold for NF-$\kappa$B to activate I-$\kappa$B.
In fact, an approximative separation criterium can be obtained directly
from Figure~\ref{fig:param-compare} as
\begin{equation}
\label{eq:param-separation}
\begin{aligned}
p \in \mathcal A \Leftrightarrow b_3 \gtrsim 0.16 + 0.21 a_4.
\end{aligned}
\end{equation}
Apoptotic cells are thus characterised by high values for $b_3$ and low values
for $a_4$, which relates well to biological intuition.
The parameter $b_2$, which is the threshold for 
caspase 3 to inhibit NF-$\kappa$B activation, seems to have little to 
no influence on the cell fate.
These results indicate that the cell fate is determined by influences from
the NF-$\kappa$B pathway to the caspase cascade, and not vice versa.

\begin{figure}[t]
\centering
\begin{tikzpicture}
\input{matlab_plots/traj_C3.pgf}
\end{tikzpicture}
\quad
\begin{tikzpicture}
\input{matlab_plots/traj_NFkB.pgf}
\end{tikzpicture}
\caption{Comparison of a few trajectories for apoptotic (dashed)
and non-apoptotic (full lines) cells.}
\label{fig:states-compare}
\end{figure}
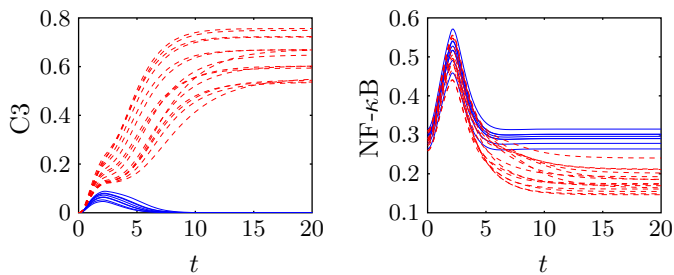

Next, we try to find early-stage markers for the cell's fate.
This is of interest because individual parameter values are 
not known when observing a single cell by e.g. live cell imaging,
yet we may want to predict the fate of a specific individual cell.
The collection of trajectories for the considered parameter sample is shown
in Figure~\ref{fig:states-compare}.
Obviously, early-stage caspase activity is a good indicator for the later fate of the cell.
However, it is quite interesting from the biological viewpoint that early-stage
NF-$\kappa$B activity seems not to be a good indicator.
In fact, the NF-$\kappa$B trajectories in Figure~\ref{fig:states-compare} separate only
for $t>10$, a time for which most apoptotic cells show already high caspase activity.

\section{Summary and Conclusions}

Heterogeneity in cell populations is an important aspect for research in systems biology.
However, computational approaches to deal with heterogeneous populations are rare.
A reasonable way to describe heterogeneity is to assume that parameter values
are stochastically distributed within the population.

In the modelling process, it is then necessary to estimate the parameter distribution functions
from suitable experimental data.
For this paper, we assume that the output distribution in the cell population is
measured at discrete sampling times.
We present an optimisation-based approach to estimate parameter distributions from
such measurements, which minimizes the prediction error based on a suitable sampling
of the parameter space.
With the suggested latin hypercube sampling, the approach scales well to
systems with a high-dimensional parameter space.

We applied the suggested estimation method to artificial data for a model of TNF signal
transduction.
For the parameters where heterogeneity was assumed, our method gives good estimates
of the parameter distribution function.
The results thus indicate that those parameters are identifiable
from the measurements used in this setup.

%

\section{Acknowledgments}

We thank Peter Scheurich for carefully explaining the peculiarities of TNF
signal transduction.
We also thank Cristian Rojas and Nicole Radde for helpful comments on a
previous version of the manuscript.
This work was supported by the BMBF in the FORSYS-Partner program, 
grant nr.\ 0315-280A.


\bibliographystyle{/home/waldherr/texmf/tex/ifac/plainnatifac}

\begin{small}
\bibliography{/home/waldherr/Forschung/Referenzen.bib}
\end{small}

\end{document}

%% file: matlab_plots/measdist_t=0.pgf
  \begin{pgfscope}
    \definecolor{matfig2pgf_color}{rgb}{1,1,1}\pgfsetfillcolor{matfig2pgf_color}
    \pgfpathrectangle{\pgfpoint{1.12889cm}{0.846667cm}}{\pgfpoint{1.90564cm}{2.08717cm}}
    \pgfusepath{fill}
  \end{pgfscope}
  \begin{pgfscope}
    \pgfsetlinewidth{0.5pt}
    \foreach \x in {2.08171}
    {
      \pgfpathmoveto{\pgfpoint{\x cm}{0.846667cm}}\pgfpathlineto{\pgfpoint{\x cm}{0.867538cm}}
      \pgfpathmoveto{\pgfpoint{\x cm}{2.93384cm}}\pgfpathlineto{\pgfpoint{\x cm}{2.91297cm}}
    }
    \foreach \y in {1.89025}
    {
      \pgfpathmoveto{\pgfpoint{1.12889cm}{\y cm}}\pgfpathlineto{\pgfpoint{1.14795cm}{\y cm}}
      \pgfpathmoveto{\pgfpoint{3.03453cm}{\y cm}}\pgfpathlineto{\pgfpoint{3.01547cm}{\y cm}}
    }
    \pgfusepath{stroke}
  \end{pgfscope}
  \begin{pgfscope}
    \pgfsetlinewidth{0.5pt}
    \pgfpathrectangle{\pgfpoint{1.12889cm}{0.846667cm}}{\pgfpoint{1.90564cm}{2.08717cm}}
    \pgfusepath{stroke}
  \end{pgfscope}
  {\small
    \pgftext[x=1.12889cm,y=0.746667cm,top]{$0$}
    \pgftext[x=2.08171cm,y=0.746667cm,top]{$0.5$}
    \pgftext[x=3.03453cm,y=0.746667cm,top]{$1$}
    \pgftext[x=1.02889cm,y=0.846667cm,right]{$0$}
    \pgftext[x=1.02889cm,y=1.89025cm,right]{$0.5$}
    \pgftext[x=1.02889cm,y=2.93384cm,right]{$1$}
  }
  \begin{pgfscope}
    \pgfsetlinewidth{0.50pt}
    \definecolor{matfig2pgf_linecolor}{rgb}{0.000,0.000,0.000}
    \pgfsetstrokecolor{matfig2pgf_linecolor}
    \pgfsetdash{}{0pt}
    \definecolor{matfig2pgf_facecolor}{rgb}{102,108,97}\pgfsetfillcolor{matfig2pgf_facecolor}
    \definecolor{matfig2pgf_facecolor}{rgb}{0,0,0.5625}\pgfsetfillcolor{matfig2pgf_facecolor}
    \pgfplothandlerlineto
\pgfplotstreamstart
\foreach \x/\y in {1.148/0.847,1.148/2.934,1.300/2.934,1.300/0.847}
{
\pgfplotstreampoint{\pgfpoint{\x cm}{\y cm}}
}
\pgfplotstreamend
    \pgfpathclose
    \pgfusepath{stroke,fill}
    \pgfplothandlerlineto
\pgfplotstreamstart
\foreach \x/\y in {1.339/0.847,1.339/0.847,1.491/0.847,1.491/0.847}
{
\pgfplotstreampoint{\pgfpoint{\x cm}{\y cm}}
}
\pgfplotstreamend
    \pgfpathclose
    \pgfusepath{stroke,fill}
    \pgfplothandlerlineto
\pgfplotstreamstart
\foreach \x/\y in {1.529/0.847,1.529/0.847,1.682/0.847,1.682/0.847}
{
\pgfplotstreampoint{\pgfpoint{\x cm}{\y cm}}
}
\pgfplotstreamend
    \pgfpathclose
    \pgfusepath{stroke,fill}
    \pgfplothandlerlineto
\pgfplotstreamstart
\foreach \x/\y in {1.720/0.847,1.720/0.847,1.872/0.847,1.872/0.847}
{
\pgfplotstreampoint{\pgfpoint{\x cm}{\y cm}}
}
\pgfplotstreamend
    \pgfpathclose
    \pgfusepath{stroke,fill}
    \pgfplothandlerlineto
\pgfplotstreamstart
\foreach \x/\y in {1.910/0.847,1.910/0.847,2.063/0.847,2.063/0.847}
{
\pgfplotstreampoint{\pgfpoint{\x cm}{\y cm}}
}
\pgfplotstreamend
    \pgfpathclose
    \pgfusepath{stroke,fill}
    \pgfplothandlerlineto
\pgfplotstreamstart
\foreach \x/\y in {2.101/0.847,2.101/0.847,2.253/0.847,2.253/0.847}
{
\pgfplotstreampoint{\pgfpoint{\x cm}{\y cm}}
}
\pgfplotstreamend
    \pgfpathclose
    \pgfusepath{stroke,fill}
    \pgfplothandlerlineto
\pgfplotstreamstart
\foreach \x/\y in {2.291/0.847,2.291/0.847,2.444/0.847,2.444/0.847}
{
\pgfplotstreampoint{\pgfpoint{\x cm}{\y cm}}
}
\pgfplotstreamend
    \pgfpathclose
    \pgfusepath{stroke,fill}
    \pgfplothandlerlineto
\pgfplotstreamstart
\foreach \x/\y in {2.482/0.847,2.482/0.847,2.634/0.847,2.634/0.847}
{
\pgfplotstreampoint{\pgfpoint{\x cm}{\y cm}}
}
\pgfplotstreamend
    \pgfpathclose
    \pgfusepath{stroke,fill}
    \pgfplothandlerlineto
\pgfplotstreamstart
\foreach \x/\y in {2.672/0.847,2.672/0.847,2.825/0.847,2.825/0.847}
{
\pgfplotstreampoint{\pgfpoint{\x cm}{\y cm}}
}
\pgfplotstreamend
    \pgfpathclose
    \pgfusepath{stroke,fill}
    \pgfplothandlerlineto
\pgfplotstreamstart
\foreach \x/\y in {2.863/0.847,2.863/0.847,3.015/0.847,3.015/0.847}
{
\pgfplotstreampoint{\pgfpoint{\x cm}{\y cm}}
}
\pgfplotstreamend
    \pgfpathclose
    \pgfusepath{stroke,fill}
  \end{pgfscope}
  \makeatletter\ifpgf@draftmode\makeatother\else
  \begin{pgfscope}
    \pgfpathrectangle{\pgfpoint{1.12889cm}{0.846667cm}}{\pgfpoint{1.90564cm}{2.08717cm}}
    \pgfusepath{clip}
    \begin{pgfscope}
      \pgfsetlinewidth{1.00pt}
      \definecolor{matfig2pgf_linecolor}{rgb}{0.000,0.000,0.000}
      \pgfsetstrokecolor{matfig2pgf_linecolor}
      \pgfsetdash{}{0pt}
      \pgfsetroundjoin
      \pgfplothandlerlineto
\pgfplotstreamstart
\foreach \x/\y in {1.129/0.847,3.035/0.847}
{
\pgfplotstreampoint{\pgfpoint{\x cm}{\y cm}}
}
\pgfplotstreamend
      \pgfusepath{stroke}
    \end{pgfscope}
  \end{pgfscope}
  \fi
    \pgftext[bottom,x=2.0677cm,y=3.06076cm,rotate=0]{t = 0}
  \makeatletter\ifpgf@draftmode\makeatother\pgftext[x=5cm,y=4.84375cm]{\Huge{DRAFT}}\fi

%% file: matlab_plots/measdist_t=2.pgf
  \begin{pgfscope}
    \definecolor{matfig2pgf_color}{rgb}{1,1,1}\pgfsetfillcolor{matfig2pgf_color}
    \pgfpathrectangle{\pgfpoint{1.12889cm}{0.846667cm}}{\pgfpoint{1.90564cm}{2.08717cm}}
    \pgfusepath{fill}
  \end{pgfscope}
  \begin{pgfscope}
    \pgfsetlinewidth{0.5pt}
    \foreach \x in {2.08171}
    {
      \pgfpathmoveto{\pgfpoint{\x cm}{0.846667cm}}\pgfpathlineto{\pgfpoint{\x cm}{0.867538cm}}
      \pgfpathmoveto{\pgfpoint{\x cm}{2.93384cm}}\pgfpathlineto{\pgfpoint{\x cm}{2.91297cm}}
    }
    \foreach \y in {1.89025}
    {
      \pgfpathmoveto{\pgfpoint{1.12889cm}{\y cm}}\pgfpathlineto{\pgfpoint{1.14795cm}{\y cm}}
      \pgfpathmoveto{\pgfpoint{3.03453cm}{\y cm}}\pgfpathlineto{\pgfpoint{3.01547cm}{\y cm}}
    }
    \pgfusepath{stroke}
  \end{pgfscope}
  \begin{pgfscope}
    \pgfsetlinewidth{0.5pt}
    \pgfpathrectangle{\pgfpoint{1.12889cm}{0.846667cm}}{\pgfpoint{1.90564cm}{2.08717cm}}
    \pgfusepath{stroke}
  \end{pgfscope}
  {\small
    \pgftext[x=1.12889cm,y=0.746667cm,top]{$0$}
    \pgftext[x=2.08171cm,y=0.746667cm,top]{$0.5$}
    \pgftext[x=3.03453cm,y=0.746667cm,top]{$1$}
    \pgftext[x=1.02889cm,y=0.846667cm,right]{$0$}
    \pgftext[x=1.02889cm,y=1.89025cm,right]{$0.5$}
    \pgftext[x=1.02889cm,y=2.93384cm,right]{$1$}
  }
  \begin{pgfscope}
    \pgfsetlinewidth{0.50pt}
    \definecolor{matfig2pgf_linecolor}{rgb}{0.000,0.000,0.000}
    \pgfsetstrokecolor{matfig2pgf_linecolor}
    \pgfsetdash{}{0pt}
    \definecolor{matfig2pgf_facecolor}{rgb}{102,108,97}\pgfsetfillcolor{matfig2pgf_facecolor}
    \definecolor{matfig2pgf_facecolor}{rgb}{0,0,0.5625}\pgfsetfillcolor{matfig2pgf_facecolor}
    \pgfplothandlerlineto
\pgfplotstreamstart
\foreach \x/\y in {1.148/0.847,1.148/1.577,1.300/1.577,1.300/0.847}
{
\pgfplotstreampoint{\pgfpoint{\x cm}{\y cm}}
}
\pgfplotstreamend
    \pgfpathclose
    \pgfusepath{stroke,fill}
    \pgfplothandlerlineto
\pgfplotstreamstart
\foreach \x/\y in {1.339/0.847,1.339/1.995,1.491/1.995,1.491/0.847}
{
\pgfplotstreampoint{\pgfpoint{\x cm}{\y cm}}
}
\pgfplotstreamend
    \pgfpathclose
    \pgfusepath{stroke,fill}
    \pgfplothandlerlineto
\pgfplotstreamstart
\foreach \x/\y in {1.529/0.847,1.529/1.055,1.682/1.055,1.682/0.847}
{
\pgfplotstreampoint{\pgfpoint{\x cm}{\y cm}}
}
\pgfplotstreamend
    \pgfpathclose
    \pgfusepath{stroke,fill}
    \pgfplothandlerlineto
\pgfplotstreamstart
\foreach \x/\y in {1.720/0.847,1.720/0.847,1.872/0.847,1.872/0.847}
{
\pgfplotstreampoint{\pgfpoint{\x cm}{\y cm}}
}
\pgfplotstreamend
    \pgfpathclose
    \pgfusepath{stroke,fill}
    \pgfplothandlerlineto
\pgfplotstreamstart
\foreach \x/\y in {1.910/0.847,1.910/0.847,2.063/0.847,2.063/0.847}
{
\pgfplotstreampoint{\pgfpoint{\x cm}{\y cm}}
}
\pgfplotstreamend
    \pgfpathclose
    \pgfusepath{stroke,fill}
    \pgfplothandlerlineto
\pgfplotstreamstart
\foreach \x/\y in {2.101/0.847,2.101/0.847,2.253/0.847,2.253/0.847}
{
\pgfplotstreampoint{\pgfpoint{\x cm}{\y cm}}
}
\pgfplotstreamend
    \pgfpathclose
    \pgfusepath{stroke,fill}
    \pgfplothandlerlineto
\pgfplotstreamstart
\foreach \x/\y in {2.291/0.847,2.291/0.847,2.444/0.847,2.444/0.847}
{
\pgfplotstreampoint{\pgfpoint{\x cm}{\y cm}}
}
\pgfplotstreamend
    \pgfpathclose
    \pgfusepath{stroke,fill}
    \pgfplothandlerlineto
\pgfplotstreamstart
\foreach \x/\y in {2.482/0.847,2.482/0.847,2.634/0.847,2.634/0.847}
{
\pgfplotstreampoint{\pgfpoint{\x cm}{\y cm}}
}
\pgfplotstreamend
    \pgfpathclose
    \pgfusepath{stroke,fill}
    \pgfplothandlerlineto
\pgfplotstreamstart
\foreach \x/\y in {2.672/0.847,2.672/0.847,2.825/0.847,2.825/0.847}
{
\pgfplotstreampoint{\pgfpoint{\x cm}{\y cm}}
}
\pgfplotstreamend
    \pgfpathclose
    \pgfusepath{stroke,fill}
    \pgfplothandlerlineto
\pgfplotstreamstart
\foreach \x/\y in {2.863/0.847,2.863/0.847,3.015/0.847,3.015/0.847}
{
\pgfplotstreampoint{\pgfpoint{\x cm}{\y cm}}
}
\pgfplotstreamend
    \pgfpathclose
    \pgfusepath{stroke,fill}
  \end{pgfscope}
  \makeatletter\ifpgf@draftmode\makeatother\else
  \begin{pgfscope}
    \pgfpathrectangle{\pgfpoint{1.12889cm}{0.846667cm}}{\pgfpoint{1.90564cm}{2.08717cm}}
    \pgfusepath{clip}
    \begin{pgfscope}
      \pgfsetlinewidth{1.00pt}
      \definecolor{matfig2pgf_linecolor}{rgb}{0.000,0.000,0.000}
      \pgfsetstrokecolor{matfig2pgf_linecolor}
      \pgfsetdash{}{0pt}
      \pgfsetroundjoin
      \pgfplothandlerlineto
\pgfplotstreamstart
\foreach \x/\y in {1.129/0.847,3.035/0.847}
{
\pgfplotstreampoint{\pgfpoint{\x cm}{\y cm}}
}
\pgfplotstreamend
      \pgfusepath{stroke}
    \end{pgfscope}
  \end{pgfscope}
  \fi
    \pgftext[bottom,x=2.0677cm,y=3.06076cm,rotate=0]{t = 2}
  \makeatletter\ifpgf@draftmode\makeatother\pgftext[x=5cm,y=4.84375cm]{\Huge{DRAFT}}\fi

%% file: matlab_plots/measdist_t=4.pgf
  \begin{pgfscope}
    \definecolor{matfig2pgf_color}{rgb}{1,1,1}\pgfsetfillcolor{matfig2pgf_color}
    \pgfpathrectangle{\pgfpoint{1.12889cm}{0.846667cm}}{\pgfpoint{1.90564cm}{2.08717cm}}
    \pgfusepath{fill}
  \end{pgfscope}
  \begin{pgfscope}
    \pgfsetlinewidth{0.5pt}
    \foreach \x in {2.08171}
    {
      \pgfpathmoveto{\pgfpoint{\x cm}{0.846667cm}}\pgfpathlineto{\pgfpoint{\x cm}{0.867538cm}}
      \pgfpathmoveto{\pgfpoint{\x cm}{2.93384cm}}\pgfpathlineto{\pgfpoint{\x cm}{2.91297cm}}
    }
    \foreach \y in {1.89025}
    {
      \pgfpathmoveto{\pgfpoint{1.12889cm}{\y cm}}\pgfpathlineto{\pgfpoint{1.14795cm}{\y cm}}
      \pgfpathmoveto{\pgfpoint{3.03453cm}{\y cm}}\pgfpathlineto{\pgfpoint{3.01547cm}{\y cm}}
    }
    \pgfusepath{stroke}
  \end{pgfscope}
  \begin{pgfscope}
    \pgfsetlinewidth{0.5pt}
    \pgfpathrectangle{\pgfpoint{1.12889cm}{0.846667cm}}{\pgfpoint{1.90564cm}{2.08717cm}}
    \pgfusepath{stroke}
  \end{pgfscope}
  {\small
    \pgftext[x=1.12889cm,y=0.746667cm,top]{$0$}
    \pgftext[x=2.08171cm,y=0.746667cm,top]{$0.5$}
    \pgftext[x=3.03453cm,y=0.746667cm,top]{$1$}
    \pgftext[x=1.02889cm,y=0.846667cm,right]{$0$}
    \pgftext[x=1.02889cm,y=1.89025cm,right]{$0.2$}
    \pgftext[x=1.02889cm,y=2.93384cm,right]{$0.4$}
  }
  \begin{pgfscope}
    \pgfsetlinewidth{0.50pt}
    \definecolor{matfig2pgf_linecolor}{rgb}{0.000,0.000,0.000}
    \pgfsetstrokecolor{matfig2pgf_linecolor}
    \pgfsetdash{}{0pt}
    \definecolor{matfig2pgf_facecolor}{rgb}{102,108,97}\pgfsetfillcolor{matfig2pgf_facecolor}
    \definecolor{matfig2pgf_facecolor}{rgb}{0,0,0.5625}\pgfsetfillcolor{matfig2pgf_facecolor}
    \pgfplothandlerlineto
\pgfplotstreamstart
\foreach \x/\y in {1.148/0.847,1.148/2.777,1.300/2.777,1.300/0.847}
{
\pgfplotstreampoint{\pgfpoint{\x cm}{\y cm}}
}
\pgfplotstreamend
    \pgfpathclose
    \pgfusepath{stroke,fill}
    \pgfplothandlerlineto
\pgfplotstreamstart
\foreach \x/\y in {1.339/0.847,1.339/2.047,1.491/2.047,1.491/0.847}
{
\pgfplotstreampoint{\pgfpoint{\x cm}{\y cm}}
}
\pgfplotstreamend
    \pgfpathclose
    \pgfusepath{stroke,fill}
    \pgfplothandlerlineto
\pgfplotstreamstart
\foreach \x/\y in {1.529/0.847,1.529/1.838,1.682/1.838,1.682/0.847}
{
\pgfplotstreampoint{\pgfpoint{\x cm}{\y cm}}
}
\pgfplotstreamend
    \pgfpathclose
    \pgfusepath{stroke,fill}
    \pgfplothandlerlineto
\pgfplotstreamstart
\foreach \x/\y in {1.720/0.847,1.720/1.682,1.872/1.682,1.872/0.847}
{
\pgfplotstreampoint{\pgfpoint{\x cm}{\y cm}}
}
\pgfplotstreamend
    \pgfpathclose
    \pgfusepath{stroke,fill}
    \pgfplothandlerlineto
\pgfplotstreamstart
\foreach \x/\y in {1.910/0.847,1.910/1.055,2.063/1.055,2.063/0.847}
{
\pgfplotstreampoint{\pgfpoint{\x cm}{\y cm}}
}
\pgfplotstreamend
    \pgfpathclose
    \pgfusepath{stroke,fill}
    \pgfplothandlerlineto
\pgfplotstreamstart
\foreach \x/\y in {2.101/0.847,2.101/0.899,2.253/0.899,2.253/0.847}
{
\pgfplotstreampoint{\pgfpoint{\x cm}{\y cm}}
}
\pgfplotstreamend
    \pgfpathclose
    \pgfusepath{stroke,fill}
    \pgfplothandlerlineto
\pgfplotstreamstart
\foreach \x/\y in {2.291/0.847,2.291/0.847,2.444/0.847,2.444/0.847}
{
\pgfplotstreampoint{\pgfpoint{\x cm}{\y cm}}
}
\pgfplotstreamend
    \pgfpathclose
    \pgfusepath{stroke,fill}
    \pgfplothandlerlineto
\pgfplotstreamstart
\foreach \x/\y in {2.482/0.847,2.482/0.847,2.634/0.847,2.634/0.847}
{
\pgfplotstreampoint{\pgfpoint{\x cm}{\y cm}}
}
\pgfplotstreamend
    \pgfpathclose
    \pgfusepath{stroke,fill}
    \pgfplothandlerlineto
\pgfplotstreamstart
\foreach \x/\y in {2.672/0.847,2.672/0.847,2.825/0.847,2.825/0.847}
{
\pgfplotstreampoint{\pgfpoint{\x cm}{\y cm}}
}
\pgfplotstreamend
    \pgfpathclose
    \pgfusepath{stroke,fill}
    \pgfplothandlerlineto
\pgfplotstreamstart
\foreach \x/\y in {2.863/0.847,2.863/0.847,3.015/0.847,3.015/0.847}
{
\pgfplotstreampoint{\pgfpoint{\x cm}{\y cm}}
}
\pgfplotstreamend
    \pgfpathclose
    \pgfusepath{stroke,fill}
  \end{pgfscope}
  \makeatletter\ifpgf@draftmode\makeatother\else
  \begin{pgfscope}
    \pgfpathrectangle{\pgfpoint{1.12889cm}{0.846667cm}}{\pgfpoint{1.90564cm}{2.08717cm}}
    \pgfusepath{clip}
    \begin{pgfscope}
      \pgfsetlinewidth{1.00pt}
      \definecolor{matfig2pgf_linecolor}{rgb}{0.000,0.000,0.000}
      \pgfsetstrokecolor{matfig2pgf_linecolor}
      \pgfsetdash{}{0pt}
      \pgfsetroundjoin
      \pgfplothandlerlineto
\pgfplotstreamstart
\foreach \x/\y in {1.129/0.847,3.035/0.847}
{
\pgfplotstreampoint{\pgfpoint{\x cm}{\y cm}}
}
\pgfplotstreamend
      \pgfusepath{stroke}
    \end{pgfscope}
  \end{pgfscope}
  \fi
    \pgftext[bottom,x=2.0677cm,y=3.06076cm,rotate=0]{t = 4}
  \makeatletter\ifpgf@draftmode\makeatother\pgftext[x=5cm,y=4.84375cm]{\Huge{DRAFT}}\fi

%% file: matlab_plots/measdist_t=6.pgf
  \begin{pgfscope}
    \definecolor{matfig2pgf_color}{rgb}{1,1,1}\pgfsetfillcolor{matfig2pgf_color}
    \pgfpathrectangle{\pgfpoint{1.12889cm}{0.846667cm}}{\pgfpoint{1.90564cm}{2.08717cm}}
    \pgfusepath{fill}
  \end{pgfscope}
  \begin{pgfscope}
    \pgfsetlinewidth{0.5pt}
    \foreach \x in {2.08171}
    {
      \pgfpathmoveto{\pgfpoint{\x cm}{0.846667cm}}\pgfpathlineto{\pgfpoint{\x cm}{0.867538cm}}
      \pgfpathmoveto{\pgfpoint{\x cm}{2.93384cm}}\pgfpathlineto{\pgfpoint{\x cm}{2.91297cm}}
    }
    \foreach \y in {1.89025}
    {
      \pgfpathmoveto{\pgfpoint{1.12889cm}{\y cm}}\pgfpathlineto{\pgfpoint{1.14795cm}{\y cm}}
      \pgfpathmoveto{\pgfpoint{3.03453cm}{\y cm}}\pgfpathlineto{\pgfpoint{3.01547cm}{\y cm}}
    }
    \pgfusepath{stroke}
  \end{pgfscope}
  \begin{pgfscope}
    \pgfsetlinewidth{0.5pt}
    \pgfpathrectangle{\pgfpoint{1.12889cm}{0.846667cm}}{\pgfpoint{1.90564cm}{2.08717cm}}
    \pgfusepath{stroke}
  \end{pgfscope}
  {\small
    \pgftext[x=1.12889cm,y=0.746667cm,top]{$0$}
    \pgftext[x=2.08171cm,y=0.746667cm,top]{$0.5$}
    \pgftext[x=3.03453cm,y=0.746667cm,top]{$1$}
    \pgftext[x=1.02889cm,y=0.846667cm,right]{$0$}
    \pgftext[x=1.02889cm,y=1.89025cm,right]{$0.2$}
    \pgftext[x=1.02889cm,y=2.93384cm,right]{$0.4$}
  }
  \begin{pgfscope}
    \pgfsetlinewidth{0.50pt}
    \definecolor{matfig2pgf_linecolor}{rgb}{0.000,0.000,0.000}
    \pgfsetstrokecolor{matfig2pgf_linecolor}
    \pgfsetdash{}{0pt}
    \definecolor{matfig2pgf_facecolor}{rgb}{102,108,97}\pgfsetfillcolor{matfig2pgf_facecolor}
    \definecolor{matfig2pgf_facecolor}{rgb}{0,0,0.5625}\pgfsetfillcolor{matfig2pgf_facecolor}
    \pgfplothandlerlineto
\pgfplotstreamstart
\foreach \x/\y in {1.148/0.847,1.148/2.777,1.300/2.777,1.300/0.847}
{
\pgfplotstreampoint{\pgfpoint{\x cm}{\y cm}}
}
\pgfplotstreamend
    \pgfpathclose
    \pgfusepath{stroke,fill}
    \pgfplothandlerlineto
\pgfplotstreamstart
\foreach \x/\y in {1.339/0.847,1.339/1.264,1.491/1.264,1.491/0.847}
{
\pgfplotstreampoint{\pgfpoint{\x cm}{\y cm}}
}
\pgfplotstreamend
    \pgfpathclose
    \pgfusepath{stroke,fill}
    \pgfplothandlerlineto
\pgfplotstreamstart
\foreach \x/\y in {1.529/0.847,1.529/1.473,1.682/1.473,1.682/0.847}
{
\pgfplotstreampoint{\pgfpoint{\x cm}{\y cm}}
}
\pgfplotstreamend
    \pgfpathclose
    \pgfusepath{stroke,fill}
    \pgfplothandlerlineto
\pgfplotstreamstart
\foreach \x/\y in {1.720/0.847,1.720/1.421,1.872/1.421,1.872/0.847}
{
\pgfplotstreampoint{\pgfpoint{\x cm}{\y cm}}
}
\pgfplotstreamend
    \pgfpathclose
    \pgfusepath{stroke,fill}
    \pgfplothandlerlineto
\pgfplotstreamstart
\foreach \x/\y in {1.910/0.847,1.910/1.629,2.063/1.629,2.063/0.847}
{
\pgfplotstreampoint{\pgfpoint{\x cm}{\y cm}}
}
\pgfplotstreamend
    \pgfpathclose
    \pgfusepath{stroke,fill}
    \pgfplothandlerlineto
\pgfplotstreamstart
\foreach \x/\y in {2.101/0.847,2.101/1.473,2.253/1.473,2.253/0.847}
{
\pgfplotstreampoint{\pgfpoint{\x cm}{\y cm}}
}
\pgfplotstreamend
    \pgfpathclose
    \pgfusepath{stroke,fill}
    \pgfplothandlerlineto
\pgfplotstreamstart
\foreach \x/\y in {2.291/0.847,2.291/1.108,2.444/1.108,2.444/0.847}
{
\pgfplotstreampoint{\pgfpoint{\x cm}{\y cm}}
}
\pgfplotstreamend
    \pgfpathclose
    \pgfusepath{stroke,fill}
    \pgfplothandlerlineto
\pgfplotstreamstart
\foreach \x/\y in {2.482/0.847,2.482/0.847,2.634/0.847,2.634/0.847}
{
\pgfplotstreampoint{\pgfpoint{\x cm}{\y cm}}
}
\pgfplotstreamend
    \pgfpathclose
    \pgfusepath{stroke,fill}
    \pgfplothandlerlineto
\pgfplotstreamstart
\foreach \x/\y in {2.672/0.847,2.672/0.847,2.825/0.847,2.825/0.847}
{
\pgfplotstreampoint{\pgfpoint{\x cm}{\y cm}}
}
\pgfplotstreamend
    \pgfpathclose
    \pgfusepath{stroke,fill}
    \pgfplothandlerlineto
\pgfplotstreamstart
\foreach \x/\y in {2.863/0.847,2.863/0.847,3.015/0.847,3.015/0.847}
{
\pgfplotstreampoint{\pgfpoint{\x cm}{\y cm}}
}
\pgfplotstreamend
    \pgfpathclose
    \pgfusepath{stroke,fill}
  \end{pgfscope}
  \makeatletter\ifpgf@draftmode\makeatother\else
  \begin{pgfscope}
    \pgfpathrectangle{\pgfpoint{1.12889cm}{0.846667cm}}{\pgfpoint{1.90564cm}{2.08717cm}}
    \pgfusepath{clip}
    \begin{pgfscope}
      \pgfsetlinewidth{1.00pt}
      \definecolor{matfig2pgf_linecolor}{rgb}{0.000,0.000,0.000}
      \pgfsetstrokecolor{matfig2pgf_linecolor}
      \pgfsetdash{}{0pt}
      \pgfsetroundjoin
      \pgfplothandlerlineto
\pgfplotstreamstart
\foreach \x/\y in {1.129/0.847,3.035/0.847}
{
\pgfplotstreampoint{\pgfpoint{\x cm}{\y cm}}
}
\pgfplotstreamend
      \pgfusepath{stroke}
    \end{pgfscope}
  \end{pgfscope}
  \fi
    \pgftext[bottom,x=2.0677cm,y=3.06076cm,rotate=0]{t = 6}
  \makeatletter\ifpgf@draftmode\makeatother\pgftext[x=5cm,y=4.84375cm]{\Huge{DRAFT}}\fi

%% file: matlab_plots/measdist_t=10.pgf
  \begin{pgfscope}
    \definecolor{matfig2pgf_color}{rgb}{1,1,1}\pgfsetfillcolor{matfig2pgf_color}
    \pgfpathrectangle{\pgfpoint{1.12889cm}{0.846667cm}}{\pgfpoint{1.90564cm}{2.09264cm}}
    \pgfusepath{fill}
  \end{pgfscope}
  \begin{pgfscope}
    \pgfsetlinewidth{0.5pt}
    \foreach \x in {2.08171}
    {
      \pgfpathmoveto{\pgfpoint{\x cm}{0.846667cm}}\pgfpathlineto{\pgfpoint{\x cm}{0.867593cm}}
      \pgfpathmoveto{\pgfpoint{\x cm}{2.93931cm}}\pgfpathlineto{\pgfpoint{\x cm}{2.91838cm}}
    }
    \foreach \y in {1.89299}
    {
      \pgfpathmoveto{\pgfpoint{1.12889cm}{\y cm}}\pgfpathlineto{\pgfpoint{1.14795cm}{\y cm}}
      \pgfpathmoveto{\pgfpoint{3.03453cm}{\y cm}}\pgfpathlineto{\pgfpoint{3.01547cm}{\y cm}}
    }
    \pgfusepath{stroke}
  \end{pgfscope}
  \begin{pgfscope}
    \pgfsetlinewidth{0.5pt}
    \pgfpathrectangle{\pgfpoint{1.12889cm}{0.846667cm}}{\pgfpoint{1.90564cm}{2.09264cm}}
    \pgfusepath{stroke}
  \end{pgfscope}
  {\small
    \pgftext[x=1.12889cm,y=0.746667cm,top]{$0$}
    \pgftext[x=2.08171cm,y=0.746667cm,top]{$0.5$}
    \pgftext[x=3.03453cm,y=0.746667cm,top]{$1$}
    \pgftext[x=1.02889cm,y=0.846667cm,right]{$0$}
    \pgftext[x=1.02889cm,y=1.89299cm,right]{$0.2$}
    \pgftext[x=1.02889cm,y=2.93931cm,right]{$0.4$}
  }
  \begin{pgfscope}
    \pgfsetlinewidth{0.50pt}
    \definecolor{matfig2pgf_linecolor}{rgb}{0.000,0.000,0.000}
    \pgfsetstrokecolor{matfig2pgf_linecolor}
    \pgfsetdash{}{0pt}
    \definecolor{matfig2pgf_facecolor}{rgb}{102,108,97}\pgfsetfillcolor{matfig2pgf_facecolor}
    \definecolor{matfig2pgf_facecolor}{rgb}{0,0,0.5625}\pgfsetfillcolor{matfig2pgf_facecolor}
    \pgfplothandlerlineto
\pgfplotstreamstart
\foreach \x/\y in {1.148/0.847,1.148/2.835,1.300/2.835,1.300/0.847}
{
\pgfplotstreampoint{\pgfpoint{\x cm}{\y cm}}
}
\pgfplotstreamend
    \pgfpathclose
    \pgfusepath{stroke,fill}
    \pgfplothandlerlineto
\pgfplotstreamstart
\foreach \x/\y in {1.339/0.847,1.339/0.847,1.491/0.847,1.491/0.847}
{
\pgfplotstreampoint{\pgfpoint{\x cm}{\y cm}}
}
\pgfplotstreamend
    \pgfpathclose
    \pgfusepath{stroke,fill}
    \pgfplothandlerlineto
\pgfplotstreamstart
\foreach \x/\y in {1.529/0.847,1.529/0.951,1.682/0.951,1.682/0.847}
{
\pgfplotstreampoint{\pgfpoint{\x cm}{\y cm}}
}
\pgfplotstreamend
    \pgfpathclose
    \pgfusepath{stroke,fill}
    \pgfplothandlerlineto
\pgfplotstreamstart
\foreach \x/\y in {1.720/0.847,1.720/1.108,1.872/1.108,1.872/0.847}
{
\pgfplotstreampoint{\pgfpoint{\x cm}{\y cm}}
}
\pgfplotstreamend
    \pgfpathclose
    \pgfusepath{stroke,fill}
    \pgfplothandlerlineto
\pgfplotstreamstart
\foreach \x/\y in {1.910/0.847,1.910/1.422,2.063/1.422,2.063/0.847}
{
\pgfplotstreampoint{\pgfpoint{\x cm}{\y cm}}
}
\pgfplotstreamend
    \pgfpathclose
    \pgfusepath{stroke,fill}
    \pgfplothandlerlineto
\pgfplotstreamstart
\foreach \x/\y in {2.101/0.847,2.101/1.631,2.253/1.631,2.253/0.847}
{
\pgfplotstreampoint{\pgfpoint{\x cm}{\y cm}}
}
\pgfplotstreamend
    \pgfpathclose
    \pgfusepath{stroke,fill}
    \pgfplothandlerlineto
\pgfplotstreamstart
\foreach \x/\y in {2.291/0.847,2.291/1.841,2.444/1.841,2.444/0.847}
{
\pgfplotstreampoint{\pgfpoint{\x cm}{\y cm}}
}
\pgfplotstreamend
    \pgfpathclose
    \pgfusepath{stroke,fill}
    \pgfplothandlerlineto
\pgfplotstreamstart
\foreach \x/\y in {2.482/0.847,2.482/1.370,2.634/1.370,2.634/0.847}
{
\pgfplotstreampoint{\pgfpoint{\x cm}{\y cm}}
}
\pgfplotstreamend
    \pgfpathclose
    \pgfusepath{stroke,fill}
    \pgfplothandlerlineto
\pgfplotstreamstart
\foreach \x/\y in {2.672/0.847,2.672/0.847,2.825/0.847,2.825/0.847}
{
\pgfplotstreampoint{\pgfpoint{\x cm}{\y cm}}
}
\pgfplotstreamend
    \pgfpathclose
    \pgfusepath{stroke,fill}
    \pgfplothandlerlineto
\pgfplotstreamstart
\foreach \x/\y in {2.863/0.847,2.863/0.847,3.015/0.847,3.015/0.847}
{
\pgfplotstreampoint{\pgfpoint{\x cm}{\y cm}}
}
\pgfplotstreamend
    \pgfpathclose
    \pgfusepath{stroke,fill}
  \end{pgfscope}
  \makeatletter\ifpgf@draftmode\makeatother\else
  \begin{pgfscope}
    \pgfpathrectangle{\pgfpoint{1.12889cm}{0.846667cm}}{\pgfpoint{1.90564cm}{2.09264cm}}
    \pgfusepath{clip}
    \begin{pgfscope}
      \pgfsetlinewidth{1.00pt}
      \definecolor{matfig2pgf_linecolor}{rgb}{0.000,0.000,0.000}
      \pgfsetstrokecolor{matfig2pgf_linecolor}
      \pgfsetdash{}{0pt}
      \pgfsetroundjoin
      \pgfplothandlerlineto
\pgfplotstreamstart
\foreach \x/\y in {1.129/0.847,3.035/0.847}
{
\pgfplotstreampoint{\pgfpoint{\x cm}{\y cm}}
}
\pgfplotstreamend
      \pgfusepath{stroke}
    \end{pgfscope}
  \end{pgfscope}
  \fi
    \pgftext[bottom,x=2.0677cm,y=3.06656cm,rotate=0]{t = 10}
  \makeatletter\ifpgf@draftmode\makeatother\pgftext[x=5cm,y=4.82456cm]{\Huge{DRAFT}}\fi

%% file: matlab_plots/measdist_t=20.pgf
  \begin{pgfscope}
    \definecolor{matfig2pgf_color}{rgb}{1,1,1}\pgfsetfillcolor{matfig2pgf_color}
    \pgfpathrectangle{\pgfpoint{1.12889cm}{0.846667cm}}{\pgfpoint{1.90564cm}{2.08717cm}}
    \pgfusepath{fill}
  \end{pgfscope}
  \begin{pgfscope}
    \pgfsetlinewidth{0.5pt}
    \foreach \x in {2.08171}
    {
      \pgfpathmoveto{\pgfpoint{\x cm}{0.846667cm}}\pgfpathlineto{\pgfpoint{\x cm}{0.867538cm}}
      \pgfpathmoveto{\pgfpoint{\x cm}{2.93384cm}}\pgfpathlineto{\pgfpoint{\x cm}{2.91297cm}}
    }
    \foreach \y in {1.89025}
    {
      \pgfpathmoveto{\pgfpoint{1.12889cm}{\y cm}}\pgfpathlineto{\pgfpoint{1.14795cm}{\y cm}}
      \pgfpathmoveto{\pgfpoint{3.03453cm}{\y cm}}\pgfpathlineto{\pgfpoint{3.01547cm}{\y cm}}
    }
    \pgfusepath{stroke}
  \end{pgfscope}
  \begin{pgfscope}
    \pgfsetlinewidth{0.5pt}
    \pgfpathrectangle{\pgfpoint{1.12889cm}{0.846667cm}}{\pgfpoint{1.90564cm}{2.08717cm}}
    \pgfusepath{stroke}
  \end{pgfscope}
  {\small
    \pgftext[x=1.12889cm,y=0.746667cm,top]{$0$}
    \pgftext[x=2.08171cm,y=0.746667cm,top]{$0.5$}
    \pgftext[x=3.03453cm,y=0.746667cm,top]{$1$}
    \pgftext[x=1.02889cm,y=0.846667cm,right]{$0$}
    \pgftext[x=1.02889cm,y=1.89025cm,right]{$0.2$}
    \pgftext[x=1.02889cm,y=2.93384cm,right]{$0.4$}
  }
  \begin{pgfscope}
    \pgfsetlinewidth{0.50pt}
    \definecolor{matfig2pgf_linecolor}{rgb}{0.000,0.000,0.000}
    \pgfsetstrokecolor{matfig2pgf_linecolor}
    \pgfsetdash{}{0pt}
    \definecolor{matfig2pgf_facecolor}{rgb}{102,108,97}\pgfsetfillcolor{matfig2pgf_facecolor}
    \definecolor{matfig2pgf_facecolor}{rgb}{0,0,0.5625}\pgfsetfillcolor{matfig2pgf_facecolor}
    \pgfplothandlerlineto
\pgfplotstreamstart
\foreach \x/\y in {1.148/0.847,1.148/2.829,1.300/2.829,1.300/0.847}
{
\pgfplotstreampoint{\pgfpoint{\x cm}{\y cm}}
}
\pgfplotstreamend
    \pgfpathclose
    \pgfusepath{stroke,fill}
    \pgfplothandlerlineto
\pgfplotstreamstart
\foreach \x/\y in {1.339/0.847,1.339/0.847,1.491/0.847,1.491/0.847}
{
\pgfplotstreampoint{\pgfpoint{\x cm}{\y cm}}
}
\pgfplotstreamend
    \pgfpathclose
    \pgfusepath{stroke,fill}
    \pgfplothandlerlineto
\pgfplotstreamstart
\foreach \x/\y in {1.529/0.847,1.529/0.847,1.682/0.847,1.682/0.847}
{
\pgfplotstreampoint{\pgfpoint{\x cm}{\y cm}}
}
\pgfplotstreamend
    \pgfpathclose
    \pgfusepath{stroke,fill}
    \pgfplothandlerlineto
\pgfplotstreamstart
\foreach \x/\y in {1.720/0.847,1.720/0.847,1.872/0.847,1.872/0.847}
{
\pgfplotstreampoint{\pgfpoint{\x cm}{\y cm}}
}
\pgfplotstreamend
    \pgfpathclose
    \pgfusepath{stroke,fill}
    \pgfplothandlerlineto
\pgfplotstreamstart
\foreach \x/\y in {1.910/0.847,1.910/1.212,2.063/1.212,2.063/0.847}
{
\pgfplotstreampoint{\pgfpoint{\x cm}{\y cm}}
}
\pgfplotstreamend
    \pgfpathclose
    \pgfusepath{stroke,fill}
    \pgfplothandlerlineto
\pgfplotstreamstart
\foreach \x/\y in {2.101/0.847,2.101/1.525,2.253/1.525,2.253/0.847}
{
\pgfplotstreampoint{\pgfpoint{\x cm}{\y cm}}
}
\pgfplotstreamend
    \pgfpathclose
    \pgfusepath{stroke,fill}
    \pgfplothandlerlineto
\pgfplotstreamstart
\foreach \x/\y in {2.291/0.847,2.291/2.151,2.444/2.151,2.444/0.847}
{
\pgfplotstreampoint{\pgfpoint{\x cm}{\y cm}}
}
\pgfplotstreamend
    \pgfpathclose
    \pgfusepath{stroke,fill}
    \pgfplothandlerlineto
\pgfplotstreamstart
\foreach \x/\y in {2.482/0.847,2.482/1.734,2.634/1.734,2.634/0.847}
{
\pgfplotstreampoint{\pgfpoint{\x cm}{\y cm}}
}
\pgfplotstreamend
    \pgfpathclose
    \pgfusepath{stroke,fill}
    \pgfplothandlerlineto
\pgfplotstreamstart
\foreach \x/\y in {2.672/0.847,2.672/0.847,2.825/0.847,2.825/0.847}
{
\pgfplotstreampoint{\pgfpoint{\x cm}{\y cm}}
}
\pgfplotstreamend
    \pgfpathclose
    \pgfusepath{stroke,fill}
    \pgfplothandlerlineto
\pgfplotstreamstart
\foreach \x/\y in {2.863/0.847,2.863/0.847,3.015/0.847,3.015/0.847}
{
\pgfplotstreampoint{\pgfpoint{\x cm}{\y cm}}
}
\pgfplotstreamend
    \pgfpathclose
    \pgfusepath{stroke,fill}
  \end{pgfscope}
  \makeatletter\ifpgf@draftmode\makeatother\else
  \begin{pgfscope}
    \pgfpathrectangle{\pgfpoint{1.12889cm}{0.846667cm}}{\pgfpoint{1.90564cm}{2.08717cm}}
    \pgfusepath{clip}
    \begin{pgfscope}
      \pgfsetlinewidth{1.00pt}
      \definecolor{matfig2pgf_linecolor}{rgb}{0.000,0.000,0.000}
      \pgfsetstrokecolor{matfig2pgf_linecolor}
      \pgfsetdash{}{0pt}
      \pgfsetroundjoin
      \pgfplothandlerlineto
\pgfplotstreamstart
\foreach \x/\y in {1.129/0.847,3.035/0.847}
{
\pgfplotstreampoint{\pgfpoint{\x cm}{\y cm}}
}
\pgfplotstreamend
      \pgfusepath{stroke}
    \end{pgfscope}
  \end{pgfscope}
  \fi
    \pgftext[bottom,x=2.0677cm,y=3.06076cm,rotate=0]{t = 20}
  \makeatletter\ifpgf@draftmode\makeatother\pgftext[x=5cm,y=4.84375cm]{\Huge{DRAFT}}\fi

%% file: matlab_plots/pdf_a_1.pgf
  \begin{pgfscope}
    \definecolor{matfig2pgf_color}{rgb}{1,1,1}\pgfsetfillcolor{matfig2pgf_color}
    \pgfpathrectangle{\pgfpoint{1.12889cm}{0.846667cm}}{\pgfpoint{3.30406cm}{2.68166cm}}
    \pgfusepath{fill}
  \end{pgfscope}
  \begin{pgfscope}
    \pgfsetlinewidth{0.5pt}
    \foreach \x in {1.9549,2.78092,3.60694}
    {
      \pgfpathmoveto{\pgfpoint{\x cm}{0.846667cm}}\pgfpathlineto{\pgfpoint{\x cm}{0.873483cm}}
      \pgfpathmoveto{\pgfpoint{\x cm}{3.52832cm}}\pgfpathlineto{\pgfpoint{\x cm}{3.50151cm}}
    }
    \foreach \y in {1.51708,2.1875,2.85791}
    {
      \pgfpathmoveto{\pgfpoint{1.12889cm}{\y cm}}\pgfpathlineto{\pgfpoint{1.16193cm}{\y cm}}
      \pgfpathmoveto{\pgfpoint{4.43295cm}{\y cm}}\pgfpathlineto{\pgfpoint{4.39991cm}{\y cm}}
    }
    \pgfusepath{stroke}
  \end{pgfscope}
  \begin{pgfscope}
    \pgfsetlinewidth{0.5pt}
    \pgfpathrectangle{\pgfpoint{1.12889cm}{0.846667cm}}{\pgfpoint{3.30406cm}{2.68166cm}}
    \pgfusepath{stroke}
  \end{pgfscope}
  {\small
    \pgftext[x=1.12889cm,y=0.746667cm,top]{$0$}
    \pgftext[x=1.9549cm,y=0.746667cm,top]{$0.25$}
    \pgftext[x=2.78092cm,y=0.746667cm,top]{$0.5$}
    \pgftext[x=3.60694cm,y=0.746667cm,top]{$0.75$}
    \pgftext[x=4.43295cm,y=0.746667cm,top]{$1$}
    \pgftext[x=1.02889cm,y=0.846667cm,right]{$0$}
    \pgftext[x=1.02889cm,y=1.51708cm,right]{$2$}
    \pgftext[x=1.02889cm,y=2.1875cm,right]{$4$}
    \pgftext[x=1.02889cm,y=2.85791cm,right]{$6$}
    \pgftext[x=1.02889cm,y=3.52832cm,right]{$8$}
  }
  \makeatletter\ifpgf@draftmode\makeatother\else
  \begin{pgfscope}
    \pgfpathrectangle{\pgfpoint{1.12889cm}{0.846667cm}}{\pgfpoint{3.30406cm}{2.68166cm}}
    \pgfusepath{clip}
    \begin{pgfscope}
      \pgfsetlinewidth{0.10pt}
      \definecolor{matfig2pgf_linecolor}{rgb}{0.000,0.000,1.000}
      \pgfsetstrokecolor{matfig2pgf_linecolor}
      \pgfsetdash{}{0pt}
      \pgfsetroundjoin
      \pgfplothandlerlineto
\pgfplotstreamstart
\foreach \x/\y in {1.130/0.847,2.339/0.847,2.356/0.848,2.365/0.849,2.379/0.849,2.398/0.851,2.402/0.852,2.413/0.852,2.418/0.853,2.436/0.856,2.441/0.857,2.448/0.859,2.450/0.861,2.458/0.861,2.465/0.863,2.466/0.864,2.468/0.865,2.474/0.867,2.478/0.869,2.481/0.870,2.486/0.873,2.493/0.875,2.496/0.877,2.498/0.877,2.504/0.882,2.508/0.883,2.509/0.884,2.512/0.885,2.516/0.887,2.522/0.889,2.524/0.889,2.531/0.895,2.536/0.899,2.539/0.901,2.546/0.909,2.549/0.910,2.552/0.912,2.557/0.917,2.559/0.918,2.562/0.923,2.564/0.923,2.567/0.928,2.574/0.935,2.577/0.937,2.580/0.943,2.584/0.947,2.587/0.952,2.588/0.953,2.590/0.957,2.595/0.965,2.597/0.966,2.600/0.971,2.602/0.972,2.608/0.980,2.610/0.981,2.615/0.991,2.620/0.999,2.621/1.004,2.623/1.006,2.628/1.021,2.631/1.028,2.633/1.029,2.641/1.047,2.646/1.059,2.651/1.069,2.655/1.077,2.659/1.092,2.661/1.098,2.664/1.107,2.666/1.109,2.671/1.128,2.674/1.135,2.683/1.158,2.686/1.171,2.693/1.186,2.694/1.194,2.696/1.197,2.707/1.233,2.712/1.250,2.719/1.270,2.726/1.293,2.732/1.322,2.735/1.334,2.745/1.376,2.747/1.383,2.752/1.409,2.760/1.446,2.767/1.468,2.772/1.492,2.778/1.519,2.785/1.548,2.792/1.586,2.793/1.591,2.798/1.620,2.802/1.635,2.810/1.678,2.820/1.727,2.825/1.754,2.828/1.775,2.835/1.802,2.840/1.828,2.848/1.877,2.851/1.900,2.858/1.939,2.868/1.992,2.873/2.024,2.883/2.082,2.886/2.096,2.894/2.140,2.899/2.170,2.902/2.184,2.904/2.196,2.907/2.212,2.912/2.237,2.919/2.268,2.921/2.280,2.924/2.295,2.925/2.304,2.929/2.314,2.935/2.353,2.942/2.380,2.945/2.401,2.950/2.421,2.952/2.431,2.955/2.444,2.960/2.458,2.965/2.483,2.970/2.502,2.975/2.525,2.978/2.538,2.983/2.559,2.988/2.570,2.995/2.590,2.997/2.592,3.000/2.607,3.001/2.610,3.003/2.618,3.005/2.620,3.008/2.631,3.010/2.637,3.013/2.645,3.015/2.651,3.018/2.653,3.020/2.661,3.025/2.677,3.026/2.679,3.030/2.693,3.031/2.697,3.033/2.697,3.036/2.711,3.039/2.721,3.044/2.729,3.048/2.738,3.054/2.749,3.056/2.749,3.058/2.752,3.063/2.755,3.066/2.755,3.068/2.757,3.069/2.757,3.071/2.755,3.073/2.754,3.074/2.756,3.076/2.756,3.077/2.758,3.079/2.757,3.081/2.753,3.084/2.752,3.086/2.752,3.089/2.756,3.092/2.755,3.099/2.767,3.101/2.767,3.104/2.774,3.106/2.776,3.107/2.774,3.109/2.773,3.111/2.770,3.114/2.770,3.117/2.767,3.119/2.763,3.120/2.761,3.122/2.757,3.124/2.755,3.125/2.758,3.127/2.751,3.129/2.747,3.130/2.746,3.135/2.733,3.137/2.732,3.139/2.729,3.140/2.727,3.142/2.724,3.144/2.723,3.145/2.721,3.147/2.717,3.152/2.709,3.153/2.700,3.157/2.695,3.158/2.688,3.160/2.684,3.162/2.682,3.163/2.683,3.165/2.683,3.167/2.680,3.168/2.675,3.170/2.673,3.172/2.669,3.175/2.655,3.182/2.630,3.185/2.625,3.186/2.614,3.190/2.600,3.191/2.596,3.193/2.588,3.195/2.586,3.200/2.571,3.201/2.561,3.203/2.557,3.208/2.534,3.215/2.509,3.216/2.506,3.224/2.481,3.231/2.449,3.234/2.438,3.236/2.427,3.238/2.421,3.239/2.417,3.244/2.396,3.251/2.373,3.258/2.338,3.261/2.329,3.264/2.317,3.267/2.299,3.274/2.280,3.277/2.258,3.284/2.227,3.291/2.202,3.302/2.147,3.307/2.119,3.309/2.113,3.314/2.088,3.315/2.077,3.320/2.058,3.322/2.050,3.324/2.045,3.329/2.024,3.332/2.005,3.334/1.999,3.335/1.985,3.338/1.974,3.342/1.954,3.345/1.940,3.347/1.930,3.350/1.919,3.360/1.870,3.362/1.861,3.365/1.851,3.372/1.813,3.383/1.765,3.386/1.750,3.388/1.746,3.393/1.725,3.396/1.710,3.398/1.706,3.401/1.693,3.403/1.689,3.410/1.662,3.414/1.648,3.419/1.624,3.421/1.614,3.423/1.610,3.426/1.597,3.433/1.564,3.436/1.542,3.441/1.524,3.443/1.521,3.448/1.500,3.451/1.492,3.454/1.481,3.457/1.464,3.459/1.461,3.462/1.447,3.464/1.445,3.477/1.398,3.484/1.376,3.486/1.375,3.494/1.347,3.499/1.327,3.507/1.303,3.512/1.293,3.517/1.275,3.519/1.274,3.522/1.268,3.525/1.261,3.528/1.247,3.533/1.235,3.540/1.222,3.543/1.211,3.550/1.195,3.552/1.188,3.553/1.184,3.557/1.174,3.562/1.163,3.568/1.151,3.570/1.149,3.571/1.143,3.573/1.140,3.580/1.125,3.581/1.119,3.588/1.106,3.590/1.104,3.593/1.099,3.595/1.092,3.598/1.087,3.603/1.077,3.606/1.070,3.611/1.062,3.613/1.061,3.616/1.055,3.618/1.051,3.621/1.044,3.623/1.044,3.629/1.032,3.631/1.031,3.634/1.025,3.637/1.022,3.639/1.020,3.642/1.014,3.644/1.012,3.647/1.011,3.654/1.004,3.656/0.999,3.659/0.994,3.664/0.990,3.666/0.986,3.667/0.986,3.674/0.980,3.675/0.978,3.677/0.976,3.679/0.976,3.680/0.974,3.682/0.973,3.684/0.970,3.685/0.969,3.689/0.966,3.690/0.963,3.692/0.962,3.697/0.955,3.699/0.953,3.700/0.952,3.702/0.949,3.704/0.949,3.710/0.942,3.712/0.940,3.713/0.939,3.715/0.937,3.718/0.935,3.720/0.931,3.725/0.928,3.728/0.928,3.733/0.923,3.737/0.920,3.738/0.920,3.743/0.916,3.745/0.914,3.753/0.910,3.755/0.908,3.756/0.907,3.758/0.906,3.765/0.903,3.766/0.902,3.775/0.899,3.778/0.896,3.781/0.895,3.783/0.893,3.786/0.891,3.791/0.890,3.794/0.890,3.803/0.887,3.808/0.886,3.811/0.885,3.813/0.884,3.816/0.882,3.826/0.879,3.832/0.877,3.834/0.876,3.841/0.875,3.844/0.874,3.851/0.870,3.862/0.867,3.865/0.866,3.867/0.865,3.874/0.864,3.875/0.864,3.880/0.862,3.887/0.862,3.890/0.862,3.900/0.861,3.903/0.861,3.907/0.860,3.920/0.860,3.927/0.860,3.930/0.859,3.932/0.858,3.943/0.857,3.946/0.857,3.950/0.856,3.953/0.856,3.956/0.855,3.966/0.854,3.970/0.853,3.978/0.853,3.983/0.852,4.013/0.852,4.026/0.851,4.069/0.850,4.089/0.849,4.102/0.849,4.275/0.847,4.432/0.847}
{
\pgfplotstreampoint{\pgfpoint{\x cm}{\y cm}}
}
\pgfplotstreamend
      \pgfusepath{stroke}
    \end{pgfscope}
  \end{pgfscope}
  \fi
  \makeatletter\ifpgf@draftmode\makeatother\else
  \begin{pgfscope}
    \pgfpathrectangle{\pgfpoint{1.12889cm}{0.846667cm}}{\pgfpoint{3.30406cm}{2.68166cm}}
    \pgfusepath{clip}
    \begin{pgfscope}
      \pgfsetlinewidth{0.10pt}
      \definecolor{matfig2pgf_linecolor}{rgb}{1.000,0.000,0.000}
      \pgfsetstrokecolor{matfig2pgf_linecolor}
      \pgfsetdash{{2.00pt}{2.00pt}}{0pt}
      \pgfsetroundjoin
      \pgfplothandlerlineto
\pgfplotstreamstart
\foreach \x/\y in {1.130/0.855,1.151/0.854,1.154/0.860,1.235/0.856,1.237/0.859,1.306/0.857,1.338/0.857,1.340/0.855,1.371/0.855,1.373/0.852,1.434/0.852,1.437/0.854,1.444/0.855,1.483/0.855,1.487/0.852,1.492/0.852,1.493/0.855,1.531/0.855,1.533/0.858,1.534/0.858,1.536/0.860,1.563/0.860,1.566/0.863,1.568/0.865,1.569/0.862,1.766/0.862,1.769/0.860,1.776/0.859,1.824/0.859,1.825/0.856,1.838/0.856,1.840/0.857,1.863/0.857,1.865/0.854,1.867/0.854,1.868/0.852,1.895/0.852,1.898/0.850,1.900/0.848,1.911/0.848,1.913/0.849,1.916/0.856,1.985/0.856,1.990/0.860,1.992/0.861,1.994/0.865,1.995/0.866,2.005/0.866,2.007/0.868,2.055/0.868,2.057/0.869,2.121/0.869,2.123/0.870,2.146/0.870,2.149/0.873,2.151/0.873,2.152/0.874,2.156/0.874,2.159/0.877,2.161/0.884,2.166/0.890,2.170/0.890,2.172/0.888,2.175/0.888,2.179/0.891,2.199/0.891,2.200/0.894,2.202/0.895,2.204/0.895,2.205/0.897,2.212/0.897,2.215/0.899,2.220/0.913,2.222/0.916,2.227/0.916,2.228/0.920,2.230/0.920,2.232/0.921,2.243/0.921,2.245/0.920,2.246/0.916,2.248/0.915,2.251/0.921,2.255/0.925,2.256/0.927,2.270/0.927,2.271/0.929,2.293/0.929,2.294/0.931,2.318/0.931,2.321/0.928,2.322/0.928,2.324/0.929,2.326/0.928,2.337/0.928,2.339/0.926,2.352/0.926,2.354/0.929,2.356/0.934,2.357/0.937,2.360/0.940,2.367/0.940,2.370/0.943,2.380/0.943,2.384/0.948,2.385/0.948,2.387/0.953,2.392/0.978,2.394/0.984,2.395/0.986,2.417/0.986,2.420/0.990,2.422/0.990,2.423/0.991,2.425/0.994,2.427/1.001,2.430/1.029,2.432/1.038,2.433/1.043,2.435/1.045,2.453/1.045,2.455/1.044,2.458/1.047,2.460/1.047,2.463/1.050,2.478/1.050,2.481/1.046,2.483/1.046,2.484/1.045,2.488/1.045,2.491/1.042,2.493/1.035,2.494/1.033,2.496/1.033,2.498/1.031,2.499/1.032,2.508/1.032,2.511/1.030,2.514/1.030,2.516/1.031,2.517/1.033,2.521/1.048,2.524/1.054,2.531/1.054,2.532/1.051,2.534/1.050,2.536/1.050,2.537/1.048,2.539/1.048,2.541/1.050,2.544/1.058,2.547/1.070,2.549/1.073,2.550/1.072,2.554/1.066,2.555/1.066,2.557/1.067,2.560/1.075,2.564/1.107,2.565/1.118,2.567/1.122,2.569/1.122,2.570/1.123,2.572/1.126,2.574/1.126,2.575/1.130,2.579/1.130,2.580/1.128,2.584/1.127,2.585/1.133,2.588/1.140,2.590/1.147,2.595/1.153,2.600/1.153,2.602/1.156,2.603/1.175,2.605/1.210,2.608/1.320,2.610/1.352,2.612/1.365,2.615/1.370,2.617/1.370,2.618/1.372,2.625/1.372,2.626/1.371,2.628/1.371,2.630/1.372,2.631/1.376,2.635/1.416,2.638/1.475,2.640/1.481,2.641/1.484,2.643/1.484,2.645/1.485,2.653/1.485,2.655/1.484,2.658/1.484,2.661/1.488,2.664/1.488,2.668/1.492,2.671/1.513,2.673/1.528,2.674/1.538,2.678/1.547,2.679/1.549,2.684/1.549,2.686/1.546,2.688/1.541,2.689/1.538,2.693/1.535,2.694/1.537,2.696/1.541,2.699/1.556,2.701/1.557,2.702/1.555,2.712/1.555,2.716/1.550,2.717/1.550,2.719/1.545,2.722/1.527,2.724/1.521,2.727/1.516,2.729/1.517,2.731/1.521,2.734/1.521,2.737/1.523,2.739/1.528,2.742/1.546,2.744/1.552,2.745/1.553,2.749/1.553,2.750/1.551,2.752/1.552,2.754/1.553,2.755/1.552,2.757/1.549,2.759/1.542,2.762/1.514,2.764/1.505,2.765/1.502,2.767/1.501,2.769/1.504,2.772/1.506,2.778/1.506,2.782/1.510,2.783/1.510,2.785/1.511,2.787/1.511,2.790/1.508,2.792/1.508,2.795/1.506,2.800/1.506,2.803/1.509,2.808/1.509,2.810/1.511,2.813/1.518,2.826/1.518,2.828/1.516,2.830/1.516,2.831/1.515,2.840/1.515,2.841/1.516,2.843/1.523,2.845/1.534,2.849/1.600,2.851/1.606,2.856/1.637,2.858/1.641,2.859/1.641,2.861/1.645,2.863/1.651,2.864/1.674,2.869/1.845,2.871/1.913,2.873/1.946,2.874/1.953,2.876/1.952,2.881/1.929,2.883/1.925,2.884/1.924,2.886/1.924,2.887/1.925,2.891/1.922,2.894/1.905,2.897/1.884,2.899/1.880,2.901/1.880,2.902/1.878,2.904/1.875,2.906/1.875,2.907/1.873,2.911/1.879,2.912/1.879,2.914/1.877,2.916/1.874,2.917/1.868,2.919/1.865,2.921/1.871,2.925/1.918,2.927/1.920,2.929/1.921,2.932/1.921,2.934/1.919,2.935/1.898,2.937/1.863,2.940/1.753,2.942/1.721,2.944/1.710,2.945/1.707,2.949/1.711,2.950/1.716,2.952/1.738,2.954/1.792,2.957/1.961,2.959/2.014,2.960/2.029,2.962/2.033,2.963/2.029,2.967/1.994,2.970/1.940,2.972/1.940,2.977/1.971,2.978/1.973,2.987/1.973,2.988/1.976,2.990/1.982,2.992/1.984,2.997/1.984,3.000/1.980,3.003/1.958,3.005/1.943,3.006/1.933,3.010/1.924,3.011/1.922,3.013/1.924,3.015/1.932,3.016/1.951,3.021/2.075,3.023/2.092,3.025/2.095,3.028/2.099,3.030/2.095,3.031/2.097,3.033/2.103,3.035/2.104,3.044/2.104,3.049/2.110,3.054/2.110,3.056/2.108,3.058/2.105,3.059/2.105,3.061/2.104,3.063/2.100,3.064/2.100,3.069/2.112,3.071/2.112,3.073/2.105,3.076/2.095,3.077/2.094,3.081/2.097,3.086/2.106,3.089/2.116,3.091/2.120,3.094/2.123,3.096/2.126,3.097/2.128,3.099/2.149,3.101/2.201,3.106/2.489,3.107/2.526,3.109/2.536,3.111/2.538,3.114/2.535,3.115/2.535,3.117/2.533,3.120/2.533,3.122/2.534,3.127/2.534,3.129/2.535,3.130/2.538,3.135/2.555,3.137/2.555,3.140/2.558,3.142/2.557,3.145/2.550,3.172/2.550,3.173/2.549,3.175/2.542,3.177/2.531,3.182/2.464,3.188/2.409,3.190/2.405,3.191/2.405,3.193/2.401,3.195/2.395,3.196/2.372,3.201/2.201,3.203/2.133,3.205/2.101,3.206/2.090,3.208/2.087,3.210/2.089,3.211/2.093,3.213/2.099,3.216/2.104,3.220/2.105,3.221/2.105,3.224/2.103,3.228/2.092,3.229/2.090,3.239/2.091,3.243/2.086,3.244/2.086,3.246/2.088,3.248/2.092,3.249/2.091,3.251/2.088,3.253/2.076,3.258/2.019,3.259/2.014,3.261/2.013,3.266/2.013,3.269/2.019,3.272/2.040,3.276/2.051,3.277/2.051,3.279/2.049,3.281/2.048,3.282/2.040,3.284/2.018,3.286/1.966,3.289/1.801,3.291/1.751,3.292/1.739,3.294/1.736,3.296/1.737,3.299/1.736,3.300/1.736,3.302/1.733,3.304/1.726,3.309/1.694,3.310/1.700,3.312/1.715,3.314/1.742,3.317/1.829,3.319/1.855,3.320/1.860,3.324/1.850,3.325/1.849,3.343/1.849,3.345/1.849,3.347/1.841,3.348/1.822,3.353/1.698,3.355/1.682,3.357/1.679,3.358/1.677,3.362/1.671,3.365/1.656,3.367/1.655,3.368/1.656,3.372/1.656,3.375/1.660,3.376/1.663,3.378/1.663,3.381/1.659,3.386/1.659,3.390/1.663,3.393/1.672,3.395/1.675,3.396/1.676,3.398/1.673,3.401/1.661,3.406/1.653,3.413/1.653,3.418/1.640,3.421/1.630,3.423/1.628,3.428/1.635,3.429/1.652,3.431/1.650,3.433/1.604,3.438/1.317,3.439/1.280,3.441/1.271,3.443/1.271,3.444/1.276,3.451/1.276,3.452/1.277,3.456/1.277,3.457/1.281,3.459/1.282,3.461/1.282,3.462/1.280,3.464/1.276,3.466/1.268,3.467/1.264,3.469/1.272,3.471/1.275,3.472/1.274,3.477/1.278,3.481/1.278,3.482/1.280,3.486/1.280,3.489/1.283,3.495/1.283,3.499/1.286,3.500/1.286,3.502/1.293,3.504/1.310,3.505/1.318,3.510/1.322,3.512/1.320,3.514/1.320,3.515/1.317,3.517/1.316,3.520/1.316,3.525/1.320,3.527/1.320,3.528/1.321,3.533/1.321,3.535/1.322,3.537/1.319,3.538/1.319,3.540/1.321,3.542/1.320,3.543/1.318,3.545/1.312,3.548/1.309,3.550/1.312,3.552/1.323,3.555/1.359,3.557/1.371,3.570/1.371,3.571/1.369,3.573/1.371,3.576/1.383,3.578/1.383,3.580/1.379,3.590/1.379,3.591/1.380,3.598/1.380,3.600/1.377,3.601/1.376,3.604/1.356,3.608/1.343,3.609/1.343,3.613/1.340,3.616/1.340,3.619/1.337,3.624/1.328,3.631/1.322,3.633/1.323,3.634/1.323,3.637/1.335,3.639/1.337,3.641/1.337,3.642/1.330,3.644/1.316,3.646/1.290,3.649/1.203,3.651/1.176,3.652/1.166,3.654/1.163,3.656/1.165,3.667/1.165,3.669/1.166,3.674/1.183,3.675/1.183,3.677/1.180,3.689/1.180,3.694/1.173,3.695/1.173,3.697/1.172,3.699/1.173,3.700/1.173,3.702/1.175,3.704/1.177,3.705/1.178,3.707/1.176,3.709/1.172,3.710/1.171,3.712/1.172,3.713/1.175,3.715/1.176,3.718/1.176,3.722/1.173,3.723/1.170,3.725/1.170,3.727/1.167,3.728/1.166,3.733/1.166,3.737/1.162,3.742/1.162,3.745/1.159,3.750/1.159,3.751/1.161,3.753/1.164,3.755/1.164,3.758/1.156,3.760/1.150,3.763/1.109,3.765/1.101,3.766/1.098,3.768/1.098,3.771/1.105,3.773/1.103,3.776/1.099,3.778/1.099,3.780/1.100,3.781/1.104,3.783/1.104,3.786/1.101,3.788/1.101,3.789/1.098,3.794/1.094,3.796/1.093,3.798/1.093,3.799/1.089,3.801/1.081,3.803/1.079,3.804/1.079,3.809/1.074,3.813/1.074,3.814/1.073,3.816/1.073,3.818/1.074,3.819/1.074,3.821/1.075,3.827/1.075,3.831/1.072,3.832/1.072,3.834/1.065,3.836/1.048,3.837/1.041,3.839/1.041,3.842/1.039,3.852/1.039,3.857/1.035,3.859/1.035,3.861/1.034,3.865/1.034,3.867/1.033,3.869/1.033,3.870/1.034,3.875/1.029,3.879/1.029,3.880/1.028,3.882/1.023,3.889/0.965,3.892/0.965,3.894/0.968,3.903/0.968,3.905/0.964,3.908/0.952,3.910/0.950,3.912/0.953,3.922/0.953,3.923/0.951,3.932/0.951,3.935/0.948,3.956/0.948,3.960/0.956,3.963/0.956,3.965/0.953,3.966/0.953,3.970/0.942,3.971/0.938,3.976/0.933,3.979/0.933,3.981/0.935,3.984/0.935,3.986/0.936,3.988/0.935,3.991/0.935,3.993/0.936,3.999/0.936,4.001/0.934,4.006/0.917,4.029/0.917,4.034/0.913,4.037/0.908,4.039/0.908,4.041/0.910,4.042/0.910,4.044/0.909,4.046/0.906,4.047/0.905,4.054/0.905,4.055/0.903,4.057/0.899,4.059/0.900,4.064/0.900,4.065/0.901,4.082/0.901,4.084/0.899,4.085/0.897,4.087/0.895,4.100/0.895,4.103/0.888,4.107/0.888,4.108/0.885,4.110/0.885,4.112/0.884,4.113/0.881,4.133/0.881,4.135/0.883,4.136/0.889,4.148/0.889,4.151/0.887,4.153/0.884,4.155/0.888,4.168/0.888,4.171/0.885,4.178/0.885,4.179/0.887,4.193/0.887,4.194/0.888,4.201/0.888,4.202/0.887,4.216/0.887,4.217/0.886,4.224/0.886,4.226/0.883,4.242/0.883,4.244/0.880,4.255/0.880,4.257/0.881,4.272/0.882,4.288/0.884,4.292/0.875,4.312/0.877,4.313/0.875,4.316/0.876,4.318/0.875,4.323/0.875,4.325/0.874,4.371/0.880,4.373/0.877,4.389/0.879,4.391/0.878,4.396/0.879,4.397/0.877,4.432/0.883}
{
\pgfplotstreampoint{\pgfpoint{\x cm}{\y cm}}
}
\pgfplotstreamend
      \pgfusepath{stroke}
    \end{pgfscope}
  \end{pgfscope}
  \fi
    \pgftext[top,x=2.75268cm,y=0.267993cm,rotate=0]{$$$a_1$$$}
    \pgftext[bottom,x=0.691171cm,y=2.15927cm,rotate=90]{$$$\varphi_{1}$$$}
  \makeatletter\ifpgf@draftmode\makeatother\pgftext[x=5cm,y=4.02174cm]{\Huge{DRAFT}}\fi

%% file: matlab_plots/pdf_a_4.pgf
  \begin{pgfscope}
    \definecolor{matfig2pgf_color}{rgb}{1,1,1}\pgfsetfillcolor{matfig2pgf_color}
    \pgfpathrectangle{\pgfpoint{1.12889cm}{0.846667cm}}{\pgfpoint{3.30406cm}{2.68166cm}}
    \pgfusepath{fill}
  \end{pgfscope}
  \begin{pgfscope}
    \pgfsetlinewidth{0.5pt}
    \foreach \x in {1.9549,2.78092,3.60694}
    {
      \pgfpathmoveto{\pgfpoint{\x cm}{0.846667cm}}\pgfpathlineto{\pgfpoint{\x cm}{0.873483cm}}
      \pgfpathmoveto{\pgfpoint{\x cm}{3.52832cm}}\pgfpathlineto{\pgfpoint{\x cm}{3.50151cm}}
    }
    \foreach \y in {1.51708,2.1875,2.85791}
    {
      \pgfpathmoveto{\pgfpoint{1.12889cm}{\y cm}}\pgfpathlineto{\pgfpoint{1.16193cm}{\y cm}}
      \pgfpathmoveto{\pgfpoint{4.43295cm}{\y cm}}\pgfpathlineto{\pgfpoint{4.39991cm}{\y cm}}
    }
    \pgfusepath{stroke}
  \end{pgfscope}
  \begin{pgfscope}
    \pgfsetlinewidth{0.5pt}
    \pgfpathrectangle{\pgfpoint{1.12889cm}{0.846667cm}}{\pgfpoint{3.30406cm}{2.68166cm}}
    \pgfusepath{stroke}
  \end{pgfscope}
  {\small
    \pgftext[x=1.12889cm,y=0.746667cm,top]{$0$}
    \pgftext[x=1.9549cm,y=0.746667cm,top]{$0.25$}
    \pgftext[x=2.78092cm,y=0.746667cm,top]{$0.5$}
    \pgftext[x=3.60694cm,y=0.746667cm,top]{$0.75$}
    \pgftext[x=4.43295cm,y=0.746667cm,top]{$1$}
    \pgftext[x=1.02889cm,y=0.846667cm,right]{$0$}
    \pgftext[x=1.02889cm,y=1.51708cm,right]{$2$}
    \pgftext[x=1.02889cm,y=2.1875cm,right]{$4$}
    \pgftext[x=1.02889cm,y=2.85791cm,right]{$6$}
    \pgftext[x=1.02889cm,y=3.52832cm,right]{$8$}
  }
  \makeatletter\ifpgf@draftmode\makeatother\else
  \begin{pgfscope}
    \pgfpathrectangle{\pgfpoint{1.12889cm}{0.846667cm}}{\pgfpoint{3.30406cm}{2.68166cm}}
    \pgfusepath{clip}
    \begin{pgfscope}
      \pgfsetlinewidth{0.10pt}
      \definecolor{matfig2pgf_linecolor}{rgb}{0.000,0.000,1.000}
      \pgfsetstrokecolor{matfig2pgf_linecolor}
      \pgfsetdash{}{0pt}
      \pgfsetroundjoin
      \pgfplothandlerlineto
\pgfplotstreamstart
\foreach \x/\y in {1.130/0.847,1.721/0.847,1.753/0.848,1.764/0.849,1.786/0.850,1.795/0.851,1.802/0.852,1.815/0.852,1.817/0.853,1.830/0.853,1.833/0.854,1.847/0.854,1.858/0.855,1.860/0.855,1.862/0.856,1.870/0.857,1.885/0.860,1.886/0.860,1.888/0.861,1.905/0.861,1.909/0.863,1.911/0.863,1.913/0.864,1.918/0.864,1.926/0.867,1.931/0.870,1.938/0.871,1.943/0.872,1.949/0.875,1.956/0.876,1.962/0.880,1.967/0.885,1.971/0.887,1.972/0.888,1.974/0.890,1.981/0.892,1.987/0.896,1.994/0.899,1.995/0.900,1.999/0.905,2.004/0.908,2.007/0.910,2.010/0.912,2.017/0.918,2.019/0.919,2.023/0.925,2.027/0.929,2.028/0.929,2.030/0.931,2.033/0.933,2.037/0.935,2.038/0.936,2.040/0.938,2.043/0.941,2.050/0.945,2.053/0.949,2.057/0.951,2.061/0.955,2.066/0.958,2.068/0.960,2.070/0.961,2.075/0.968,2.076/0.971,2.083/0.977,2.086/0.979,2.096/0.992,2.098/0.993,2.099/0.996,2.103/0.998,2.108/1.007,2.111/1.011,2.116/1.019,2.118/1.021,2.123/1.029,2.124/1.030,2.126/1.033,2.131/1.039,2.132/1.041,2.134/1.043,2.136/1.046,2.137/1.048,2.142/1.057,2.144/1.058,2.147/1.064,2.149/1.064,2.157/1.074,2.161/1.077,2.166/1.088,2.167/1.090,2.172/1.101,2.175/1.106,2.177/1.110,2.179/1.111,2.180/1.115,2.182/1.117,2.189/1.131,2.194/1.137,2.195/1.141,2.197/1.143,2.199/1.147,2.200/1.149,2.204/1.154,2.207/1.162,2.208/1.165,2.210/1.169,2.212/1.171,2.217/1.182,2.218/1.184,2.223/1.196,2.228/1.205,2.233/1.217,2.235/1.223,2.237/1.225,2.238/1.232,2.243/1.242,2.245/1.247,2.250/1.258,2.255/1.264,2.258/1.271,2.266/1.288,2.268/1.293,2.270/1.295,2.275/1.310,2.278/1.317,2.281/1.325,2.283/1.328,2.286/1.336,2.288/1.342,2.293/1.347,2.296/1.355,2.299/1.361,2.304/1.374,2.306/1.374,2.309/1.380,2.314/1.397,2.318/1.403,2.319/1.404,2.321/1.409,2.322/1.411,2.329/1.434,2.337/1.455,2.341/1.466,2.346/1.478,2.352/1.501,2.354/1.505,2.356/1.511,2.364/1.525,2.365/1.529,2.367/1.532,2.372/1.545,2.380/1.561,2.382/1.566,2.384/1.567,2.387/1.574,2.394/1.598,2.397/1.609,2.400/1.616,2.402/1.624,2.407/1.632,2.408/1.632,2.415/1.647,2.418/1.659,2.423/1.672,2.427/1.676,2.430/1.687,2.432/1.690,2.435/1.700,2.436/1.702,2.438/1.703,2.441/1.711,2.443/1.717,2.450/1.730,2.455/1.742,2.460/1.761,2.465/1.770,2.471/1.790,2.473/1.793,2.479/1.812,2.481/1.818,2.483/1.822,2.484/1.823,2.489/1.832,2.496/1.843,2.498/1.843,2.503/1.855,2.504/1.857,2.508/1.865,2.509/1.867,2.516/1.882,2.517/1.885,2.519/1.890,2.521/1.889,2.522/1.896,2.529/1.911,2.531/1.912,2.534/1.922,2.536/1.929,2.539/1.932,2.541/1.937,2.542/1.937,2.544/1.944,2.546/1.947,2.549/1.951,2.552/1.960,2.554/1.962,2.557/1.971,2.559/1.974,2.565/1.981,2.567/1.980,2.569/1.982,2.570/1.982,2.572/1.983,2.575/1.988,2.577/1.988,2.582/1.998,2.587/2.004,2.588/2.003,2.590/2.006,2.592/2.005,2.593/2.003,2.595/2.002,2.597/2.002,2.600/2.001,2.602/2.002,2.605/2.000,2.608/2.004,2.610/2.005,2.613/2.003,2.615/2.008,2.618/2.010,2.620/2.011,2.623/2.016,2.625/2.021,2.630/2.022,2.631/2.024,2.633/2.024,2.635/2.025,2.636/2.026,2.638/2.029,2.640/2.030,2.641/2.033,2.643/2.035,2.645/2.034,2.646/2.032,2.650/2.034,2.651/2.037,2.655/2.040,2.658/2.039,2.661/2.034,2.663/2.034,2.664/2.036,2.666/2.036,2.668/2.038,2.669/2.038,2.671/2.042,2.673/2.040,2.674/2.042,2.678/2.044,2.679/2.041,2.681/2.042,2.683/2.040,2.684/2.035,2.688/2.034,2.691/2.038,2.693/2.042,2.696/2.045,2.699/2.046,2.701/2.045,2.706/2.048,2.707/2.047,2.709/2.048,2.712/2.055,2.714/2.054,2.716/2.054,2.717/2.052,2.719/2.053,2.722/2.046,2.724/2.044,2.726/2.045,2.729/2.050,2.732/2.049,2.734/2.045,2.735/2.043,2.737/2.043,2.739/2.044,2.740/2.053,2.742/2.054,2.744/2.053,2.747/2.052,2.752/2.046,2.754/2.042,2.759/2.047,2.760/2.042,2.764/2.040,2.767/2.037,2.769/2.034,2.770/2.034,2.772/2.033,2.773/2.029,2.775/2.027,2.777/2.027,2.778/2.025,2.782/2.025,2.790/2.008,2.792/2.006,2.793/2.005,2.797/2.002,2.802/1.990,2.803/1.988,2.808/1.987,2.810/1.983,2.811/1.982,2.813/1.977,2.815/1.976,2.816/1.979,2.820/1.977,2.821/1.978,2.825/1.977,2.828/1.974,2.831/1.972,2.833/1.969,2.836/1.966,2.840/1.962,2.841/1.958,2.845/1.956,2.851/1.945,2.853/1.946,2.859/1.935,2.861/1.934,2.863/1.935,2.866/1.928,2.868/1.921,2.871/1.917,2.873/1.912,2.874/1.911,2.876/1.906,2.878/1.904,2.879/1.900,2.881/1.897,2.887/1.880,2.891/1.874,2.896/1.871,2.899/1.867,2.901/1.866,2.902/1.863,2.904/1.862,2.906/1.862,2.911/1.853,2.912/1.848,2.914/1.846,2.922/1.843,2.925/1.844,2.927/1.845,2.929/1.842,2.930/1.845,2.932/1.842,2.934/1.842,2.935/1.840,2.937/1.840,2.939/1.838,2.940/1.834,2.942/1.833,2.944/1.834,2.945/1.834,2.947/1.830,2.950/1.827,2.952/1.823,2.954/1.824,2.955/1.822,2.960/1.811,2.962/1.810,2.965/1.804,2.967/1.804,2.970/1.800,2.975/1.790,2.980/1.784,2.982/1.783,2.990/1.767,2.993/1.763,3.000/1.750,3.001/1.748,3.003/1.739,3.006/1.733,3.008/1.732,3.013/1.723,3.016/1.721,3.020/1.712,3.025/1.703,3.026/1.699,3.028/1.697,3.030/1.693,3.035/1.686,3.036/1.681,3.041/1.674,3.044/1.663,3.048/1.660,3.049/1.659,3.051/1.655,3.054/1.653,3.056/1.649,3.061/1.628,3.063/1.628,3.064/1.625,3.069/1.620,3.071/1.616,3.073/1.609,3.074/1.604,3.076/1.604,3.079/1.598,3.084/1.592,3.086/1.592,3.087/1.586,3.089/1.584,3.091/1.579,3.097/1.569,3.099/1.568,3.102/1.563,3.104/1.559,3.106/1.558,3.107/1.553,3.112/1.544,3.114/1.540,3.115/1.540,3.117/1.538,3.119/1.537,3.122/1.533,3.124/1.529,3.125/1.527,3.127/1.527,3.129/1.525,3.132/1.523,3.134/1.520,3.137/1.514,3.139/1.508,3.140/1.505,3.142/1.503,3.144/1.498,3.145/1.497,3.150/1.486,3.152/1.485,3.153/1.481,3.158/1.474,3.160/1.470,3.162/1.470,3.163/1.467,3.167/1.463,3.168/1.462,3.170/1.462,3.172/1.458,3.175/1.455,3.177/1.450,3.178/1.450,3.182/1.445,3.186/1.434,3.188/1.435,3.190/1.433,3.196/1.417,3.201/1.411,3.203/1.410,3.208/1.404,3.210/1.400,3.211/1.401,3.215/1.397,3.216/1.394,3.218/1.393,3.223/1.387,3.226/1.382,3.228/1.378,3.233/1.371,3.236/1.368,3.238/1.363,3.243/1.354,3.244/1.353,3.251/1.339,3.253/1.339,3.256/1.332,3.258/1.330,3.259/1.326,3.261/1.327,3.262/1.323,3.264/1.322,3.271/1.314,3.272/1.313,3.274/1.310,3.276/1.305,3.277/1.305,3.284/1.296,3.286/1.292,3.292/1.284,3.297/1.281,3.299/1.275,3.302/1.271,3.304/1.270,3.307/1.266,3.309/1.265,3.310/1.262,3.312/1.261,3.317/1.256,3.319/1.255,3.325/1.248,3.329/1.244,3.330/1.244,3.334/1.238,3.335/1.237,3.337/1.239,3.338/1.235,3.340/1.232,3.343/1.230,3.348/1.222,3.352/1.219,3.355/1.214,3.357/1.210,3.358/1.211,3.365/1.204,3.370/1.198,3.373/1.196,3.375/1.194,3.376/1.193,3.380/1.189,3.385/1.185,3.386/1.182,3.388/1.181,3.391/1.181,3.393/1.180,3.396/1.175,3.398/1.173,3.400/1.173,3.401/1.170,3.403/1.168,3.405/1.165,3.406/1.164,3.410/1.160,3.411/1.160,3.413/1.158,3.416/1.153,3.419/1.151,3.421/1.148,3.423/1.147,3.424/1.147,3.426/1.145,3.428/1.145,3.429/1.142,3.431/1.141,3.433/1.141,3.436/1.140,3.439/1.139,3.441/1.136,3.443/1.136,3.444/1.135,3.446/1.135,3.448/1.133,3.451/1.130,3.454/1.124,3.456/1.124,3.457/1.123,3.461/1.118,3.464/1.115,3.467/1.114,3.469/1.111,3.471/1.111,3.474/1.108,3.481/1.104,3.482/1.104,3.484/1.101,3.487/1.100,3.489/1.097,3.492/1.096,3.494/1.094,3.495/1.095,3.497/1.094,3.499/1.090,3.500/1.088,3.502/1.085,3.504/1.084,3.507/1.080,3.509/1.079,3.512/1.075,3.517/1.073,3.519/1.072,3.520/1.068,3.525/1.065,3.527/1.064,3.528/1.064,3.530/1.063,3.533/1.063,3.535/1.060,3.542/1.056,3.543/1.053,3.548/1.051,3.553/1.048,3.557/1.046,3.558/1.045,3.562/1.043,3.563/1.041,3.568/1.037,3.573/1.034,3.575/1.035,3.576/1.033,3.580/1.031,3.583/1.031,3.586/1.029,3.591/1.028,3.593/1.026,3.596/1.025,3.598/1.023,3.603/1.020,3.604/1.020,3.606/1.018,3.608/1.017,3.609/1.015,3.619/1.007,3.624/1.006,3.629/1.001,3.631/1.001,3.636/0.997,3.641/0.994,3.642/0.994,3.646/0.990,3.649/0.989,3.651/0.989,3.652/0.987,3.657/0.985,3.662/0.982,3.664/0.982,3.667/0.980,3.669/0.978,3.671/0.979,3.675/0.974,3.677/0.974,3.685/0.970,3.687/0.971,3.692/0.968,3.695/0.967,3.699/0.965,3.705/0.962,3.707/0.961,3.710/0.963,3.713/0.961,3.715/0.961,3.718/0.959,3.720/0.959,3.723/0.956,3.728/0.955,3.732/0.953,3.740/0.951,3.743/0.949,3.745/0.947,3.748/0.947,3.751/0.944,3.758/0.943,3.760/0.941,3.761/0.942,3.763/0.940,3.766/0.938,3.768/0.938,3.770/0.936,3.773/0.936,3.775/0.934,3.776/0.934,3.783/0.931,3.786/0.931,3.793/0.929,3.794/0.929,3.796/0.927,3.806/0.924,3.808/0.925,3.814/0.921,3.816/0.921,3.823/0.918,3.826/0.916,3.827/0.913,3.831/0.913,3.832/0.911,3.834/0.911,3.836/0.910,3.839/0.910,3.844/0.910,3.849/0.907,3.852/0.907,3.854/0.906,3.857/0.906,3.865/0.903,3.869/0.903,3.870/0.902,3.872/0.902,3.880/0.901,3.884/0.901,3.892/0.898,3.899/0.898,3.905/0.897,3.908/0.897,3.915/0.897,3.918/0.895,3.922/0.895,3.927/0.893,3.932/0.892,3.933/0.891,3.935/0.892,3.938/0.890,3.945/0.890,3.946/0.889,3.948/0.890,3.951/0.889,3.955/0.889,3.970/0.888,3.984/0.885,3.989/0.884,3.991/0.884,3.998/0.883,3.999/0.883,4.003/0.881,4.006/0.880,4.008/0.880,4.013/0.879,4.016/0.879,4.019/0.878,4.029/0.877,4.039/0.876,4.042/0.874,4.046/0.874,4.057/0.872,4.069/0.872,4.072/0.871,4.075/0.871,4.080/0.870,4.085/0.870,4.087/0.869,4.092/0.870,4.098/0.870,4.102/0.869,4.115/0.868,4.120/0.866,4.133/0.865,4.136/0.866,4.143/0.865,4.169/0.865,4.176/0.863,4.191/0.863,4.196/0.863,4.201/0.862,4.207/0.862,4.219/0.861,4.237/0.861,4.245/0.860,4.249/0.859,4.259/0.859,4.262/0.859,4.265/0.858,4.267/0.861,4.277/0.861,4.280/0.861,4.287/0.859,4.305/0.860,4.316/0.860,4.325/0.860,4.338/0.859,4.346/0.859,4.353/0.860,4.359/0.860,4.364/0.860,4.369/0.859,4.376/0.860,4.381/0.859,4.402/0.861,4.406/0.860,4.419/0.861,4.422/0.859,4.430/0.860,4.432/0.859}
{
\pgfplotstreampoint{\pgfpoint{\x cm}{\y cm}}
}
\pgfplotstreamend
      \pgfusepath{stroke}
    \end{pgfscope}
  \end{pgfscope}
  \fi
  \makeatletter\ifpgf@draftmode\makeatother\else
  \begin{pgfscope}
    \pgfpathrectangle{\pgfpoint{1.12889cm}{0.846667cm}}{\pgfpoint{3.30406cm}{2.68166cm}}
    \pgfusepath{clip}
    \begin{pgfscope}
      \pgfsetlinewidth{0.10pt}
      \definecolor{matfig2pgf_linecolor}{rgb}{1.000,0.000,0.000}
      \pgfsetstrokecolor{matfig2pgf_linecolor}
      \pgfsetdash{{2.00pt}{2.00pt}}{0pt}
      \pgfsetroundjoin
      \pgfplothandlerlineto
\pgfplotstreamstart
\foreach \x/\y in {1.130/0.847,1.668/0.847,1.670/0.848,1.718/0.848,1.719/0.850,1.888/0.850,1.890/0.851,1.895/0.851,1.896/0.854,1.900/0.857,1.901/0.858,1.903/0.862,1.911/0.862,1.913/0.863,1.928/0.863,1.931/0.869,1.936/0.873,1.938/0.878,1.941/0.881,1.949/0.881,1.952/0.884,1.956/0.897,1.959/0.902,1.982/0.902,1.984/0.904,2.000/0.904,2.002/0.902,2.020/0.902,2.022/0.903,2.023/0.908,2.028/0.918,2.030/0.918,2.032/0.920,2.037/0.936,2.038/0.939,2.050/0.939,2.052/0.938,2.057/0.938,2.060/0.940,2.063/0.951,2.065/0.958,2.066/0.962,2.070/0.965,2.071/0.969,2.073/0.970,2.085/0.970,2.088/0.972,2.091/0.979,2.093/0.979,2.095/0.981,2.096/0.987,2.103/1.026,2.104/1.033,2.108/1.039,2.118/1.039,2.119/1.040,2.121/1.040,2.124/1.044,2.134/1.044,2.136/1.047,2.141/1.091,2.142/1.096,2.151/1.096,2.152/1.098,2.154/1.098,2.156/1.100,2.157/1.100,2.159/1.101,2.161/1.104,2.164/1.104,2.166/1.105,2.172/1.105,2.174/1.107,2.175/1.110,2.177/1.112,2.180/1.112,2.182/1.116,2.185/1.121,2.187/1.122,2.202/1.122,2.205/1.124,2.208/1.137,2.213/1.142,2.218/1.142,2.220/1.147,2.222/1.159,2.227/1.227,2.228/1.238,2.230/1.245,2.232/1.244,2.233/1.244,2.235/1.246,2.237/1.256,2.240/1.290,2.243/1.303,2.246/1.304,2.248/1.304,2.251/1.306,2.260/1.306,2.263/1.301,2.268/1.297,2.270/1.291,2.273/1.289,2.281/1.289,2.283/1.287,2.284/1.291,2.288/1.286,2.291/1.284,2.293/1.288,2.298/1.323,2.299/1.327,2.303/1.331,2.306/1.331,2.309/1.347,2.314/1.398,2.318/1.414,2.319/1.420,2.321/1.423,2.326/1.427,2.329/1.442,2.331/1.443,2.334/1.443,2.337/1.459,2.341/1.485,2.342/1.491,2.344/1.491,2.346/1.493,2.349/1.493,2.351/1.495,2.352/1.502,2.354/1.504,2.356/1.509,2.357/1.511,2.360/1.504,2.362/1.506,2.364/1.506,2.365/1.505,2.370/1.525,2.374/1.532,2.377/1.532,2.379/1.535,2.380/1.544,2.387/1.566,2.389/1.575,2.392/1.586,2.394/1.585,2.395/1.580,2.397/1.580,2.398/1.589,2.402/1.619,2.405/1.625,2.408/1.628,2.412/1.634,2.417/1.634,2.420/1.632,2.423/1.625,2.425/1.625,2.427/1.622,2.428/1.617,2.432/1.600,2.433/1.594,2.436/1.600,2.438/1.605,2.440/1.604,2.443/1.613,2.445/1.621,2.448/1.652,2.451/1.698,2.458/1.859,2.461/1.900,2.465/1.912,2.466/1.917,2.468/1.917,2.473/1.873,2.474/1.868,2.476/1.869,2.479/1.869,2.481/1.870,2.484/1.878,2.486/1.890,2.491/1.909,2.496/1.921,2.498/1.923,2.499/1.928,2.503/1.934,2.504/1.934,2.506/1.932,2.508/1.929,2.509/1.927,2.512/1.927,2.514/1.923,2.517/1.919,2.519/1.917,2.522/1.917,2.524/1.921,2.526/1.923,2.534/1.923,2.537/1.920,2.541/1.912,2.542/1.914,2.544/1.920,2.547/1.926,2.549/1.928,2.550/1.928,2.552/1.924,2.554/1.910,2.559/1.843,2.560/1.830,2.562/1.825,2.565/1.822,2.567/1.818,2.569/1.807,2.572/1.775,2.574/1.775,2.575/1.779,2.579/1.798,2.580/1.802,2.582/1.803,2.585/1.809,2.587/1.813,2.593/1.899,2.597/1.931,2.600/2.003,2.603/2.092,2.605/2.123,2.607/2.138,2.610/2.151,2.613/2.157,2.615/2.161,2.617/2.157,2.618/2.154,2.620/2.149,2.623/2.146,2.625/2.142,2.630/2.107,2.631/2.103,2.635/2.099,2.638/2.099,2.641/2.083,2.646/2.032,2.648/2.021,2.651/2.009,2.653/2.006,2.656/2.008,2.658/2.012,2.659/2.014,2.661/2.012,2.666/2.019,2.668/2.018,2.669/2.018,2.673/2.005,2.674/2.001,2.676/2.001,2.678/1.999,2.679/1.999,2.683/2.007,2.684/2.005,2.686/2.004,2.688/1.996,2.689/1.993,2.693/1.993,2.694/1.996,2.696/2.004,2.699/2.035,2.706/2.056,2.711/2.088,2.712/2.089,2.714/2.088,2.717/2.079,2.721/2.063,2.722/2.058,2.724/2.055,2.726/2.060,2.727/2.072,2.729/2.079,2.731/2.071,2.734/2.045,2.735/2.038,2.737/2.034,2.740/2.038,2.745/2.055,2.752/2.055,2.754/2.063,2.755/2.076,2.760/2.146,2.762/2.153,2.764/2.151,2.765/2.146,2.770/2.118,2.775/2.107,2.777/2.101,2.783/2.040,2.790/1.890,2.793/1.848,2.795/1.841,2.797/1.840,2.798/1.843,2.802/1.874,2.805/1.910,2.807/1.921,2.808/1.927,2.810/1.929,2.811/1.931,2.813/1.930,2.815/1.926,2.816/1.920,2.818/1.909,2.823/1.888,2.825/1.883,2.830/1.875,2.831/1.870,2.833/1.870,2.838/1.877,2.846/1.877,2.848/1.879,2.854/1.879,2.856/1.875,2.858/1.875,2.863/1.884,2.866/1.892,2.868/1.895,2.869/1.896,2.871/1.899,2.874/1.909,2.876/1.911,2.878/1.915,2.881/1.935,2.883/1.939,2.884/1.941,2.887/1.941,2.889/1.942,2.892/1.944,2.896/1.952,2.901/1.968,2.902/1.975,2.904/1.976,2.906/1.971,2.909/1.950,2.911/1.943,2.912/1.941,2.914/1.953,2.919/1.968,2.921/1.956,2.924/1.914,2.925/1.899,2.927/1.895,2.929/1.896,2.930/1.902,2.934/1.854,2.937/1.793,2.939/1.783,2.942/1.772,2.944/1.770,2.945/1.771,2.947/1.770,2.950/1.773,2.954/1.773,2.955/1.775,2.965/1.775,2.968/1.778,2.977/1.778,2.982/1.782,2.985/1.786,2.988/1.780,2.992/1.771,2.993/1.763,2.995/1.762,2.997/1.763,2.998/1.763,3.003/1.739,3.006/1.744,3.011/1.780,3.013/1.785,3.015/1.788,3.020/1.785,3.021/1.783,3.025/1.783,3.028/1.771,3.033/1.726,3.038/1.703,3.044/1.658,3.046/1.653,3.048/1.652,3.051/1.663,3.054/1.677,3.056/1.677,3.058/1.670,3.061/1.643,3.063/1.637,3.064/1.635,3.066/1.636,3.068/1.639,3.069/1.641,3.071/1.639,3.073/1.636,3.076/1.617,3.077/1.612,3.081/1.618,3.084/1.618,3.086/1.612,3.087/1.599,3.092/1.533,3.094/1.528,3.096/1.527,3.106/1.527,3.109/1.534,3.111/1.544,3.112/1.559,3.114/1.565,3.115/1.564,3.117/1.562,3.120/1.554,3.122/1.555,3.124/1.559,3.125/1.560,3.127/1.565,3.129/1.561,3.130/1.553,3.137/1.484,3.139/1.472,3.140/1.465,3.142/1.463,3.144/1.461,3.147/1.465,3.148/1.469,3.150/1.471,3.153/1.471,3.157/1.467,3.158/1.467,3.160/1.466,3.162/1.468,3.163/1.472,3.165/1.472,3.167/1.468,3.168/1.467,3.170/1.465,3.178/1.465,3.180/1.463,3.188/1.463,3.191/1.459,3.193/1.456,3.195/1.456,3.196/1.454,3.198/1.453,3.200/1.450,3.201/1.449,3.203/1.445,3.210/1.410,3.213/1.385,3.215/1.381,3.216/1.380,3.218/1.380,3.220/1.381,3.221/1.380,3.223/1.382,3.224/1.381,3.226/1.380,3.229/1.377,3.231/1.375,3.234/1.363,3.236/1.362,3.239/1.363,3.241/1.361,3.243/1.359,3.244/1.361,3.246/1.350,3.248/1.350,3.249/1.350,3.253/1.339,3.256/1.331,3.258/1.330,3.259/1.324,3.261/1.308,3.264/1.250,3.266/1.230,3.267/1.225,3.269/1.224,3.271/1.220,3.272/1.218,3.274/1.219,3.277/1.216,3.279/1.214,3.281/1.213,3.282/1.214,3.286/1.214,3.287/1.213,3.297/1.213,3.300/1.209,3.302/1.212,3.307/1.234,3.309/1.239,3.310/1.241,3.312/1.244,3.314/1.246,3.315/1.244,3.317/1.248,3.319/1.254,3.322/1.275,3.324/1.282,3.325/1.283,3.329/1.278,3.330/1.276,3.334/1.276,3.335/1.274,3.338/1.264,3.342/1.244,3.347/1.228,3.348/1.224,3.350/1.222,3.355/1.222,3.358/1.225,3.362/1.221,3.365/1.205,3.368/1.197,3.373/1.197,3.375/1.199,3.376/1.199,3.378/1.204,3.380/1.206,3.383/1.197,3.386/1.181,3.388/1.178,3.390/1.177,3.395/1.177,3.396/1.176,3.398/1.175,3.400/1.171,3.403/1.168,3.410/1.168,3.413/1.162,3.416/1.162,3.418/1.160,3.419/1.160,3.421/1.158,3.426/1.162,3.429/1.166,3.433/1.173,3.436/1.173,3.438/1.176,3.439/1.175,3.441/1.171,3.443/1.160,3.446/1.129,3.448/1.123,3.449/1.119,3.452/1.119,3.454/1.117,3.456/1.113,3.457/1.112,3.459/1.108,3.477/1.108,3.481/1.106,3.482/1.106,3.484/1.109,3.487/1.113,3.489/1.113,3.490/1.111,3.492/1.111,3.494/1.109,3.495/1.104,3.497/1.102,3.499/1.102,3.500/1.100,3.504/1.100,3.507/1.103,3.509/1.107,3.515/1.107,3.524/1.114,3.525/1.117,3.527/1.113,3.530/1.109,3.545/1.109,3.547/1.108,3.548/1.108,3.550/1.109,3.553/1.110,3.555/1.106,3.557/1.106,3.560/1.097,3.562/1.093,3.563/1.091,3.565/1.091,3.570/1.086,3.575/1.086,3.578/1.079,3.581/1.062,3.585/1.057,3.588/1.057,3.591/1.052,3.593/1.049,3.598/1.049,3.600/1.048,3.601/1.044,3.603/1.043,3.604/1.043,3.606/1.041,3.608/1.041,3.609/1.039,3.611/1.041,3.613/1.041,3.614/1.041,3.616/1.044,3.623/1.044,3.624/1.045,3.628/1.050,3.629/1.055,3.633/1.055,3.634/1.053,3.639/1.031,3.641/1.025,3.646/1.014,3.647/1.014,3.651/1.003,3.656/0.977,3.657/0.976,3.672/0.976,3.674/0.973,3.675/0.966,3.679/0.958,3.687/0.958,3.690/0.955,3.692/0.953,3.694/0.952,3.695/0.953,3.705/0.953,3.707/0.951,3.709/0.951,3.715/0.946,3.727/0.946,3.730/0.943,3.745/0.943,3.747/0.944,3.750/0.952,3.753/0.952,3.755/0.951,3.760/0.941,3.761/0.940,3.765/0.933,3.768/0.933,3.771/0.927,3.773/0.926,3.808/0.926,3.811/0.923,3.816/0.915,3.819/0.911,3.829/0.911,3.831/0.910,3.836/0.910,3.839/0.906,3.841/0.903,3.847/0.903,3.856/0.895,3.857/0.892,3.880/0.892,3.884/0.889,3.887/0.889,3.889/0.887,3.894/0.887,3.895/0.888,3.897/0.895,3.937/0.895,3.938/0.898,3.941/0.898,3.943/0.895,3.945/0.894,3.948/0.888,3.955/0.888,3.956/0.887,3.960/0.882,3.961/0.877,3.981/0.877,3.983/0.876,3.984/0.873,3.986/0.872,4.024/0.872,4.027/0.869,4.039/0.869,4.041/0.871,4.042/0.868,4.055/0.868,4.059/0.872,4.077/0.872,4.079/0.870,4.082/0.863,4.123/0.863,4.125/0.865,4.178/0.865,4.179/0.867,4.226/0.867,4.227/0.866,4.229/0.860,4.234/0.860,4.236/0.861,4.269/0.861,4.270/0.858,4.361/0.862,4.371/0.863,4.373/0.861,4.388/0.862,4.391/0.856,4.432/0.858}
{
\pgfplotstreampoint{\pgfpoint{\x cm}{\y cm}}
}
\pgfplotstreamend
      \pgfusepath{stroke}
    \end{pgfscope}
  \end{pgfscope}
  \fi
    \pgftext[top,x=2.75268cm,y=0.267993cm,rotate=0]{$$$a_4$$$}
    \pgftext[bottom,x=0.691171cm,y=2.15927cm,rotate=90]{$$$\varphi_{2}$$$}
  \makeatletter\ifpgf@draftmode\makeatother\pgftext[x=5cm,y=4.02174cm]{\Huge{DRAFT}}\fi

%% file: matlab_plots/pdf_b_2.pgf
  \begin{pgfscope}
    \definecolor{matfig2pgf_color}{rgb}{1,1,1}\pgfsetfillcolor{matfig2pgf_color}
    \pgfpathrectangle{\pgfpoint{1.12889cm}{0.846667cm}}{\pgfpoint{3.30406cm}{2.68166cm}}
    \pgfusepath{fill}
  \end{pgfscope}
  \begin{pgfscope}
    \pgfsetlinewidth{0.5pt}
    \foreach \x in {1.9549,2.78092,3.60694}
    {
      \pgfpathmoveto{\pgfpoint{\x cm}{0.846667cm}}\pgfpathlineto{\pgfpoint{\x cm}{0.873483cm}}
      \pgfpathmoveto{\pgfpoint{\x cm}{3.52832cm}}\pgfpathlineto{\pgfpoint{\x cm}{3.50151cm}}
    }
    \foreach \y in {1.51708,2.1875,2.85791}
    {
      \pgfpathmoveto{\pgfpoint{1.12889cm}{\y cm}}\pgfpathlineto{\pgfpoint{1.16193cm}{\y cm}}
      \pgfpathmoveto{\pgfpoint{4.43295cm}{\y cm}}\pgfpathlineto{\pgfpoint{4.39991cm}{\y cm}}
    }
    \pgfusepath{stroke}
  \end{pgfscope}
  \begin{pgfscope}
    \pgfsetlinewidth{0.5pt}
    \pgfpathrectangle{\pgfpoint{1.12889cm}{0.846667cm}}{\pgfpoint{3.30406cm}{2.68166cm}}
    \pgfusepath{stroke}
  \end{pgfscope}
  {\small
    \pgftext[x=1.12889cm,y=0.746667cm,top]{$0$}
    \pgftext[x=1.9549cm,y=0.746667cm,top]{$0.25$}
    \pgftext[x=2.78092cm,y=0.746667cm,top]{$0.5$}
    \pgftext[x=3.60694cm,y=0.746667cm,top]{$0.75$}
    \pgftext[x=4.43295cm,y=0.746667cm,top]{$1$}
    \pgftext[x=1.02889cm,y=0.846667cm,right]{$0$}
    \pgftext[x=1.02889cm,y=1.51708cm,right]{$2$}
    \pgftext[x=1.02889cm,y=2.1875cm,right]{$4$}
    \pgftext[x=1.02889cm,y=2.85791cm,right]{$6$}
    \pgftext[x=1.02889cm,y=3.52832cm,right]{$8$}
  }
  \makeatletter\ifpgf@draftmode\makeatother\else
  \begin{pgfscope}
    \pgfpathrectangle{\pgfpoint{1.12889cm}{0.846667cm}}{\pgfpoint{3.30406cm}{2.68166cm}}
    \pgfusepath{clip}
    \begin{pgfscope}
      \pgfsetlinewidth{0.10pt}
      \definecolor{matfig2pgf_linecolor}{rgb}{0.000,0.000,1.000}
      \pgfsetstrokecolor{matfig2pgf_linecolor}
      \pgfsetdash{}{0pt}
      \pgfsetroundjoin
      \pgfplothandlerlineto
\pgfplotstreamstart
\foreach \x/\y in {1.130/0.847,2.521/0.847,2.531/0.847,2.534/0.848,2.539/0.848,2.542/0.849,2.587/0.850,2.607/0.851,2.621/0.851,2.651/0.853,2.656/0.853,2.658/0.854,2.666/0.854,2.668/0.855,2.686/0.857,2.694/0.860,2.696/0.861,2.701/0.861,2.704/0.862,2.712/0.863,2.724/0.866,2.727/0.866,2.740/0.869,2.744/0.871,2.747/0.872,2.749/0.873,2.752/0.874,2.754/0.876,2.757/0.877,2.764/0.881,2.772/0.884,2.777/0.886,2.785/0.890,2.788/0.892,2.793/0.894,2.800/0.899,2.803/0.903,2.807/0.905,2.808/0.907,2.811/0.909,2.816/0.913,2.820/0.914,2.826/0.919,2.830/0.923,2.831/0.926,2.835/0.929,2.836/0.930,2.843/0.936,2.846/0.940,2.848/0.941,2.849/0.943,2.851/0.944,2.853/0.946,2.859/0.956,2.864/0.960,2.871/0.968,2.873/0.968,2.876/0.973,2.879/0.975,2.881/0.976,2.884/0.982,2.887/0.989,2.891/0.995,2.894/0.999,2.896/1.000,2.902/1.011,2.912/1.029,2.914/1.030,2.919/1.042,2.924/1.047,2.929/1.056,2.930/1.060,2.934/1.065,2.935/1.073,2.940/1.084,2.944/1.093,2.945/1.094,2.950/1.104,2.954/1.115,2.955/1.118,2.957/1.123,2.960/1.130,2.967/1.149,2.968/1.151,2.970/1.158,2.973/1.166,2.975/1.169,2.982/1.190,2.985/1.195,2.990/1.207,2.992/1.213,2.997/1.224,3.008/1.261,3.010/1.265,3.018/1.297,3.020/1.301,3.023/1.315,3.025/1.319,3.028/1.332,3.033/1.348,3.035/1.351,3.036/1.356,3.039/1.370,3.046/1.403,3.048/1.406,3.053/1.428,3.054/1.431,3.059/1.447,3.061/1.450,3.071/1.489,3.074/1.500,3.089/1.548,3.094/1.567,3.096/1.570,3.097/1.580,3.101/1.591,3.104/1.609,3.107/1.621,3.112/1.648,3.119/1.677,3.120/1.682,3.124/1.698,3.127/1.711,3.132/1.736,3.134/1.742,3.135/1.752,3.140/1.769,3.162/1.863,3.165/1.872,3.168/1.891,3.170/1.897,3.172/1.901,3.173/1.909,3.175/1.913,3.178/1.931,3.180/1.937,3.182/1.940,3.185/1.955,3.190/1.970,3.193/1.983,3.195/1.994,3.196/1.998,3.208/2.047,3.213/2.075,3.221/2.099,3.224/2.109,3.228/2.124,3.229/2.129,3.231/2.140,3.234/2.151,3.238/2.159,3.241/2.177,3.244/2.190,3.246/2.200,3.251/2.216,3.254/2.227,3.256/2.236,3.259/2.250,3.261/2.257,3.264/2.263,3.266/2.270,3.267/2.270,3.269/2.278,3.274/2.291,3.277/2.304,3.279/2.307,3.282/2.320,3.287/2.331,3.289/2.338,3.294/2.350,3.300/2.372,3.304/2.380,3.307/2.392,3.319/2.424,3.320/2.426,3.322/2.430,3.324/2.429,3.325/2.434,3.329/2.437,3.330/2.436,3.334/2.439,3.335/2.442,3.337/2.444,3.338/2.443,3.340/2.444,3.342/2.449,3.345/2.450,3.348/2.451,3.350/2.452,3.352/2.455,3.353/2.455,3.355/2.452,3.357/2.455,3.360/2.454,3.363/2.462,3.365/2.464,3.368/2.474,3.370/2.474,3.372/2.478,3.375/2.480,3.376/2.478,3.380/2.487,3.381/2.486,3.383/2.487,3.385/2.485,3.386/2.489,3.388/2.489,3.390/2.492,3.393/2.499,3.396/2.496,3.401/2.501,3.405/2.501,3.410/2.505,3.411/2.507,3.413/2.507,3.416/2.512,3.423/2.517,3.424/2.517,3.426/2.518,3.428/2.518,3.429/2.514,3.431/2.516,3.433/2.519,3.438/2.514,3.439/2.514,3.443/2.508,3.444/2.508,3.446/2.506,3.448/2.505,3.454/2.497,3.456/2.495,3.457/2.497,3.459/2.494,3.461/2.489,3.464/2.483,3.466/2.482,3.467/2.478,3.471/2.481,3.472/2.478,3.477/2.472,3.487/2.455,3.489/2.454,3.492/2.446,3.495/2.441,3.497/2.442,3.499/2.440,3.500/2.434,3.502/2.430,3.504/2.431,3.505/2.426,3.507/2.426,3.510/2.415,3.514/2.410,3.515/2.404,3.522/2.393,3.524/2.387,3.525/2.386,3.527/2.378,3.528/2.378,3.530/2.376,3.533/2.363,3.535/2.363,3.537/2.359,3.538/2.352,3.543/2.337,3.548/2.324,3.553/2.311,3.555/2.309,3.558/2.300,3.563/2.282,3.565/2.278,3.566/2.272,3.570/2.267,3.578/2.228,3.580/2.222,3.581/2.220,3.583/2.212,3.586/2.206,3.588/2.198,3.590/2.193,3.595/2.174,3.596/2.172,3.600/2.163,3.601/2.156,3.604/2.149,3.608/2.134,3.613/2.123,3.618/2.107,3.619/2.103,3.621/2.095,3.626/2.078,3.631/2.055,3.636/2.041,3.647/1.994,3.652/1.978,3.654/1.971,3.656/1.968,3.657/1.961,3.659/1.957,3.662/1.953,3.666/1.944,3.667/1.936,3.674/1.920,3.679/1.907,3.684/1.890,3.689/1.880,3.694/1.864,3.702/1.831,3.704/1.823,3.705/1.819,3.712/1.790,3.717/1.778,3.722/1.760,3.725/1.751,3.727/1.747,3.728/1.741,3.730/1.737,3.740/1.703,3.743/1.689,3.747/1.677,3.751/1.656,3.755/1.641,3.760/1.629,3.763/1.617,3.766/1.603,3.771/1.589,3.775/1.576,3.781/1.556,3.783/1.549,3.785/1.546,3.789/1.527,3.793/1.519,3.794/1.512,3.796/1.510,3.803/1.483,3.811/1.457,3.818/1.445,3.823/1.427,3.827/1.413,3.831/1.401,3.834/1.393,3.839/1.378,3.844/1.366,3.847/1.362,3.852/1.348,3.856/1.343,3.861/1.330,3.864/1.321,3.865/1.319,3.867/1.313,3.872/1.303,3.877/1.288,3.885/1.273,3.889/1.266,3.890/1.263,3.894/1.259,3.895/1.253,3.897/1.250,3.899/1.248,3.903/1.236,3.908/1.231,3.912/1.224,3.913/1.218,3.915/1.216,3.918/1.209,3.920/1.206,3.923/1.198,3.928/1.190,3.932/1.182,3.937/1.170,3.938/1.168,3.941/1.161,3.943/1.158,3.945/1.152,3.948/1.145,3.955/1.134,3.956/1.133,3.965/1.120,3.970/1.109,3.973/1.105,3.976/1.097,3.981/1.092,3.984/1.087,3.991/1.082,3.994/1.077,3.998/1.073,3.999/1.072,4.004/1.064,4.011/1.052,4.013/1.052,4.016/1.048,4.017/1.044,4.019/1.043,4.021/1.040,4.024/1.037,4.029/1.029,4.031/1.027,4.032/1.022,4.036/1.020,4.037/1.017,4.039/1.016,4.041/1.015,4.049/1.001,4.051/1.000,4.054/0.997,4.057/0.991,4.060/0.989,4.065/0.981,4.067/0.980,4.070/0.977,4.075/0.975,4.077/0.973,4.080/0.973,4.084/0.969,4.085/0.969,4.089/0.966,4.092/0.962,4.095/0.960,4.100/0.954,4.102/0.953,4.103/0.951,4.105/0.950,4.110/0.944,4.112/0.943,4.115/0.944,4.120/0.942,4.123/0.938,4.126/0.937,4.128/0.934,4.130/0.934,4.135/0.930,4.136/0.930,4.141/0.926,4.143/0.925,4.145/0.922,4.151/0.917,4.155/0.917,4.158/0.915,4.161/0.914,4.163/0.912,4.164/0.911,4.171/0.910,4.173/0.908,4.174/0.908,4.178/0.907,4.181/0.904,4.184/0.903,4.193/0.897,4.198/0.896,4.201/0.894,4.202/0.894,4.204/0.893,4.206/0.893,4.212/0.890,4.219/0.888,4.222/0.887,4.224/0.885,4.229/0.883,4.232/0.882,4.237/0.880,4.239/0.879,4.249/0.877,4.255/0.875,4.265/0.874,4.267/0.876,4.280/0.875,4.282/0.874,4.287/0.874,4.288/0.872,4.293/0.872,4.308/0.872,4.310/0.871,4.315/0.870,4.335/0.870,4.343/0.870,4.345/0.869,4.346/0.869,4.348/0.868,4.354/0.868,4.361/0.868,4.366/0.868,4.369/0.868,4.371/0.867,4.378/0.867,4.383/0.866,4.389/0.865,4.396/0.866,4.399/0.866,4.402/0.865,4.404/0.864,4.406/0.864,4.407/0.863,4.414/0.863,4.422/0.863,4.426/0.863,4.429/0.863,4.432/0.863}
{
\pgfplotstreampoint{\pgfpoint{\x cm}{\y cm}}
}
\pgfplotstreamend
      \pgfusepath{stroke}
    \end{pgfscope}
  \end{pgfscope}
  \fi
  \makeatletter\ifpgf@draftmode\makeatother\else
  \begin{pgfscope}
    \pgfpathrectangle{\pgfpoint{1.12889cm}{0.846667cm}}{\pgfpoint{3.30406cm}{2.68166cm}}
    \pgfusepath{clip}
    \begin{pgfscope}
      \pgfsetlinewidth{0.10pt}
      \definecolor{matfig2pgf_linecolor}{rgb}{1.000,0.000,0.000}
      \pgfsetstrokecolor{matfig2pgf_linecolor}
      \pgfsetdash{{2.00pt}{2.00pt}}{0pt}
      \pgfsetroundjoin
      \pgfplothandlerlineto
\pgfplotstreamstart
\foreach \x/\y in {1.130/0.847,2.567/0.847,2.569/0.848,2.570/0.848,2.574/0.852,2.582/0.852,2.584/0.854,2.602/0.854,2.603/0.855,2.610/0.855,2.612/0.856,2.613/0.856,2.615/0.858,2.633/0.858,2.635/0.860,2.638/0.868,2.640/0.869,2.656/0.869,2.658/0.871,2.706/0.871,2.709/0.876,2.712/0.879,2.726/0.879,2.727/0.880,2.729/0.880,2.731/0.881,2.760/0.881,2.762/0.883,2.787/0.883,2.788/0.886,2.792/0.897,2.793/0.901,2.795/0.903,2.811/0.903,2.816/0.909,2.818/0.911,2.821/0.914,2.823/0.914,2.825/0.915,2.826/0.918,2.835/0.918,2.841/0.931,2.843/0.933,2.848/0.933,2.849/0.935,2.858/0.935,2.859/0.937,2.864/0.952,2.868/0.970,2.869/0.975,2.871/0.978,2.873/0.978,2.876/0.982,2.878/0.985,2.886/0.985,2.887/0.988,2.897/0.988,2.901/0.995,2.904/1.008,2.906/1.011,2.907/1.011,2.911/1.015,2.912/1.020,2.914/1.021,2.916/1.019,2.917/1.019,2.922/1.026,2.925/1.041,2.929/1.050,2.930/1.051,2.934/1.051,2.935/1.050,2.944/1.050,2.945/1.053,2.947/1.060,2.950/1.087,2.954/1.128,2.957/1.202,2.959/1.219,2.960/1.227,2.962/1.229,2.963/1.233,2.968/1.234,2.970/1.230,2.972/1.229,2.982/1.229,2.983/1.230,2.990/1.230,2.992/1.232,2.995/1.235,2.998/1.235,3.000/1.236,3.005/1.246,3.008/1.246,3.010/1.247,3.011/1.255,3.013/1.276,3.018/1.386,3.020/1.401,3.021/1.406,3.031/1.406,3.033/1.409,3.035/1.413,3.038/1.413,3.041/1.408,3.043/1.407,3.044/1.410,3.046/1.420,3.049/1.454,3.051/1.460,3.053/1.463,3.058/1.463,3.059/1.462,3.061/1.462,3.063/1.460,3.092/1.460,3.094/1.458,3.097/1.458,3.099/1.460,3.101/1.463,3.102/1.470,3.104/1.486,3.107/1.551,3.112/1.747,3.114/1.775,3.115/1.785,3.117/1.790,3.120/1.814,3.122/1.815,3.125/1.813,3.127/1.815,3.137/1.815,3.140/1.821,3.142/1.823,3.145/1.837,3.148/1.856,3.150/1.858,3.152/1.858,3.153/1.857,3.155/1.857,3.157/1.856,3.158/1.853,3.167/1.853,3.168/1.850,3.170/1.849,3.172/1.857,3.180/1.943,3.182/1.970,3.186/2.075,3.188/2.089,3.190/2.092,3.191/2.092,3.193/2.089,3.195/2.087,3.198/2.080,3.200/2.072,3.201/2.067,3.203/2.064,3.205/2.066,3.206/2.066,3.210/2.081,3.213/2.139,3.216/2.209,3.218/2.226,3.221/2.226,3.223/2.229,3.226/2.232,3.229/2.232,3.234/2.220,3.238/2.210,3.239/2.210,3.243/2.207,3.244/2.202,3.246/2.200,3.249/2.204,3.254/2.233,3.256/2.233,3.259/2.224,3.261/2.221,3.262/2.220,3.271/2.220,3.272/2.223,3.274/2.230,3.276/2.230,3.277/2.230,3.279/2.228,3.281/2.229,3.284/2.245,3.286/2.240,3.289/2.172,3.291/2.158,3.292/2.151,3.294/2.149,3.296/2.145,3.297/2.146,3.299/2.145,3.300/2.140,3.312/2.140,3.315/2.142,3.317/2.146,3.320/2.146,3.324/2.142,3.325/2.142,3.327/2.141,3.330/2.150,3.332/2.153,3.335/2.150,3.337/2.156,3.338/2.179,3.342/2.329,3.343/2.424,3.345/2.470,3.347/2.455,3.350/2.389,3.352/2.375,3.353/2.370,3.355/2.370,3.358/2.373,3.363/2.373,3.367/2.371,3.368/2.374,3.375/2.395,3.376/2.390,3.380/2.364,3.381/2.350,3.383/2.354,3.385/2.369,3.386/2.395,3.390/2.471,3.393/2.492,3.395/2.498,3.396/2.514,3.401/2.627,3.403/2.645,3.405/2.646,3.410/2.646,3.413/2.649,3.414/2.652,3.416/2.659,3.418/2.663,3.419/2.665,3.423/2.665,3.426/2.669,3.428/2.674,3.429/2.675,3.431/2.673,3.433/2.670,3.434/2.663,3.436/2.647,3.439/2.582,3.444/2.391,3.446/2.365,3.451/2.345,3.452/2.350,3.454/2.372,3.457/2.444,3.459/2.458,3.467/2.493,3.469/2.498,3.471/2.498,3.474/2.494,3.479/2.480,3.481/2.481,3.486/2.499,3.487/2.500,3.490/2.507,3.494/2.507,3.495/2.509,3.500/2.509,3.502/2.507,3.504/2.501,3.507/2.496,3.509/2.497,3.512/2.463,3.519/2.331,3.520/2.319,3.522/2.316,3.525/2.313,3.527/2.310,3.528/2.309,3.530/2.304,3.532/2.302,3.533/2.302,3.535/2.303,3.538/2.300,3.540/2.290,3.542/2.284,3.545/2.230,3.547/2.194,3.548/2.171,3.550/2.156,3.552/2.153,3.553/2.153,3.555/2.152,3.557/2.152,3.562/2.158,3.565/2.158,3.566/2.154,3.568/2.153,3.570/2.153,3.573/2.156,3.575/2.159,3.576/2.160,3.578/2.160,3.581/2.156,3.583/2.147,3.585/2.141,3.586/2.144,3.590/2.156,3.595/2.198,3.598/2.201,3.600/2.204,3.601/2.209,3.603/2.210,3.604/2.207,3.606/2.201,3.608/2.200,3.609/2.206,3.611/2.217,3.613/2.223,3.614/2.212,3.616/2.185,3.618/2.168,3.619/2.163,3.621/2.162,3.623/2.159,3.624/2.160,3.631/2.212,3.633/2.217,3.634/2.217,3.641/2.222,3.644/2.228,3.646/2.227,3.647/2.227,3.649/2.223,3.656/2.223,3.659/2.221,3.661/2.218,3.664/2.209,3.667/2.206,3.669/2.196,3.671/2.172,3.674/2.022,3.677/1.852,3.679/1.828,3.680/1.827,3.682/1.831,3.684/1.830,3.685/1.831,3.687/1.831,3.690/1.828,3.695/1.828,3.700/1.823,3.704/1.814,3.707/1.820,3.709/1.830,3.710/1.834,3.713/1.832,3.715/1.822,3.718/1.779,3.720/1.733,3.722/1.706,3.725/1.686,3.727/1.683,3.728/1.667,3.733/1.559,3.735/1.543,3.737/1.544,3.742/1.544,3.745/1.542,3.747/1.539,3.748/1.532,3.750/1.528,3.751/1.525,3.755/1.525,3.758/1.522,3.760/1.517,3.763/1.517,3.765/1.518,3.768/1.530,3.770/1.533,3.771/1.533,3.773/1.532,3.775/1.529,3.778/1.536,3.780/1.535,3.781/1.538,3.783/1.534,3.785/1.519,3.789/1.409,3.791/1.394,3.794/1.385,3.799/1.364,3.804/1.357,3.806/1.357,3.808/1.355,3.809/1.351,3.816/1.312,3.818/1.307,3.819/1.306,3.823/1.299,3.824/1.300,3.826/1.304,3.827/1.304,3.829/1.306,3.832/1.306,3.836/1.302,3.841/1.252,3.842/1.249,3.849/1.249,3.851/1.248,3.852/1.245,3.861/1.245,3.862/1.247,3.865/1.247,3.867/1.246,3.872/1.246,3.874/1.249,3.875/1.260,3.877/1.264,3.879/1.259,3.880/1.260,3.882/1.258,3.885/1.258,3.889/1.253,3.892/1.248,3.894/1.246,3.895/1.246,3.897/1.245,3.899/1.245,3.900/1.244,3.902/1.244,3.905/1.241,3.907/1.238,3.908/1.237,3.910/1.240,3.913/1.240,3.915/1.238,3.917/1.231,3.920/1.207,3.922/1.199,3.925/1.168,3.927/1.158,3.930/1.155,3.932/1.152,3.933/1.147,3.935/1.147,3.937/1.148,3.940/1.148,3.941/1.146,3.945/1.109,3.946/1.098,3.948/1.101,3.951/1.101,3.955/1.104,3.956/1.101,3.963/1.047,3.965/1.042,3.966/1.042,3.973/1.037,3.976/1.031,3.978/1.031,3.979/1.029,3.988/1.029,3.991/1.033,3.999/1.033,4.001/1.036,4.011/1.036,4.013/1.034,4.014/1.028,4.016/1.028,4.017/1.027,4.019/1.027,4.021/1.029,4.022/1.031,4.029/1.031,4.032/1.028,4.034/1.028,4.036/1.027,4.041/1.000,4.042/0.996,4.044/0.996,4.046/0.994,4.051/0.993,4.052/0.997,4.054/0.994,4.055/0.993,4.057/0.993,4.059/0.991,4.060/0.991,4.062/0.990,4.069/0.981,4.070/0.981,4.072/0.982,4.089/0.982,4.093/0.995,4.095/0.997,4.097/0.996,4.098/0.992,4.100/0.985,4.102/0.982,4.105/0.982,4.107/0.984,4.110/0.974,4.113/0.969,4.117/0.967,4.122/0.967,4.123/0.964,4.125/0.963,4.126/0.960,4.131/0.960,4.133/0.957,4.135/0.958,4.155/0.958,4.156/0.957,4.158/0.953,4.161/0.949,4.163/0.949,4.164/0.951,4.166/0.962,4.168/0.959,4.174/0.959,4.176/0.969,4.181/0.969,4.183/0.972,4.193/0.972,4.194/0.970,4.196/0.970,4.198/0.971,4.202/0.971,4.204/0.972,4.206/0.965,4.207/0.953,4.209/0.946,4.211/0.946,4.212/0.940,4.216/0.940,4.217/0.941,4.227/0.941,4.229/0.943,4.240/0.943,4.242/0.940,4.249/0.940,4.252/0.937,4.254/0.935,4.255/0.937,4.259/0.935,4.265/0.935,4.269/0.933,4.270/0.933,4.272/0.932,4.274/0.927,4.275/0.926,4.278/0.927,4.280/0.924,4.283/0.925,4.287/0.923,4.320/0.932,4.323/0.929,4.325/0.928,4.331/0.930,4.333/0.927,4.351/0.933,4.353/0.930,4.354/0.928,4.383/0.939,4.384/0.933,4.402/0.941,4.404/0.940,4.421/0.948,4.424/0.933,4.426/0.923,4.427/0.921,4.430/0.919,4.432/0.920}
{
\pgfplotstreampoint{\pgfpoint{\x cm}{\y cm}}
}
\pgfplotstreamend
      \pgfusepath{stroke}
    \end{pgfscope}
  \end{pgfscope}
  \fi
    \pgftext[top,x=2.75268cm,y=0.267993cm,rotate=0]{$$$b_2$$$}
    \pgftext[bottom,x=0.691171cm,y=2.15927cm,rotate=90]{$$$\varphi_{3}$$$}
  \makeatletter\ifpgf@draftmode\makeatother\pgftext[x=5cm,y=4.02174cm]{\Huge{DRAFT}}\fi

%% file: matlab_plots/pdf_b_3.pgf
  \begin{pgfscope}
    \definecolor{matfig2pgf_color}{rgb}{1,1,1}\pgfsetfillcolor{matfig2pgf_color}
    \pgfpathrectangle{\pgfpoint{1.12889cm}{0.846667cm}}{\pgfpoint{3.30406cm}{2.68166cm}}
    \pgfusepath{fill}
  \end{pgfscope}
  \begin{pgfscope}
    \pgfsetlinewidth{0.5pt}
    \foreach \x in {1.9549,2.78092,3.60694}
    {
      \pgfpathmoveto{\pgfpoint{\x cm}{0.846667cm}}\pgfpathlineto{\pgfpoint{\x cm}{0.873483cm}}
      \pgfpathmoveto{\pgfpoint{\x cm}{3.52832cm}}\pgfpathlineto{\pgfpoint{\x cm}{3.50151cm}}
    }
    \foreach \y in {1.51708,2.1875,2.85791}
    {
      \pgfpathmoveto{\pgfpoint{1.12889cm}{\y cm}}\pgfpathlineto{\pgfpoint{1.16193cm}{\y cm}}
      \pgfpathmoveto{\pgfpoint{4.43295cm}{\y cm}}\pgfpathlineto{\pgfpoint{4.39991cm}{\y cm}}
    }
    \pgfusepath{stroke}
  \end{pgfscope}
  \begin{pgfscope}
    \pgfsetlinewidth{0.5pt}
    \pgfpathrectangle{\pgfpoint{1.12889cm}{0.846667cm}}{\pgfpoint{3.30406cm}{2.68166cm}}
    \pgfusepath{stroke}
  \end{pgfscope}
  {\small
    \pgftext[x=1.12889cm,y=0.746667cm,top]{$0$}
    \pgftext[x=1.9549cm,y=0.746667cm,top]{$0.25$}
    \pgftext[x=2.78092cm,y=0.746667cm,top]{$0.5$}
    \pgftext[x=3.60694cm,y=0.746667cm,top]{$0.75$}
    \pgftext[x=4.43295cm,y=0.746667cm,top]{$1$}
    \pgftext[x=1.02889cm,y=0.846667cm,right]{$0$}
    \pgftext[x=1.02889cm,y=1.51708cm,right]{$2$}
    \pgftext[x=1.02889cm,y=2.1875cm,right]{$4$}
    \pgftext[x=1.02889cm,y=2.85791cm,right]{$6$}
    \pgftext[x=1.02889cm,y=3.52832cm,right]{$8$}
  }
  \makeatletter\ifpgf@draftmode\makeatother\else
  \begin{pgfscope}
    \pgfpathrectangle{\pgfpoint{1.12889cm}{0.846667cm}}{\pgfpoint{3.30406cm}{2.68166cm}}
    \pgfusepath{clip}
    \begin{pgfscope}
      \pgfsetlinewidth{0.10pt}
      \definecolor{matfig2pgf_linecolor}{rgb}{0.000,0.000,1.000}
      \pgfsetstrokecolor{matfig2pgf_linecolor}
      \pgfsetdash{}{0pt}
      \pgfsetroundjoin
      \pgfplothandlerlineto
\pgfplotstreamstart
\foreach \x/\y in {1.130/0.847,1.401/0.847,1.424/0.848,1.434/0.848,1.450/0.849,1.455/0.850,1.462/0.851,1.473/0.851,1.480/0.852,1.492/0.853,1.496/0.854,1.498/0.856,1.503/0.857,1.505/0.858,1.510/0.859,1.513/0.861,1.515/0.861,1.520/0.863,1.526/0.863,1.530/0.866,1.533/0.868,1.538/0.872,1.539/0.872,1.549/0.879,1.551/0.881,1.554/0.883,1.561/0.888,1.566/0.890,1.571/0.894,1.577/0.903,1.582/0.908,1.586/0.911,1.587/0.914,1.589/0.915,1.596/0.924,1.597/0.925,1.599/0.927,1.602/0.928,1.607/0.936,1.609/0.940,1.612/0.945,1.614/0.946,1.617/0.951,1.619/0.952,1.622/0.960,1.629/0.970,1.630/0.971,1.632/0.975,1.634/0.977,1.637/0.983,1.639/0.984,1.640/0.989,1.642/0.991,1.650/1.012,1.655/1.022,1.658/1.035,1.668/1.063,1.672/1.070,1.678/1.090,1.683/1.111,1.685/1.114,1.686/1.121,1.688/1.123,1.693/1.141,1.698/1.161,1.701/1.169,1.705/1.183,1.708/1.194,1.711/1.210,1.723/1.255,1.726/1.268,1.729/1.276,1.733/1.291,1.736/1.304,1.739/1.321,1.748/1.351,1.749/1.356,1.754/1.381,1.756/1.394,1.769/1.454,1.774/1.481,1.777/1.493,1.782/1.517,1.787/1.547,1.791/1.561,1.799/1.607,1.804/1.628,1.809/1.656,1.825/1.748,1.829/1.762,1.830/1.766,1.842/1.835,1.850/1.878,1.855/1.909,1.862/1.941,1.880/2.041,1.886/2.083,1.891/2.106,1.895/2.126,1.905/2.171,1.908/2.186,1.913/2.212,1.921/2.244,1.924/2.263,1.926/2.268,1.936/2.317,1.938/2.324,1.939/2.327,1.943/2.340,1.947/2.373,1.949/2.380,1.951/2.392,1.952/2.398,1.954/2.401,1.957/2.417,1.959/2.421,1.962/2.435,1.964/2.437,1.971/2.468,1.972/2.473,1.974/2.483,1.976/2.489,1.979/2.495,1.984/2.522,1.987/2.533,1.994/2.546,1.995/2.553,1.997/2.555,2.000/2.564,2.005/2.575,2.010/2.587,2.015/2.595,2.017/2.598,2.019/2.600,2.020/2.607,2.022/2.611,2.023/2.614,2.025/2.618,2.028/2.618,2.033/2.633,2.035/2.634,2.037/2.637,2.038/2.643,2.040/2.645,2.042/2.645,2.043/2.647,2.045/2.651,2.047/2.652,2.048/2.652,2.053/2.658,2.057/2.660,2.061/2.670,2.063/2.669,2.065/2.670,2.066/2.671,2.068/2.672,2.070/2.670,2.073/2.671,2.075/2.673,2.076/2.672,2.078/2.673,2.081/2.679,2.083/2.676,2.085/2.677,2.086/2.680,2.088/2.677,2.093/2.679,2.095/2.676,2.096/2.676,2.098/2.673,2.099/2.674,2.103/2.669,2.104/2.669,2.106/2.671,2.108/2.675,2.109/2.675,2.111/2.674,2.113/2.672,2.114/2.671,2.116/2.669,2.118/2.668,2.119/2.664,2.121/2.664,2.124/2.660,2.126/2.656,2.128/2.653,2.129/2.652,2.131/2.653,2.132/2.652,2.134/2.650,2.136/2.649,2.137/2.646,2.139/2.645,2.141/2.642,2.142/2.642,2.144/2.637,2.146/2.638,2.147/2.634,2.149/2.633,2.152/2.627,2.156/2.621,2.157/2.618,2.159/2.618,2.161/2.615,2.162/2.616,2.166/2.607,2.167/2.600,2.172/2.589,2.175/2.575,2.179/2.568,2.184/2.551,2.185/2.548,2.189/2.540,2.192/2.532,2.194/2.523,2.197/2.514,2.200/2.506,2.207/2.479,2.208/2.476,2.212/2.465,2.213/2.456,2.217/2.447,2.218/2.439,2.220/2.437,2.227/2.417,2.230/2.408,2.238/2.389,2.243/2.363,2.253/2.336,2.256/2.319,2.260/2.309,2.263/2.296,2.268/2.278,2.270/2.270,2.271/2.267,2.275/2.254,2.280/2.225,2.281/2.219,2.283/2.209,2.286/2.201,2.288/2.191,2.291/2.184,2.294/2.170,2.296/2.166,2.301/2.145,2.308/2.117,2.311/2.108,2.316/2.077,2.318/2.074,2.326/2.039,2.327/2.030,2.329/2.028,2.332/2.014,2.336/2.006,2.342/1.983,2.347/1.960,2.352/1.942,2.357/1.919,2.360/1.910,2.365/1.887,2.369/1.872,2.372/1.856,2.379/1.827,2.380/1.823,2.392/1.775,2.398/1.753,2.400/1.746,2.402/1.742,2.403/1.733,2.405/1.729,2.417/1.687,2.420/1.668,2.423/1.654,2.428/1.638,2.430/1.634,2.435/1.619,2.441/1.586,2.446/1.569,2.450/1.553,2.453/1.546,2.458/1.530,2.463/1.506,2.468/1.491,2.470/1.484,2.473/1.475,2.474/1.468,2.476/1.463,2.478/1.454,2.481/1.444,2.488/1.419,2.496/1.392,2.498/1.390,2.499/1.388,2.501/1.380,2.504/1.370,2.508/1.366,2.516/1.343,2.519/1.332,2.524/1.322,2.526/1.320,2.532/1.297,2.539/1.286,2.541/1.279,2.546/1.269,2.549/1.259,2.550/1.256,2.554/1.246,2.560/1.227,2.564/1.221,2.567/1.213,2.570/1.205,2.572/1.204,2.580/1.191,2.588/1.173,2.590/1.172,2.600/1.154,2.603/1.151,2.605/1.149,2.607/1.147,2.610/1.141,2.612/1.139,2.615/1.133,2.618/1.124,2.620/1.121,2.621/1.116,2.625/1.111,2.626/1.109,2.630/1.104,2.631/1.100,2.636/1.094,2.638/1.090,2.645/1.079,2.648/1.074,2.650/1.070,2.655/1.063,2.658/1.061,2.659/1.060,2.661/1.055,2.666/1.049,2.671/1.041,2.674/1.034,2.678/1.032,2.681/1.029,2.686/1.025,2.688/1.023,2.689/1.021,2.693/1.016,2.694/1.015,2.697/1.011,2.699/1.010,2.701/1.008,2.704/1.006,2.707/1.003,2.709/1.002,2.716/0.996,2.717/0.994,2.722/0.991,2.727/0.985,2.729/0.984,2.731/0.982,2.732/0.981,2.734/0.979,2.735/0.979,2.739/0.976,2.742/0.974,2.744/0.972,2.747/0.970,2.752/0.966,2.754/0.964,2.755/0.963,2.757/0.961,2.762/0.958,2.769/0.953,2.770/0.953,2.772/0.951,2.773/0.951,2.782/0.946,2.785/0.942,2.788/0.941,2.797/0.939,2.798/0.938,2.800/0.937,2.802/0.936,2.805/0.932,2.808/0.930,2.810/0.930,2.813/0.928,2.818/0.927,2.820/0.925,2.821/0.925,2.826/0.923,2.828/0.921,2.831/0.919,2.836/0.918,2.841/0.915,2.848/0.913,2.854/0.910,2.856/0.909,2.859/0.909,2.864/0.907,2.866/0.906,2.873/0.905,2.874/0.903,2.879/0.902,2.883/0.900,2.897/0.895,2.906/0.889,2.921/0.886,2.922/0.885,2.929/0.884,2.932/0.882,2.939/0.879,2.940/0.879,2.944/0.877,2.955/0.877,2.965/0.877,2.972/0.875,2.982/0.874,2.987/0.872,2.993/0.870,3.000/0.870,3.006/0.870,3.013/0.867,3.018/0.866,3.025/0.866,3.028/0.866,3.031/0.865,3.038/0.864,3.041/0.863,3.049/0.863,3.054/0.862,3.071/0.861,3.074/0.861,3.089/0.860,3.096/0.859,3.099/0.859,3.117/0.858,3.120/0.858,3.122/0.856,3.142/0.856,3.148/0.855,3.167/0.855,3.173/0.855,3.190/0.855,3.216/0.854,3.233/0.854,3.249/0.854,3.256/0.853,3.259/0.853,3.287/0.852,3.296/0.852,3.332/0.852,3.342/0.851,3.352/0.851,3.358/0.850,3.484/0.849,3.515/0.848,3.628/0.847,4.432/0.847}
{
\pgfplotstreampoint{\pgfpoint{\x cm}{\y cm}}
}
\pgfplotstreamend
      \pgfusepath{stroke}
    \end{pgfscope}
  \end{pgfscope}
  \fi
  \makeatletter\ifpgf@draftmode\makeatother\else
  \begin{pgfscope}
    \pgfpathrectangle{\pgfpoint{1.12889cm}{0.846667cm}}{\pgfpoint{3.30406cm}{2.68166cm}}
    \pgfusepath{clip}
    \begin{pgfscope}
      \pgfsetlinewidth{0.10pt}
      \definecolor{matfig2pgf_linecolor}{rgb}{1.000,0.000,0.000}
      \pgfsetstrokecolor{matfig2pgf_linecolor}
      \pgfsetdash{{2.00pt}{2.00pt}}{0pt}
      \pgfsetroundjoin
      \pgfplothandlerlineto
\pgfplotstreamstart
\foreach \x/\y in {1.130/0.847,1.406/0.847,1.407/0.848,1.457/0.848,1.458/0.849,1.516/0.849,1.518/0.852,1.533/0.852,1.534/0.853,1.546/0.853,1.548/0.855,1.553/0.855,1.554/0.860,1.558/0.864,1.559/0.868,1.561/0.869,1.563/0.869,1.564/0.870,1.566/0.870,1.568/0.873,1.586/0.873,1.591/0.878,1.592/0.879,1.597/0.879,1.599/0.883,1.602/0.910,1.605/0.922,1.607/0.922,1.609/0.925,1.615/0.931,1.617/0.931,1.619/0.932,1.622/0.932,1.624/0.936,1.630/0.955,1.637/0.955,1.640/0.961,1.642/0.962,1.643/0.966,1.647/0.969,1.648/0.969,1.652/0.978,1.653/0.979,1.655/0.979,1.657/0.980,1.660/0.980,1.662/0.983,1.665/0.983,1.675/1.009,1.678/1.012,1.685/1.012,1.686/1.014,1.690/1.018,1.691/1.023,1.693/1.031,1.698/1.104,1.700/1.118,1.701/1.122,1.706/1.126,1.708/1.129,1.713/1.146,1.715/1.151,1.724/1.151,1.726/1.156,1.728/1.170,1.729/1.191,1.736/1.340,1.738/1.351,1.743/1.362,1.744/1.363,1.748/1.363,1.751/1.370,1.754/1.374,1.756/1.375,1.759/1.379,1.761/1.379,1.762/1.391,1.767/1.517,1.769/1.535,1.771/1.540,1.772/1.542,1.774/1.542,1.776/1.544,1.781/1.557,1.782/1.559,1.786/1.559,1.787/1.560,1.791/1.566,1.792/1.573,1.794/1.586,1.795/1.594,1.797/1.598,1.805/1.607,1.807/1.610,1.809/1.611,1.812/1.617,1.815/1.628,1.820/1.675,1.822/1.683,1.824/1.687,1.829/1.742,1.830/1.753,1.833/1.763,1.835/1.767,1.837/1.778,1.840/1.839,1.842/1.855,1.843/1.862,1.845/1.864,1.847/1.865,1.848/1.867,1.850/1.865,1.853/1.865,1.860/1.878,1.865/1.903,1.868/1.908,1.870/1.908,1.873/1.915,1.875/1.925,1.880/1.989,1.885/2.026,1.890/2.084,1.891/2.095,1.893/2.099,1.895/2.105,1.898/2.127,1.900/2.133,1.901/2.136,1.903/2.136,1.906/2.142,1.908/2.143,1.909/2.148,1.911/2.161,1.913/2.196,1.916/2.307,1.918/2.334,1.921/2.355,1.924/2.366,1.928/2.393,1.929/2.402,1.931/2.405,1.933/2.397,1.934/2.382,1.938/2.370,1.939/2.370,1.941/2.367,1.943/2.365,1.944/2.367,1.946/2.382,1.951/2.462,1.952/2.477,1.954/2.483,1.956/2.483,1.961/2.468,1.962/2.464,1.966/2.471,1.967/2.472,1.969/2.472,1.971/2.471,1.972/2.468,1.974/2.467,1.976/2.463,1.977/2.461,1.979/2.461,1.982/2.465,1.984/2.469,1.990/2.527,1.992/2.536,1.994/2.540,1.997/2.553,1.999/2.554,2.000/2.554,2.002/2.552,2.004/2.552,2.005/2.555,2.007/2.566,2.012/2.638,2.014/2.649,2.015/2.653,2.017/2.655,2.019/2.655,2.022/2.661,2.023/2.671,2.025/2.675,2.028/2.650,2.030/2.645,2.032/2.646,2.035/2.661,2.037/2.665,2.038/2.666,2.040/2.663,2.043/2.650,2.047/2.645,2.048/2.653,2.050/2.668,2.055/2.750,2.058/2.828,2.060/2.852,2.061/2.851,2.063/2.830,2.068/2.723,2.070/2.715,2.071/2.713,2.073/2.712,2.075/2.713,2.078/2.724,2.080/2.732,2.083/2.733,2.085/2.733,2.086/2.730,2.088/2.729,2.090/2.730,2.091/2.732,2.093/2.739,2.095/2.734,2.099/2.629,2.101/2.613,2.103/2.609,2.106/2.609,2.108/2.609,2.113/2.595,2.114/2.593,2.118/2.593,2.119/2.592,2.123/2.588,2.124/2.594,2.126/2.607,2.131/2.719,2.132/2.735,2.134/2.739,2.136/2.739,2.137/2.738,2.139/2.736,2.141/2.737,2.142/2.740,2.146/2.751,2.147/2.753,2.152/2.727,2.154/2.721,2.156/2.718,2.161/2.662,2.162/2.652,2.164/2.646,2.166/2.644,2.167/2.645,2.169/2.641,2.170/2.622,2.172/2.587,2.174/2.571,2.175/2.564,2.177/2.562,2.180/2.560,2.184/2.568,2.185/2.569,2.187/2.566,2.189/2.565,2.192/2.569,2.195/2.570,2.199/2.581,2.202/2.581,2.205/2.575,2.207/2.565,2.210/2.539,2.215/2.513,2.217/2.502,2.222/2.437,2.223/2.423,2.227/2.414,2.230/2.395,2.232/2.389,2.233/2.386,2.240/2.386,2.242/2.382,2.243/2.370,2.245/2.335,2.248/2.224,2.250/2.198,2.251/2.190,2.256/2.214,2.260/2.192,2.263/2.178,2.265/2.177,2.266/2.185,2.271/2.228,2.273/2.236,2.275/2.241,2.276/2.241,2.278/2.227,2.281/2.179,2.283/2.164,2.284/2.160,2.289/2.208,2.291/2.213,2.294/2.216,2.296/2.217,2.298/2.213,2.299/2.212,2.301/2.213,2.306/2.213,2.308/2.218,2.314/2.252,2.316/2.248,2.322/2.189,2.324/2.180,2.331/2.156,2.336/2.145,2.337/2.137,2.339/2.122,2.342/2.066,2.344/2.046,2.346/2.035,2.349/2.041,2.351/2.043,2.354/2.036,2.365/1.938,2.369/1.925,2.370/1.924,2.374/1.930,2.375/1.930,2.379/1.927,2.380/1.918,2.382/1.903,2.387/1.822,2.392/1.701,2.394/1.680,2.395/1.669,2.398/1.661,2.402/1.656,2.403/1.653,2.405/1.652,2.412/1.626,2.415/1.620,2.417/1.620,2.425/1.631,2.427/1.628,2.432/1.609,2.433/1.607,2.436/1.604,2.438/1.604,2.440/1.603,2.453/1.603,2.455/1.599,2.456/1.586,2.458/1.560,2.461/1.467,2.463/1.435,2.465/1.417,2.466/1.411,2.470/1.408,2.471/1.408,2.473/1.406,2.474/1.400,2.481/1.362,2.484/1.351,2.488/1.349,2.496/1.349,2.498/1.350,2.499/1.348,2.501/1.348,2.503/1.355,2.504/1.367,2.506/1.374,2.508/1.375,2.511/1.375,2.512/1.373,2.516/1.366,2.517/1.364,2.519/1.364,2.521/1.363,2.524/1.353,2.529/1.328,2.531/1.324,2.532/1.324,2.534/1.325,2.537/1.325,2.539/1.324,2.541/1.321,2.544/1.295,2.546/1.288,2.547/1.285,2.550/1.285,2.554/1.282,2.555/1.282,2.557/1.283,2.559/1.283,2.560/1.281,2.562/1.284,2.564/1.282,2.567/1.282,2.570/1.276,2.572/1.274,2.574/1.274,2.575/1.273,2.580/1.273,2.582/1.272,2.584/1.265,2.588/1.217,2.592/1.212,2.593/1.211,2.595/1.211,2.597/1.208,2.600/1.188,2.603/1.156,2.605/1.148,2.607/1.148,2.608/1.150,2.610/1.151,2.612/1.150,2.615/1.133,2.621/1.083,2.623/1.080,2.626/1.076,2.628/1.072,2.631/1.072,2.635/1.070,2.638/1.070,2.640/1.064,2.645/1.036,2.646/1.022,2.648/1.016,2.650/1.015,2.668/1.015,2.669/1.018,2.678/1.018,2.679/1.011,2.684/1.001,2.686/1.000,2.701/1.000,2.702/0.998,2.706/0.992,2.716/0.992,2.717/0.994,2.721/0.997,2.745/0.997,2.750/0.991,2.752/0.989,2.755/0.977,2.757/0.968,2.759/0.965,2.760/0.963,2.828/0.963,2.831/0.956,2.833/0.949,2.836/0.925,2.838/0.918,2.840/0.917,2.841/0.919,2.854/0.919,2.856/0.917,2.858/0.917,2.859/0.915,2.864/0.902,2.866/0.901,2.887/0.901,2.889/0.900,2.891/0.897,2.892/0.896,2.894/0.893,2.937/0.893,2.940/0.882,2.942/0.878,2.949/0.878,2.952/0.863,2.954/0.859,2.955/0.857,3.000/0.857,3.001/0.854,3.048/0.854,3.049/0.852,3.053/0.849,3.167/0.849,3.168/0.848,3.172/0.848,3.173/0.847,4.432/0.847}
{
\pgfplotstreampoint{\pgfpoint{\x cm}{\y cm}}
}
\pgfplotstreamend
      \pgfusepath{stroke}
    \end{pgfscope}
  \end{pgfscope}
  \fi
    \pgftext[top,x=2.75268cm,y=0.267993cm,rotate=0]{$$$b_3$$$}
    \pgftext[bottom,x=0.691171cm,y=2.15927cm,rotate=90]{$$$\varphi_{4}$$$}
  \makeatletter\ifpgf@draftmode\makeatother\pgftext[x=5cm,y=4.02174cm]{\Huge{DRAFT}}\fi

%% file: matlab_plots/a4_b3.pgf
  \begin{pgfscope}
    \definecolor{matfig2pgf_color}{rgb}{1,1,1}\pgfsetfillcolor{matfig2pgf_color}
    \pgfpathrectangle{\pgfpoint{1.12889cm}{0.846667cm}}{\pgfpoint{3.10429cm}{2.59279cm}}
    \pgfusepath{fill}
  \end{pgfscope}
  \begin{pgfscope}
    \pgfsetlinewidth{0.5pt}
    \foreach \x in {1.90496,2.68103,3.4571}
    {
      \pgfpathmoveto{\pgfpoint{\x cm}{0.846667cm}}\pgfpathlineto{\pgfpoint{\x cm}{0.872595cm}}
      \pgfpathmoveto{\pgfpoint{\x cm}{3.43946cm}}\pgfpathlineto{\pgfpoint{\x cm}{3.41353cm}}
    }
    \foreach \y in {1.36523,1.88378,2.40234,2.9209}
    {
      \pgfpathmoveto{\pgfpoint{1.12889cm}{\y cm}}\pgfpathlineto{\pgfpoint{1.15993cm}{\y cm}}
      \pgfpathmoveto{\pgfpoint{4.23318cm}{\y cm}}\pgfpathlineto{\pgfpoint{4.20213cm}{\y cm}}
    }
    \pgfusepath{stroke}
  \end{pgfscope}
  \begin{pgfscope}
    \pgfsetlinewidth{0.5pt}
    \pgfpathrectangle{\pgfpoint{1.12889cm}{0.846667cm}}{\pgfpoint{3.10429cm}{2.59279cm}}
    \pgfusepath{stroke}
  \end{pgfscope}
  {\small
    \pgftext[x=1.12889cm,y=0.746667cm,top]{$0.2$}
    \pgftext[x=1.90496cm,y=0.746667cm,top]{$0.4$}
    \pgftext[x=2.68103cm,y=0.746667cm,top]{$0.6$}
    \pgftext[x=3.4571cm,y=0.746667cm,top]{$0.8$}
    \pgftext[x=4.23318cm,y=0.746667cm,top]{$1$}
    \pgftext[x=1.02889cm,y=0.846667cm,right]{$0.1$}
    \pgftext[x=1.02889cm,y=1.36523cm,right]{$0.2$}
    \pgftext[x=1.02889cm,y=1.88378cm,right]{$0.3$}
    \pgftext[x=1.02889cm,y=2.40234cm,right]{$0.4$}
    \pgftext[x=1.02889cm,y=2.9209cm,right]{$0.5$}
    \pgftext[x=1.02889cm,y=3.43946cm,right]{$0.6$}
  }
  \makeatletter\ifpgf@draftmode\makeatother\else
  \begin{pgfscope}
    \pgfpathrectangle{\pgfpoint{1.12889cm}{0.846667cm}}{\pgfpoint{3.10429cm}{2.59279cm}}
    \pgfusepath{clip}
    \begin{pgfscope}
      \pgfsetlinewidth{0.10pt}
      \definecolor{matfig2pgf_edgecolor}{rgb}{0.000,0.000,1.000}
      \pgfsetstrokecolor{matfig2pgf_edgecolor}
      \pgfplothandlermark{\pgfpathcircle{\pgfpointorigin}{1.00pt}\pgfusepath{stroke}}
\pgfplotstreamstart
\foreach \x/\y in {2.572/1.438,2.046/1.421,2.561/1.504,2.305/1.425,2.955/1.387,2.295/1.230,1.791/1.427,2.434/1.536,2.069/1.182,2.204/1.251,2.060/1.537,2.340/1.676,2.005/1.512,2.827/1.755,2.675/1.768,2.334/1.275,1.958/1.549,3.089/1.563,2.764/1.702,2.121/1.504,2.356/1.399,1.937/1.463,2.565/1.394,2.479/1.745,2.708/1.693,3.275/1.916,2.743/1.565,2.795/1.691,2.920/1.811,2.859/1.771,2.510/1.499,2.607/1.754,2.283/1.682,2.401/1.577,2.319/1.597,2.886/1.590,2.498/1.447,2.501/1.769}
{
\pgfplotstreampoint{\pgfpoint{\x cm}{\y cm}}
}
\pgfplotstreamend
    \end{pgfscope}
  \end{pgfscope}
  \fi
  \makeatletter\ifpgf@draftmode\makeatother\else
  \begin{pgfscope}
    \pgfpathrectangle{\pgfpoint{1.12889cm}{0.846667cm}}{\pgfpoint{3.10429cm}{2.59279cm}}
    \pgfusepath{clip}
    \begin{pgfscope}
      \pgfsetlinewidth{0.10pt}
      \definecolor{matfig2pgf_edgecolor}{rgb}{1.000,0.000,0.000}
      \pgfsetstrokecolor{matfig2pgf_edgecolor}
      \pgfplothandlermark{\pgfpathmoveto{\pgfpoint{-0.71pt}{-0.71pt}}\pgfpathlineto{\pgfpoint{0.71pt}{0.71pt}}\pgfpathmoveto{\pgfpoint{-0.71pt}{0.71pt}}\pgfpathlineto{\pgfpoint{0.71pt}{-0.71pt}}\pgfusepath{stroke}}
\pgfplotstreamstart
\foreach \x/\y in {1.947/2.098,1.728/1.915,1.709/1.741,2.305/2.576,2.837/2.013,2.285/2.446,2.774/2.099,2.802/2.022,2.641/2.532,2.362/1.850,2.168/1.673,2.154/2.374,2.221/1.806,2.773/2.008,2.048/1.897,2.383/2.640,1.792/1.554,1.981/2.269,2.345/2.260,2.624/2.230,2.384/1.751,1.969/1.639,2.292/2.035,2.256/2.298,2.351/1.937,1.550/2.101,2.167/1.771,3.127/2.124,2.488/2.046,2.216/2.228,2.470/2.250,2.020/1.700,2.936/2.356,1.932/2.108,2.319/1.962,2.486/2.338,2.470/2.060,2.168/1.674,1.873/1.786,1.963/1.968,2.105/1.806,1.890/2.804,3.400/2.050,1.697/1.765,3.063/2.136,1.806/1.669,1.951/2.113,1.998/2.748,1.615/1.475,1.829/2.241,2.574/2.613,2.314/1.735,1.584/1.787,2.631/2.271,2.040/2.375,2.670/2.476,2.006/1.568,3.102/2.392,2.127/2.561,2.046/1.736,2.692/1.922,1.994/1.925}
{
\pgfplotstreampoint{\pgfpoint{\x cm}{\y cm}}
}
\pgfplotstreamend
    \end{pgfscope}
  \end{pgfscope}
  \fi
  \makeatletter\ifpgf@draftmode\makeatother\else
  \begin{pgfscope}
    \pgfpathrectangle{\pgfpoint{1.12889cm}{0.846667cm}}{\pgfpoint{3.10429cm}{2.59279cm}}
    \pgfusepath{clip}
    \begin{pgfscope}
      \pgfsetlinewidth{0.10pt}
      \definecolor{matfig2pgf_linecolor}{rgb}{0.000,0.000,0.000}
      \pgfsetstrokecolor{matfig2pgf_linecolor}
      \pgfsetdash{}{0pt}
      \pgfsetroundjoin
      \pgfplothandlerlineto
\pgfplotstreamstart
\foreach \x/\y in {1.517/1.484,3.845/2.138}
{
\pgfplotstreampoint{\pgfpoint{\x cm}{\y cm}}
}
\pgfplotstreamend
      \pgfusepath{stroke}
    \end{pgfscope}
  \end{pgfscope}
  \fi
    \pgftext[top,x=2.66692cm,y=0.268925cm,rotate=0]{$$$a_4$$$}
    \pgftext[bottom,x=0.4657cm,y=2.10079cm,rotate=90]{$$$b_3$$$}
  \makeatletter\ifpgf@draftmode\makeatother\pgftext[x=5cm,y=4.09091cm]{\Huge{DRAFT}}\fi

%% file: matlab_plots/b2_b3.pgf
  \begin{pgfscope}
    \definecolor{matfig2pgf_color}{rgb}{1,1,1}\pgfsetfillcolor{matfig2pgf_color}
    \pgfpathrectangle{\pgfpoint{1.12889cm}{0.846667cm}}{\pgfpoint{3.10429cm}{2.59279cm}}
    \pgfusepath{fill}
  \end{pgfscope}
  \begin{pgfscope}
    \pgfsetlinewidth{0.5pt}
    \foreach \x in {1.74975,2.3706,2.99146,3.61232}
    {
      \pgfpathmoveto{\pgfpoint{\x cm}{0.846667cm}}\pgfpathlineto{\pgfpoint{\x cm}{0.872595cm}}
      \pgfpathmoveto{\pgfpoint{\x cm}{3.43946cm}}\pgfpathlineto{\pgfpoint{\x cm}{3.41353cm}}
    }
    \foreach \y in {1.36523,1.88378,2.40234,2.9209}
    {
      \pgfpathmoveto{\pgfpoint{1.12889cm}{\y cm}}\pgfpathlineto{\pgfpoint{1.15993cm}{\y cm}}
      \pgfpathmoveto{\pgfpoint{4.23318cm}{\y cm}}\pgfpathlineto{\pgfpoint{4.20213cm}{\y cm}}
    }
    \pgfusepath{stroke}
  \end{pgfscope}
  \begin{pgfscope}
    \pgfsetlinewidth{0.5pt}
    \pgfpathrectangle{\pgfpoint{1.12889cm}{0.846667cm}}{\pgfpoint{3.10429cm}{2.59279cm}}
    \pgfusepath{stroke}
  \end{pgfscope}
  {\small
    \pgftext[x=1.12889cm,y=0.746667cm,top]{$0.5$}
    \pgftext[x=1.74975cm,y=0.746667cm,top]{$0.6$}
    \pgftext[x=2.3706cm,y=0.746667cm,top]{$0.7$}
    \pgftext[x=2.99146cm,y=0.746667cm,top]{$0.8$}
    \pgftext[x=3.61232cm,y=0.746667cm,top]{$0.9$}
    \pgftext[x=4.23318cm,y=0.746667cm,top]{$1$}
    \pgftext[x=1.02889cm,y=0.846667cm,right]{$0.1$}
    \pgftext[x=1.02889cm,y=1.36523cm,right]{$0.2$}
    \pgftext[x=1.02889cm,y=1.88378cm,right]{$0.3$}
    \pgftext[x=1.02889cm,y=2.40234cm,right]{$0.4$}
    \pgftext[x=1.02889cm,y=2.9209cm,right]{$0.5$}
    \pgftext[x=1.02889cm,y=3.43946cm,right]{$0.6$}
  }
  \makeatletter\ifpgf@draftmode\makeatother\else
  \begin{pgfscope}
    \pgfpathrectangle{\pgfpoint{1.12889cm}{0.846667cm}}{\pgfpoint{3.10429cm}{2.59279cm}}
    \pgfusepath{clip}
    \begin{pgfscope}
      \pgfsetlinewidth{0.10pt}
      \definecolor{matfig2pgf_edgecolor}{rgb}{0.000,0.000,1.000}
      \pgfsetstrokecolor{matfig2pgf_edgecolor}
      \pgfplothandlermark{\pgfpathcircle{\pgfpointorigin}{1.00pt}\pgfusepath{stroke}}
\pgfplotstreamstart
\foreach \x/\y in {2.537/1.438,3.263/1.421,2.564/1.504,2.337/1.425,2.637/1.387,1.764/1.230,1.918/1.427,2.448/1.536,2.288/1.182,2.100/1.251,2.608/1.537,2.676/1.676,2.707/1.512,2.484/1.755,2.534/1.768,2.760/1.275,2.735/1.549,2.340/1.563,1.840/1.702,2.504/1.504,2.022/1.399,2.892/1.463,2.341/1.394,3.311/1.745,3.017/1.693,1.818/1.916,2.314/1.565,1.435/1.691,3.047/1.811,2.377/1.771,3.076/1.499,2.029/1.754,2.413/1.682,1.691/1.577,3.088/1.597,2.589/1.590,2.138/1.447,2.233/1.769}
{
\pgfplotstreampoint{\pgfpoint{\x cm}{\y cm}}
}
\pgfplotstreamend
    \end{pgfscope}
  \end{pgfscope}
  \fi
  \makeatletter\ifpgf@draftmode\makeatother\else
  \begin{pgfscope}
    \pgfpathrectangle{\pgfpoint{1.12889cm}{0.846667cm}}{\pgfpoint{3.10429cm}{2.59279cm}}
    \pgfusepath{clip}
    \begin{pgfscope}
      \pgfsetlinewidth{0.10pt}
      \definecolor{matfig2pgf_edgecolor}{rgb}{1.000,0.000,0.000}
      \pgfsetstrokecolor{matfig2pgf_edgecolor}
      \pgfplothandlermark{\pgfpathmoveto{\pgfpoint{-0.71pt}{-0.71pt}}\pgfpathlineto{\pgfpoint{0.71pt}{0.71pt}}\pgfpathmoveto{\pgfpoint{-0.71pt}{0.71pt}}\pgfpathlineto{\pgfpoint{0.71pt}{-0.71pt}}\pgfusepath{stroke}}
\pgfplotstreamstart
\foreach \x/\y in {2.438/2.098,2.014/1.915,2.112/1.741,2.017/2.576,2.386/2.013,1.920/2.446,2.247/2.099,3.334/2.022,2.449/2.532,2.580/1.850,1.832/1.673,2.246/2.374,2.592/1.806,1.807/2.008,2.533/1.897,2.328/2.640,2.132/1.554,2.320/2.269,2.683/2.260,2.384/2.230,3.278/1.751,2.767/1.639,2.723/2.035,3.001/2.298,1.827/1.937,1.825/2.101,2.922/1.771,1.974/2.124,1.477/2.046,2.006/2.228,2.811/2.250,1.316/1.700,2.219/2.356,2.221/2.108,2.752/1.962,2.878/2.338,1.732/2.060,3.034/1.674,3.688/1.786,2.561/1.968,2.539/1.806,2.277/2.804,2.113/2.050,2.679/1.765,2.153/2.136,3.015/1.669,2.784/2.113,2.318/2.748,2.118/1.475,2.746/2.241,1.529/2.613,1.696/1.735,2.252/1.787,1.691/2.271,2.363/2.375,2.249/2.476,2.578/1.568,2.717/2.392,2.503/2.561,2.302/1.736,3.106/1.922,2.549/1.925}
{
\pgfplotstreampoint{\pgfpoint{\x cm}{\y cm}}
}
\pgfplotstreamend
    \end{pgfscope}
  \end{pgfscope}
  \fi
    \pgftext[top,x=2.66692cm,y=0.268925cm,rotate=0]{$$$b_2$$$}
    \pgftext[bottom,x=0.4657cm,y=2.10079cm,rotate=90]{$$$b_3$$$}
  \makeatletter\ifpgf@draftmode\makeatother\pgftext[x=5cm,y=4.09091cm]{\Huge{DRAFT}}\fi

%% file: matlab_plots/traj_C3.pgf
  \begin{pgfscope}
    \definecolor{matfig2pgf_color}{rgb}{1,1,1}\pgfsetfillcolor{matfig2pgf_color}
    \pgfpathrectangle{\pgfpoint{1.12889cm}{0.846667cm}}{\pgfpoint{3.10429cm}{2.59279cm}}
    \pgfusepath{fill}
  \end{pgfscope}
  \begin{pgfscope}
    \pgfsetlinewidth{0.5pt}
    \foreach \x in {1.90496,2.68103,3.4571}
    {
      \pgfpathmoveto{\pgfpoint{\x cm}{0.846667cm}}\pgfpathlineto{\pgfpoint{\x cm}{0.872595cm}}
      \pgfpathmoveto{\pgfpoint{\x cm}{3.43946cm}}\pgfpathlineto{\pgfpoint{\x cm}{3.41353cm}}
    }
    \foreach \y in {1.49487,2.14306,2.79126}
    {
      \pgfpathmoveto{\pgfpoint{1.12889cm}{\y cm}}\pgfpathlineto{\pgfpoint{1.15993cm}{\y cm}}
      \pgfpathmoveto{\pgfpoint{4.23318cm}{\y cm}}\pgfpathlineto{\pgfpoint{4.20213cm}{\y cm}}
    }
    \pgfusepath{stroke}
  \end{pgfscope}
  \begin{pgfscope}
    \pgfsetlinewidth{0.5pt}
    \pgfpathrectangle{\pgfpoint{1.12889cm}{0.846667cm}}{\pgfpoint{3.10429cm}{2.59279cm}}
    \pgfusepath{stroke}
  \end{pgfscope}
  {\small
    \pgftext[x=1.12889cm,y=0.746667cm,top]{$0$}
    \pgftext[x=1.90496cm,y=0.746667cm,top]{$5$}
    \pgftext[x=2.68103cm,y=0.746667cm,top]{$10$}
    \pgftext[x=3.4571cm,y=0.746667cm,top]{$15$}
    \pgftext[x=4.23318cm,y=0.746667cm,top]{$20$}
    \pgftext[x=1.02889cm,y=0.846667cm,right]{$0$}
    \pgftext[x=1.02889cm,y=1.49487cm,right]{$0.2$}
    \pgftext[x=1.02889cm,y=2.14306cm,right]{$0.4$}
    \pgftext[x=1.02889cm,y=2.79126cm,right]{$0.6$}
    \pgftext[x=1.02889cm,y=3.43946cm,right]{$0.8$}
  }
  \makeatletter\ifpgf@draftmode\makeatother\else
  \begin{pgfscope}
    \pgfpathrectangle{\pgfpoint{1.12889cm}{0.846667cm}}{\pgfpoint{3.10429cm}{2.59279cm}}
    \pgfusepath{clip}
    \begin{pgfscope}
      \pgfsetlinewidth{0.10pt}
      \definecolor{matfig2pgf_linecolor}{rgb}{0.000,0.000,1.000}
      \pgfsetstrokecolor{matfig2pgf_linecolor}
      \pgfsetdash{}{0pt}
      \pgfsetroundjoin
      \pgfplothandlerlineto
\pgfplotstreamstart
\foreach \x/\y in {1.129/0.847,1.145/0.847,1.162/0.852,1.178/0.861,1.194/0.874,1.211/0.891,1.260/0.949,1.276/0.967,1.292/0.983,1.309/0.998,1.325/1.010,1.341/1.021,1.358/1.030,1.374/1.038,1.390/1.044,1.407/1.049,1.423/1.053,1.439/1.055,1.455/1.056,1.471/1.056,1.486/1.054,1.502/1.050,1.533/1.042,1.580/1.027,1.642/1.005,1.798/0.943,1.861/0.921,1.908/0.907,1.954/0.894,2.001/0.883,2.048/0.875,2.110/0.866,2.173/0.860,2.251/0.855,2.360/0.851,2.516/0.848,2.844/0.847,4.233/0.847}
{
\pgfplotstreampoint{\pgfpoint{\x cm}{\y cm}}
}
\pgfplotstreamend
      \pgfusepath{stroke}
    \end{pgfscope}
  \end{pgfscope}
  \fi
  \makeatletter\ifpgf@draftmode\makeatother\else
  \begin{pgfscope}
    \pgfpathrectangle{\pgfpoint{1.12889cm}{0.846667cm}}{\pgfpoint{3.10429cm}{2.59279cm}}
    \pgfusepath{clip}
    \begin{pgfscope}
      \pgfsetlinewidth{0.10pt}
      \definecolor{matfig2pgf_linecolor}{rgb}{0.000,0.000,1.000}
      \pgfsetstrokecolor{matfig2pgf_linecolor}
      \pgfsetdash{}{0pt}
      \pgfsetroundjoin
      \pgfplothandlerlineto
\pgfplotstreamstart
\foreach \x/\y in {1.129/0.847,1.145/0.847,1.162/0.852,1.178/0.863,1.194/0.878,1.211/0.897,1.260/0.964,1.276/0.985,1.292/1.004,1.309/1.022,1.325/1.037,1.341/1.051,1.358/1.063,1.374/1.072,1.390/1.081,1.407/1.087,1.423/1.093,1.439/1.097,1.455/1.100,1.471/1.100,1.486/1.099,1.502/1.097,1.533/1.091,1.580/1.079,1.642/1.059,1.705/1.037,1.767/1.012,1.908/0.953,1.970/0.928,2.017/0.912,2.064/0.899,2.110/0.887,2.157/0.878,2.204/0.870,2.267/0.863,2.345/0.857,2.438/0.852,2.563/0.849,2.782/0.847,3.796/0.847,4.233/0.847}
{
\pgfplotstreampoint{\pgfpoint{\x cm}{\y cm}}
}
\pgfplotstreamend
      \pgfusepath{stroke}
    \end{pgfscope}
  \end{pgfscope}
  \fi
  \makeatletter\ifpgf@draftmode\makeatother\else
  \begin{pgfscope}
    \pgfpathrectangle{\pgfpoint{1.12889cm}{0.846667cm}}{\pgfpoint{3.10429cm}{2.59279cm}}
    \pgfusepath{clip}
    \begin{pgfscope}
      \pgfsetlinewidth{0.10pt}
      \definecolor{matfig2pgf_linecolor}{rgb}{0.000,0.000,1.000}
      \pgfsetstrokecolor{matfig2pgf_linecolor}
      \pgfsetdash{}{0pt}
      \pgfsetroundjoin
      \pgfplothandlerlineto
\pgfplotstreamstart
\foreach \x/\y in {1.129/0.847,1.145/0.847,1.162/0.852,1.178/0.862,1.194/0.877,1.211/0.895,1.260/0.958,1.276/0.978,1.292/0.996,1.309/1.012,1.325/1.027,1.341/1.039,1.358/1.050,1.374/1.058,1.390/1.066,1.407/1.072,1.423/1.076,1.439/1.080,1.455/1.082,1.471/1.081,1.486/1.080,1.502/1.077,1.533/1.069,1.580/1.055,1.642/1.034,1.720/1.004,1.861/0.946,1.908/0.928,1.954/0.913,2.001/0.899,2.048/0.887,2.095/0.878,2.142/0.871,2.204/0.863,2.282/0.857,2.376/0.852,2.501/0.849,2.719/0.847,3.702/0.847,4.233/0.847}
{
\pgfplotstreampoint{\pgfpoint{\x cm}{\y cm}}
}
\pgfplotstreamend
      \pgfusepath{stroke}
    \end{pgfscope}
  \end{pgfscope}
  \fi
  \makeatletter\ifpgf@draftmode\makeatother\else
  \begin{pgfscope}
    \pgfpathrectangle{\pgfpoint{1.12889cm}{0.846667cm}}{\pgfpoint{3.10429cm}{2.59279cm}}
    \pgfusepath{clip}
    \begin{pgfscope}
      \pgfsetlinewidth{0.10pt}
      \definecolor{matfig2pgf_linecolor}{rgb}{0.000,0.000,1.000}
      \pgfsetstrokecolor{matfig2pgf_linecolor}
      \pgfsetdash{}{0pt}
      \pgfsetroundjoin
      \pgfplothandlerlineto
\pgfplotstreamstart
\foreach \x/\y in {1.129/0.847,1.145/0.847,1.162/0.851,1.178/0.859,1.194/0.872,1.211/0.887,1.260/0.942,1.276/0.959,1.292/0.975,1.309/0.989,1.325/1.002,1.341/1.013,1.358/1.022,1.374/1.030,1.390/1.037,1.407/1.042,1.423/1.046,1.439/1.050,1.455/1.051,1.471/1.051,1.486/1.049,1.502/1.046,1.533/1.038,1.580/1.023,1.642/1.001,1.798/0.940,1.861/0.918,1.908/0.904,1.954/0.892,2.001/0.882,2.048/0.874,2.110/0.865,2.173/0.860,2.251/0.855,2.360/0.851,2.516/0.848,2.860/0.847,4.233/0.847}
{
\pgfplotstreampoint{\pgfpoint{\x cm}{\y cm}}
}
\pgfplotstreamend
      \pgfusepath{stroke}
    \end{pgfscope}
  \end{pgfscope}
  \fi
  \makeatletter\ifpgf@draftmode\makeatother\else
  \begin{pgfscope}
    \pgfpathrectangle{\pgfpoint{1.12889cm}{0.846667cm}}{\pgfpoint{3.10429cm}{2.59279cm}}
    \pgfusepath{clip}
    \begin{pgfscope}
      \pgfsetlinewidth{0.10pt}
      \definecolor{matfig2pgf_linecolor}{rgb}{0.000,0.000,1.000}
      \pgfsetstrokecolor{matfig2pgf_linecolor}
      \pgfsetdash{}{0pt}
      \pgfsetroundjoin
      \pgfplothandlerlineto
\pgfplotstreamstart
\foreach \x/\y in {1.129/0.847,1.145/0.847,1.162/0.851,1.178/0.859,1.194/0.871,1.211/0.885,1.260/0.934,1.276/0.949,1.292/0.963,1.309/0.975,1.325/0.985,1.341/0.994,1.358/1.001,1.374/1.006,1.390/1.011,1.407/1.014,1.423/1.016,1.439/1.018,1.455/1.018,1.471/1.016,1.486/1.014,1.517/1.006,1.564/0.992,1.736/0.931,1.798/0.911,1.845/0.898,1.892/0.887,1.939/0.878,1.986/0.871,2.048/0.863,2.126/0.857,2.220/0.853,2.345/0.849,2.563/0.847,3.500/0.847,4.233/0.847}
{
\pgfplotstreampoint{\pgfpoint{\x cm}{\y cm}}
}
\pgfplotstreamend
      \pgfusepath{stroke}
    \end{pgfscope}
  \end{pgfscope}
  \fi
  \makeatletter\ifpgf@draftmode\makeatother\else
  \begin{pgfscope}
    \pgfpathrectangle{\pgfpoint{1.12889cm}{0.846667cm}}{\pgfpoint{3.10429cm}{2.59279cm}}
    \pgfusepath{clip}
    \begin{pgfscope}
      \pgfsetlinewidth{0.10pt}
      \definecolor{matfig2pgf_linecolor}{rgb}{0.000,0.000,1.000}
      \pgfsetstrokecolor{matfig2pgf_linecolor}
      \pgfsetdash{}{0pt}
      \pgfsetroundjoin
      \pgfplothandlerlineto
\pgfplotstreamstart
\foreach \x/\y in {1.129/0.847,1.145/0.847,1.162/0.850,1.178/0.857,1.194/0.867,1.211/0.880,1.260/0.922,1.276/0.935,1.292/0.947,1.309/0.958,1.325/0.967,1.341/0.975,1.358/0.981,1.374/0.986,1.390/0.990,1.407/0.993,1.439/0.997,1.455/0.997,1.471/0.996,1.486/0.994,1.517/0.987,1.564/0.974,1.720/0.923,1.783/0.905,1.830/0.893,1.876/0.883,1.923/0.875,1.986/0.866,2.048/0.860,2.126/0.855,2.235/0.851,2.391/0.848,2.719/0.847,4.233/0.847}
{
\pgfplotstreampoint{\pgfpoint{\x cm}{\y cm}}
}
\pgfplotstreamend
      \pgfusepath{stroke}
    \end{pgfscope}
  \end{pgfscope}
  \fi
  \makeatletter\ifpgf@draftmode\makeatother\else
  \begin{pgfscope}
    \pgfpathrectangle{\pgfpoint{1.12889cm}{0.846667cm}}{\pgfpoint{3.10429cm}{2.59279cm}}
    \pgfusepath{clip}
    \begin{pgfscope}
      \pgfsetlinewidth{0.10pt}
      \definecolor{matfig2pgf_linecolor}{rgb}{0.000,0.000,1.000}
      \pgfsetstrokecolor{matfig2pgf_linecolor}
      \pgfsetdash{}{0pt}
      \pgfsetroundjoin
      \pgfplothandlerlineto
\pgfplotstreamstart
\foreach \x/\y in {1.129/0.847,1.145/0.848,1.162/0.852,1.178/0.863,1.194/0.880,1.211/0.900,1.243/0.947,1.276/0.994,1.292/1.016,1.309/1.036,1.325/1.053,1.341/1.069,1.358/1.083,1.374/1.094,1.390/1.105,1.407/1.113,1.423/1.121,1.439/1.127,1.455/1.131,1.471/1.133,1.486/1.133,1.502/1.132,1.533/1.127,1.580/1.119,1.642/1.105,1.705/1.088,1.751/1.074,1.798/1.058,1.845/1.040,1.923/1.006,2.032/0.958,2.095/0.932,2.142/0.916,2.189/0.901,2.235/0.889,2.282/0.879,2.329/0.872,2.391/0.864,2.454/0.858,2.532/0.854,2.641/0.850,2.813/0.848,3.234/0.847,4.233/0.847}
{
\pgfplotstreampoint{\pgfpoint{\x cm}{\y cm}}
}
\pgfplotstreamend
      \pgfusepath{stroke}
    \end{pgfscope}
  \end{pgfscope}
  \fi
  \makeatletter\ifpgf@draftmode\makeatother\else
  \begin{pgfscope}
    \pgfpathrectangle{\pgfpoint{1.12889cm}{0.846667cm}}{\pgfpoint{3.10429cm}{2.59279cm}}
    \pgfusepath{clip}
    \begin{pgfscope}
      \pgfsetlinewidth{0.10pt}
      \definecolor{matfig2pgf_linecolor}{rgb}{0.000,0.000,1.000}
      \pgfsetstrokecolor{matfig2pgf_linecolor}
      \pgfsetdash{}{0pt}
      \pgfsetroundjoin
      \pgfplothandlerlineto
\pgfplotstreamstart
\foreach \x/\y in {1.129/0.847,1.145/0.847,1.162/0.852,1.178/0.863,1.194/0.879,1.211/0.899,1.260/0.967,1.276/0.988,1.292/1.008,1.309/1.026,1.325/1.042,1.341/1.056,1.358/1.068,1.374/1.078,1.390/1.087,1.407/1.094,1.423/1.099,1.439/1.104,1.455/1.106,1.471/1.107,1.486/1.106,1.502/1.104,1.533/1.097,1.580/1.085,1.642/1.066,1.705/1.045,1.767/1.020,1.861/0.981,1.939/0.948,1.986/0.930,2.032/0.914,2.079/0.900,2.126/0.888,2.173/0.878,2.220/0.871,2.282/0.863,2.360/0.857,2.454/0.852,2.579/0.849,2.797/0.847,3.765/0.847,4.233/0.847}
{
\pgfplotstreampoint{\pgfpoint{\x cm}{\y cm}}
}
\pgfplotstreamend
      \pgfusepath{stroke}
    \end{pgfscope}
  \end{pgfscope}
  \fi
  \makeatletter\ifpgf@draftmode\makeatother\else
  \begin{pgfscope}
    \pgfpathrectangle{\pgfpoint{1.12889cm}{0.846667cm}}{\pgfpoint{3.10429cm}{2.59279cm}}
    \pgfusepath{clip}
    \begin{pgfscope}
      \pgfsetlinewidth{0.10pt}
      \definecolor{matfig2pgf_linecolor}{rgb}{1.000,0.000,0.000}
      \pgfsetstrokecolor{matfig2pgf_linecolor}
      \pgfsetdash{{2.00pt}{2.00pt}}{0pt}
      \pgfsetroundjoin
      \pgfplothandlerlineto
\pgfplotstreamstart
\foreach \x/\y in {1.129/0.847,1.145/0.848,1.162/0.856,1.178/0.874,1.194/0.902,1.211/0.937,1.227/0.978,1.292/1.155,1.309/1.198,1.325/1.238,1.341/1.277,1.358/1.312,1.374/1.345,1.390/1.376,1.407/1.404,1.423/1.430,1.439/1.453,1.455/1.473,1.471/1.489,1.486/1.504,1.533/1.546,1.549/1.561,1.564/1.577,1.580/1.595,1.595/1.613,1.611/1.634,1.627/1.655,1.642/1.678,1.673/1.728,1.705/1.783,1.736/1.842,1.783/1.936,1.892/2.162,1.939/2.255,1.970/2.313,2.001/2.369,2.032/2.422,2.064/2.471,2.095/2.518,2.126/2.561,2.157/2.601,2.189/2.638,2.220/2.672,2.251/2.704,2.282/2.732,2.313/2.759,2.345/2.783,2.376/2.804,2.407/2.824,2.454/2.851,2.501/2.874,2.547/2.894,2.594/2.911,2.641/2.926,2.688/2.939,2.750/2.953,2.813/2.965,2.875/2.974,2.953/2.983,3.047/2.992,3.156/2.999,3.297/3.004,3.468/3.009,3.718/3.012,4.186/3.013,4.233/3.013}
{
\pgfplotstreampoint{\pgfpoint{\x cm}{\y cm}}
}
\pgfplotstreamend
      \pgfusepath{stroke}
    \end{pgfscope}
  \end{pgfscope}
  \fi
  \makeatletter\ifpgf@draftmode\makeatother\else
  \begin{pgfscope}
    \pgfpathrectangle{\pgfpoint{1.12889cm}{0.846667cm}}{\pgfpoint{3.10429cm}{2.59279cm}}
    \pgfusepath{clip}
    \begin{pgfscope}
      \pgfsetlinewidth{0.10pt}
      \definecolor{matfig2pgf_linecolor}{rgb}{1.000,0.000,0.000}
      \pgfsetstrokecolor{matfig2pgf_linecolor}
      \pgfsetdash{{2.00pt}{2.00pt}}{0pt}
      \pgfsetroundjoin
      \pgfplothandlerlineto
\pgfplotstreamstart
\foreach \x/\y in {1.129/0.847,1.145/0.848,1.162/0.855,1.178/0.871,1.194/0.896,1.211/0.928,1.227/0.964,1.292/1.123,1.309/1.161,1.325/1.197,1.341/1.231,1.358/1.263,1.374/1.292,1.390/1.320,1.407/1.345,1.423/1.368,1.439/1.389,1.455/1.406,1.471/1.421,1.486/1.433,1.533/1.469,1.564/1.495,1.580/1.510,1.595/1.526,1.611/1.543,1.627/1.562,1.642/1.582,1.658/1.603,1.673/1.625,1.705/1.675,1.736/1.729,1.767/1.786,1.814/1.878,1.939/2.133,1.986/2.223,2.032/2.309,2.064/2.363,2.095/2.414,2.126/2.462,2.157/2.506,2.189/2.548,2.220/2.587,2.251/2.624,2.282/2.658,2.313/2.689,2.345/2.717,2.376/2.743,2.407/2.768,2.438/2.790,2.469/2.811,2.501/2.830,2.547/2.855,2.594/2.876,2.641/2.896,2.688/2.912,2.735/2.927,2.782/2.940,2.844/2.954,2.906/2.965,2.985/2.977,3.063/2.986,3.156/2.994,3.265/3.001,3.390/3.007,3.562/3.011,3.812/3.015,4.233/3.017}
{
\pgfplotstreampoint{\pgfpoint{\x cm}{\y cm}}
}
\pgfplotstreamend
      \pgfusepath{stroke}
    \end{pgfscope}
  \end{pgfscope}
  \fi
  \makeatletter\ifpgf@draftmode\makeatother\else
  \begin{pgfscope}
    \pgfpathrectangle{\pgfpoint{1.12889cm}{0.846667cm}}{\pgfpoint{3.10429cm}{2.59279cm}}
    \pgfusepath{clip}
    \begin{pgfscope}
      \pgfsetlinewidth{0.10pt}
      \definecolor{matfig2pgf_linecolor}{rgb}{1.000,0.000,0.000}
      \pgfsetstrokecolor{matfig2pgf_linecolor}
      \pgfsetdash{{2.00pt}{2.00pt}}{0pt}
      \pgfsetroundjoin
      \pgfplothandlerlineto
\pgfplotstreamstart
\foreach \x/\y in {1.129/0.847,1.145/0.848,1.162/0.856,1.178/0.872,1.194/0.897,1.211/0.929,1.227/0.965,1.276/1.079,1.292/1.115,1.309/1.149,1.325/1.181,1.341/1.210,1.358/1.237,1.374/1.262,1.390/1.283,1.407/1.303,1.423/1.321,1.439/1.336,1.455/1.349,1.471/1.359,1.486/1.367,1.533/1.388,1.564/1.404,1.580/1.413,1.595/1.423,1.611/1.434,1.627/1.446,1.642/1.459,1.658/1.473,1.673/1.489,1.689/1.505,1.720/1.541,1.751/1.582,1.783/1.626,1.814/1.673,1.861/1.749,1.986/1.961,2.048/2.063,2.095/2.135,2.142/2.201,2.189/2.263,2.220/2.302,2.251/2.338,2.282/2.372,2.313/2.404,2.360/2.447,2.407/2.487,2.438/2.511,2.469/2.534,2.516/2.564,2.563/2.592,2.610/2.616,2.657/2.638,2.704/2.658,2.750/2.675,2.797/2.690,2.860/2.707,2.922/2.722,2.985/2.735,3.063/2.748,3.156/2.760,3.250/2.769,3.359/2.777,3.500/2.785,3.656/2.790,3.859/2.794,4.171/2.797,4.233/2.798}
{
\pgfplotstreampoint{\pgfpoint{\x cm}{\y cm}}
}
\pgfplotstreamend
      \pgfusepath{stroke}
    \end{pgfscope}
  \end{pgfscope}
  \fi
  \makeatletter\ifpgf@draftmode\makeatother\else
  \begin{pgfscope}
    \pgfpathrectangle{\pgfpoint{1.12889cm}{0.846667cm}}{\pgfpoint{3.10429cm}{2.59279cm}}
    \pgfusepath{clip}
    \begin{pgfscope}
      \pgfsetlinewidth{0.10pt}
      \definecolor{matfig2pgf_linecolor}{rgb}{1.000,0.000,0.000}
      \pgfsetstrokecolor{matfig2pgf_linecolor}
      \pgfsetdash{{2.00pt}{2.00pt}}{0pt}
      \pgfsetroundjoin
      \pgfplothandlerlineto
\pgfplotstreamstart
\foreach \x/\y in {1.129/0.847,1.145/0.848,1.162/0.855,1.178/0.871,1.194/0.895,1.211/0.927,1.227/0.964,1.243/1.005,1.325/1.226,1.358/1.311,1.390/1.389,1.407/1.426,1.423/1.461,1.439/1.494,1.455/1.522,1.471/1.548,1.502/1.594,1.533/1.640,1.564/1.690,1.580/1.716,1.595/1.745,1.627/1.806,1.658/1.873,1.689/1.946,1.736/2.061,1.830/2.299,1.876/2.413,1.923/2.522,1.954/2.590,1.986/2.654,2.017/2.714,2.048/2.769,2.079/2.820,2.110/2.867,2.142/2.910,2.173/2.949,2.204/2.985,2.235/3.017,2.267/3.047,2.298/3.073,2.329/3.098,2.360/3.119,2.391/3.138,2.423/3.156,2.454/3.171,2.501/3.191,2.547/3.209,2.594/3.223,2.641/3.235,2.688/3.245,2.750/3.257,2.813/3.265,2.891/3.274,2.985/3.281,3.094/3.287,3.234/3.292,3.437/3.295,3.781/3.297,4.233/3.298}
{
\pgfplotstreampoint{\pgfpoint{\x cm}{\y cm}}
}
\pgfplotstreamend
      \pgfusepath{stroke}
    \end{pgfscope}
  \end{pgfscope}
  \fi
  \makeatletter\ifpgf@draftmode\makeatother\else
  \begin{pgfscope}
    \pgfpathrectangle{\pgfpoint{1.12889cm}{0.846667cm}}{\pgfpoint{3.10429cm}{2.59279cm}}
    \pgfusepath{clip}
    \begin{pgfscope}
      \pgfsetlinewidth{0.10pt}
      \definecolor{matfig2pgf_linecolor}{rgb}{1.000,0.000,0.000}
      \pgfsetstrokecolor{matfig2pgf_linecolor}
      \pgfsetdash{{2.00pt}{2.00pt}}{0pt}
      \pgfsetroundjoin
      \pgfplothandlerlineto
\pgfplotstreamstart
\foreach \x/\y in {1.129/0.847,1.145/0.848,1.162/0.854,1.178/0.868,1.194/0.890,1.211/0.917,1.227/0.948,1.276/1.046,1.292/1.077,1.309/1.106,1.325/1.132,1.341/1.157,1.358/1.178,1.374/1.198,1.390/1.215,1.407/1.231,1.423/1.244,1.439/1.256,1.455/1.265,1.471/1.272,1.486/1.276,1.549/1.290,1.580/1.298,1.595/1.302,1.627/1.314,1.658/1.329,1.673/1.338,1.689/1.348,1.705/1.358,1.720/1.370,1.736/1.382,1.751/1.395,1.767/1.409,1.798/1.441,1.830/1.476,1.861/1.514,1.892/1.557,1.939/1.626,1.986/1.700,2.110/1.905,2.173/2.003,2.220/2.073,2.267/2.139,2.313/2.200,2.360/2.257,2.407/2.309,2.454/2.357,2.501/2.400,2.547/2.440,2.594/2.477,2.641/2.510,2.688/2.540,2.735/2.567,2.782/2.591,2.828/2.613,2.875/2.633,2.922/2.651,2.985/2.672,3.047/2.690,3.109/2.706,3.172/2.720,3.250/2.734,3.344/2.748,3.437/2.760,3.546/2.770,3.687/2.780,3.827/2.786,4.015/2.793,4.233/2.797}
{
\pgfplotstreampoint{\pgfpoint{\x cm}{\y cm}}
}
\pgfplotstreamend
      \pgfusepath{stroke}
    \end{pgfscope}
  \end{pgfscope}
  \fi
  \makeatletter\ifpgf@draftmode\makeatother\else
  \begin{pgfscope}
    \pgfpathrectangle{\pgfpoint{1.12889cm}{0.846667cm}}{\pgfpoint{3.10429cm}{2.59279cm}}
    \pgfusepath{clip}
    \begin{pgfscope}
      \pgfsetlinewidth{0.10pt}
      \definecolor{matfig2pgf_linecolor}{rgb}{1.000,0.000,0.000}
      \pgfsetstrokecolor{matfig2pgf_linecolor}
      \pgfsetdash{{2.00pt}{2.00pt}}{0pt}
      \pgfsetroundjoin
      \pgfplothandlerlineto
\pgfplotstreamstart
\foreach \x/\y in {1.129/0.847,1.145/0.848,1.162/0.856,1.178/0.874,1.194/0.901,1.211/0.937,1.227/0.978,1.260/1.069,1.309/1.210,1.341/1.299,1.358/1.341,1.374/1.380,1.390/1.417,1.407/1.452,1.423/1.485,1.439/1.515,1.455/1.541,1.471/1.564,1.502/1.605,1.533/1.646,1.564/1.691,1.580/1.715,1.595/1.741,1.611/1.768,1.642/1.827,1.673/1.892,1.705/1.961,1.751/2.071,1.845/2.297,1.892/2.406,1.939/2.509,1.970/2.574,2.001/2.635,2.032/2.692,2.064/2.745,2.095/2.794,2.126/2.840,2.157/2.881,2.189/2.920,2.220/2.954,2.251/2.986,2.282/3.015,2.313/3.042,2.345/3.066,2.376/3.087,2.407/3.106,2.438/3.124,2.485/3.147,2.532/3.166,2.579/3.183,2.626/3.197,2.672/3.209,2.735/3.222,2.797/3.233,2.860/3.241,2.938/3.249,3.031/3.256,3.141/3.262,3.281/3.267,3.484/3.270,3.812/3.273,4.233/3.273}
{
\pgfplotstreampoint{\pgfpoint{\x cm}{\y cm}}
}
\pgfplotstreamend
      \pgfusepath{stroke}
    \end{pgfscope}
  \end{pgfscope}
  \fi
  \makeatletter\ifpgf@draftmode\makeatother\else
  \begin{pgfscope}
    \pgfpathrectangle{\pgfpoint{1.12889cm}{0.846667cm}}{\pgfpoint{3.10429cm}{2.59279cm}}
    \pgfusepath{clip}
    \begin{pgfscope}
      \pgfsetlinewidth{0.10pt}
      \definecolor{matfig2pgf_linecolor}{rgb}{1.000,0.000,0.000}
      \pgfsetstrokecolor{matfig2pgf_linecolor}
      \pgfsetdash{{2.00pt}{2.00pt}}{0pt}
      \pgfsetroundjoin
      \pgfplothandlerlineto
\pgfplotstreamstart
\foreach \x/\y in {1.129/0.847,1.145/0.848,1.162/0.855,1.178/0.869,1.194/0.892,1.211/0.920,1.227/0.953,1.276/1.058,1.292/1.091,1.309/1.122,1.325/1.152,1.341/1.179,1.358/1.203,1.374/1.226,1.390/1.246,1.407/1.264,1.423/1.280,1.439/1.295,1.455/1.306,1.471/1.315,1.486/1.322,1.533/1.339,1.564/1.352,1.580/1.359,1.595/1.368,1.611/1.377,1.627/1.387,1.642/1.399,1.658/1.411,1.673/1.425,1.689/1.440,1.705/1.456,1.720/1.473,1.736/1.492,1.767/1.532,1.798/1.577,1.830/1.625,1.861/1.677,1.908/1.760,2.048/2.019,2.095/2.101,2.142/2.179,2.189/2.252,2.220/2.298,2.251/2.341,2.282/2.382,2.329/2.439,2.376/2.492,2.407/2.524,2.438/2.554,2.469/2.582,2.516/2.621,2.563/2.656,2.610/2.688,2.657/2.717,2.704/2.742,2.750/2.765,2.797/2.785,2.844/2.803,2.891/2.820,2.938/2.834,3.000/2.851,3.078/2.868,3.156/2.883,3.219/2.893,3.312/2.904,3.422/2.915,3.515/2.922,3.656/2.929,3.812/2.935,4.030/2.940,4.233/2.943}
{
\pgfplotstreampoint{\pgfpoint{\x cm}{\y cm}}
}
\pgfplotstreamend
      \pgfusepath{stroke}
    \end{pgfscope}
  \end{pgfscope}
  \fi
  \makeatletter\ifpgf@draftmode\makeatother\else
  \begin{pgfscope}
    \pgfpathrectangle{\pgfpoint{1.12889cm}{0.846667cm}}{\pgfpoint{3.10429cm}{2.59279cm}}
    \pgfusepath{clip}
    \begin{pgfscope}
      \pgfsetlinewidth{0.10pt}
      \definecolor{matfig2pgf_linecolor}{rgb}{1.000,0.000,0.000}
      \pgfsetstrokecolor{matfig2pgf_linecolor}
      \pgfsetdash{{2.00pt}{2.00pt}}{0pt}
      \pgfsetroundjoin
      \pgfplothandlerlineto
\pgfplotstreamstart
\foreach \x/\y in {1.129/0.847,1.145/0.848,1.162/0.854,1.178/0.867,1.194/0.886,1.211/0.912,1.227/0.941,1.276/1.033,1.292/1.063,1.309/1.090,1.325/1.115,1.341/1.138,1.358/1.159,1.374/1.178,1.390/1.195,1.407/1.210,1.423/1.223,1.439/1.235,1.455/1.244,1.471/1.250,1.486/1.254,1.517/1.260,1.564/1.268,1.595/1.275,1.627/1.285,1.658/1.297,1.673/1.305,1.705/1.322,1.736/1.342,1.767/1.365,1.798/1.392,1.830/1.422,1.861/1.457,1.892/1.494,1.923/1.535,1.954/1.579,2.001/1.649,2.126/1.845,2.173/1.916,2.220/1.983,2.267/2.046,2.313/2.103,2.345/2.139,2.376/2.173,2.407/2.204,2.438/2.233,2.469/2.261,2.501/2.286,2.547/2.321,2.594/2.352,2.641/2.380,2.688/2.405,2.735/2.427,2.782/2.446,2.828/2.463,2.875/2.479,2.922/2.492,2.985/2.507,3.047/2.521,3.125/2.534,3.203/2.545,3.297/2.555,3.406/2.564,3.515/2.571,3.671/2.577,3.859/2.582,4.124/2.585,4.233/2.586}
{
\pgfplotstreampoint{\pgfpoint{\x cm}{\y cm}}
}
\pgfplotstreamend
      \pgfusepath{stroke}
    \end{pgfscope}
  \end{pgfscope}
  \fi
  \makeatletter\ifpgf@draftmode\makeatother\else
  \begin{pgfscope}
    \pgfpathrectangle{\pgfpoint{1.12889cm}{0.846667cm}}{\pgfpoint{3.10429cm}{2.59279cm}}
    \pgfusepath{clip}
    \begin{pgfscope}
      \pgfsetlinewidth{0.10pt}
      \definecolor{matfig2pgf_linecolor}{rgb}{1.000,0.000,0.000}
      \pgfsetstrokecolor{matfig2pgf_linecolor}
      \pgfsetdash{{2.00pt}{2.00pt}}{0pt}
      \pgfsetroundjoin
      \pgfplothandlerlineto
\pgfplotstreamstart
\foreach \x/\y in {1.129/0.847,1.145/0.848,1.162/0.856,1.178/0.872,1.194/0.898,1.211/0.932,1.227/0.971,1.260/1.057,1.309/1.190,1.341/1.272,1.358/1.311,1.374/1.347,1.390/1.381,1.407/1.413,1.423/1.443,1.439/1.471,1.455/1.494,1.471/1.514,1.502/1.550,1.533/1.585,1.564/1.625,1.580/1.646,1.595/1.669,1.611/1.694,1.627/1.720,1.658/1.776,1.689/1.838,1.720/1.905,1.767/2.010,1.861/2.227,1.908/2.331,1.954/2.431,1.986/2.493,2.017/2.551,2.048/2.606,2.079/2.657,2.110/2.704,2.142/2.748,2.173/2.789,2.204/2.825,2.235/2.859,2.267/2.891,2.298/2.919,2.329/2.945,2.360/2.969,2.391/2.990,2.423/3.009,2.454/3.027,2.501/3.050,2.547/3.070,2.594/3.087,2.641/3.101,2.688/3.114,2.750/3.128,2.813/3.139,2.875/3.148,2.953/3.157,3.047/3.165,3.156/3.171,3.297/3.177,3.484/3.181,3.765/3.183,4.233/3.185}
{
\pgfplotstreampoint{\pgfpoint{\x cm}{\y cm}}
}
\pgfplotstreamend
      \pgfusepath{stroke}
    \end{pgfscope}
  \end{pgfscope}
  \fi
  \makeatletter\ifpgf@draftmode\makeatother\else
  \begin{pgfscope}
    \pgfpathrectangle{\pgfpoint{1.12889cm}{0.846667cm}}{\pgfpoint{3.10429cm}{2.59279cm}}
    \pgfusepath{clip}
    \begin{pgfscope}
      \pgfsetlinewidth{0.10pt}
      \definecolor{matfig2pgf_linecolor}{rgb}{1.000,0.000,0.000}
      \pgfsetstrokecolor{matfig2pgf_linecolor}
      \pgfsetdash{{2.00pt}{2.00pt}}{0pt}
      \pgfsetroundjoin
      \pgfplothandlerlineto
\pgfplotstreamstart
\foreach \x/\y in {1.129/0.847,1.145/0.848,1.162/0.854,1.178/0.868,1.194/0.889,1.211/0.916,1.227/0.946,1.276/1.041,1.292/1.071,1.309/1.099,1.325/1.125,1.341/1.148,1.358/1.169,1.374/1.187,1.390/1.204,1.407/1.218,1.423/1.231,1.439/1.243,1.455/1.251,1.471/1.257,1.486/1.261,1.564/1.275,1.595/1.282,1.627/1.292,1.658/1.304,1.689/1.318,1.720/1.336,1.751/1.357,1.783/1.381,1.814/1.408,1.845/1.438,1.876/1.472,1.908/1.508,1.939/1.548,1.986/1.611,2.048/1.701,2.157/1.860,2.204/1.925,2.251/1.987,2.298/2.046,2.345/2.100,2.391/2.150,2.438/2.196,2.485/2.238,2.532/2.276,2.579/2.311,2.626/2.343,2.672/2.371,2.719/2.397,2.766/2.420,2.813/2.441,2.860/2.460,2.906/2.477,2.953/2.492,3.016/2.510,3.078/2.525,3.141/2.538,3.203/2.549,3.281/2.561,3.375/2.572,3.468/2.581,3.593/2.590,3.734/2.598,3.921/2.604,4.140/2.608,4.233/2.609}
{
\pgfplotstreampoint{\pgfpoint{\x cm}{\y cm}}
}
\pgfplotstreamend
      \pgfusepath{stroke}
    \end{pgfscope}
  \end{pgfscope}
  \fi
  \makeatletter\ifpgf@draftmode\makeatother\else
  \begin{pgfscope}
    \pgfpathrectangle{\pgfpoint{1.12889cm}{0.846667cm}}{\pgfpoint{3.10429cm}{2.59279cm}}
    \pgfusepath{clip}
    \begin{pgfscope}
      \pgfsetlinewidth{0.10pt}
      \definecolor{matfig2pgf_linecolor}{rgb}{1.000,0.000,0.000}
      \pgfsetstrokecolor{matfig2pgf_linecolor}
      \pgfsetdash{{2.00pt}{2.00pt}}{0pt}
      \pgfsetroundjoin
      \pgfplothandlerlineto
\pgfplotstreamstart
\foreach \x/\y in {1.129/0.847,1.145/0.848,1.162/0.855,1.178/0.869,1.194/0.892,1.211/0.919,1.243/0.982,1.276/1.045,1.292/1.074,1.309/1.100,1.325/1.124,1.341/1.145,1.358/1.164,1.374/1.181,1.390/1.195,1.407/1.207,1.423/1.218,1.439/1.227,1.455/1.234,1.471/1.238,1.486/1.241,1.517/1.243,1.580/1.248,1.611/1.252,1.642/1.258,1.673/1.265,1.705/1.274,1.736/1.285,1.767/1.298,1.798/1.312,1.830/1.328,1.861/1.347,1.892/1.367,1.923/1.390,1.954/1.414,1.986/1.442,2.017/1.471,2.064/1.519,2.110/1.570,2.157/1.624,2.345/1.850,2.407/1.922,2.469/1.990,2.516/2.039,2.563/2.085,2.610/2.128,2.657/2.169,2.704/2.207,2.750/2.242,2.797/2.275,2.844/2.306,2.891/2.335,2.938/2.361,3.000/2.393,3.063/2.422,3.109/2.441,3.156/2.459,3.219/2.480,3.297/2.504,3.359/2.521,3.422/2.535,3.500/2.551,3.609/2.569,3.687/2.580,3.796/2.593,3.937/2.605,4.061/2.614,4.233/2.622}
{
\pgfplotstreampoint{\pgfpoint{\x cm}{\y cm}}
}
\pgfplotstreamend
      \pgfusepath{stroke}
    \end{pgfscope}
  \end{pgfscope}
  \fi
  \makeatletter\ifpgf@draftmode\makeatother\else
  \begin{pgfscope}
    \pgfpathrectangle{\pgfpoint{1.12889cm}{0.846667cm}}{\pgfpoint{3.10429cm}{2.59279cm}}
    \pgfusepath{clip}
    \begin{pgfscope}
      \pgfsetlinewidth{0.10pt}
      \definecolor{matfig2pgf_linecolor}{rgb}{1.000,0.000,0.000}
      \pgfsetstrokecolor{matfig2pgf_linecolor}
      \pgfsetdash{{2.00pt}{2.00pt}}{0pt}
      \pgfsetroundjoin
      \pgfplothandlerlineto
\pgfplotstreamstart
\foreach \x/\y in {1.129/0.847,1.145/0.848,1.162/0.857,1.178/0.877,1.194/0.908,1.211/0.947,1.227/0.992,1.260/1.092,1.309/1.244,1.325/1.291,1.341/1.337,1.358/1.380,1.374/1.421,1.390/1.458,1.407/1.493,1.423/1.525,1.439/1.555,1.455/1.580,1.471/1.602,1.502/1.641,1.533/1.680,1.564/1.723,1.580/1.747,1.595/1.771,1.611/1.797,1.642/1.854,1.673/1.916,1.705/1.981,1.751/2.085,1.845/2.297,1.892/2.399,1.939/2.495,1.970/2.555,2.001/2.611,2.032/2.663,2.064/2.712,2.095/2.757,2.126/2.798,2.157/2.836,2.189/2.871,2.220/2.903,2.251/2.931,2.282/2.958,2.313/2.982,2.345/3.003,2.376/3.023,2.407/3.040,2.454/3.063,2.501/3.082,2.547/3.099,2.594/3.114,2.641/3.126,2.704/3.139,2.766/3.149,2.844/3.160,2.922/3.168,3.016/3.174,3.141/3.181,3.297/3.185,3.515/3.188,3.905/3.190,4.233/3.191}
{
\pgfplotstreampoint{\pgfpoint{\x cm}{\y cm}}
}
\pgfplotstreamend
      \pgfusepath{stroke}
    \end{pgfscope}
  \end{pgfscope}
  \fi
  \makeatletter\ifpgf@draftmode\makeatother\else
  \begin{pgfscope}
    \pgfpathrectangle{\pgfpoint{1.12889cm}{0.846667cm}}{\pgfpoint{3.10429cm}{2.59279cm}}
    \pgfusepath{clip}
    \begin{pgfscope}
      \pgfsetlinewidth{0.10pt}
      \definecolor{matfig2pgf_linecolor}{rgb}{1.000,0.000,0.000}
      \pgfsetstrokecolor{matfig2pgf_linecolor}
      \pgfsetdash{{2.00pt}{2.00pt}}{0pt}
      \pgfsetroundjoin
      \pgfplothandlerlineto
\pgfplotstreamstart
\foreach \x/\y in {1.129/0.847,1.145/0.848,1.162/0.854,1.178/0.866,1.194/0.886,1.211/0.911,1.227/0.939,1.276/1.030,1.292/1.058,1.309/1.084,1.325/1.109,1.341/1.131,1.358/1.151,1.374/1.169,1.390/1.185,1.407/1.199,1.423/1.212,1.439/1.223,1.455/1.231,1.471/1.237,1.486/1.241,1.517/1.246,1.580/1.255,1.611/1.261,1.642/1.270,1.673/1.281,1.705/1.294,1.736/1.310,1.767/1.328,1.798/1.349,1.830/1.373,1.861/1.400,1.892/1.430,1.923/1.463,1.954/1.498,1.986/1.537,2.032/1.599,2.110/1.709,2.204/1.842,2.251/1.906,2.298/1.967,2.345/2.024,2.391/2.077,2.438/2.127,2.485/2.172,2.532/2.214,2.579/2.251,2.626/2.286,2.672/2.317,2.719/2.345,2.766/2.371,2.813/2.394,2.860/2.415,2.906/2.433,2.953/2.450,3.016/2.469,3.078/2.486,3.141/2.501,3.203/2.513,3.281/2.526,3.375/2.538,3.453/2.547,3.562/2.556,3.687/2.563,3.843/2.570,4.030/2.575,4.233/2.578}
{
\pgfplotstreampoint{\pgfpoint{\x cm}{\y cm}}
}
\pgfplotstreamend
      \pgfusepath{stroke}
    \end{pgfscope}
  \end{pgfscope}
  \fi
  \makeatletter\ifpgf@draftmode\makeatother\else
  \begin{pgfscope}
    \pgfpathrectangle{\pgfpoint{1.12889cm}{0.846667cm}}{\pgfpoint{3.10429cm}{2.59279cm}}
    \pgfusepath{clip}
    \begin{pgfscope}
      \pgfsetlinewidth{0.10pt}
      \definecolor{matfig2pgf_linecolor}{rgb}{1.000,0.000,0.000}
      \pgfsetstrokecolor{matfig2pgf_linecolor}
      \pgfsetdash{{2.00pt}{2.00pt}}{0pt}
      \pgfsetroundjoin
      \pgfplothandlerlineto
\pgfplotstreamstart
\foreach \x/\y in {1.129/0.847,1.145/0.848,1.162/0.854,1.178/0.866,1.194/0.886,1.211/0.911,1.227/0.940,1.276/1.031,1.292/1.060,1.309/1.088,1.325/1.113,1.341/1.136,1.358/1.157,1.374/1.176,1.390/1.193,1.407/1.208,1.423/1.221,1.439/1.234,1.455/1.243,1.471/1.249,1.486/1.254,1.517/1.260,1.564/1.270,1.595/1.278,1.627/1.288,1.658/1.301,1.673/1.309,1.689/1.318,1.705/1.327,1.720/1.337,1.736/1.348,1.751/1.360,1.767/1.373,1.798/1.402,1.814/1.418,1.845/1.453,1.876/1.492,1.908/1.535,1.939/1.581,1.970/1.630,2.017/1.709,2.173/1.986,2.220/2.067,2.267/2.143,2.313/2.216,2.360/2.285,2.407/2.350,2.454/2.410,2.501/2.465,2.547/2.517,2.594/2.565,2.626/2.595,2.672/2.636,2.719/2.673,2.766/2.708,2.813/2.739,2.860/2.768,2.906/2.793,2.953/2.816,3.000/2.837,3.047/2.857,3.094/2.874,3.141/2.889,3.203/2.907,3.281/2.926,3.344/2.939,3.406/2.951,3.500/2.964,3.609/2.977,3.687/2.984,3.827/2.994,3.952/3.001,4.155/3.007,4.233/3.009}
{
\pgfplotstreampoint{\pgfpoint{\x cm}{\y cm}}
}
\pgfplotstreamend
      \pgfusepath{stroke}
    \end{pgfscope}
  \end{pgfscope}
  \fi
  \makeatletter\ifpgf@draftmode\makeatother\else
  \begin{pgfscope}
    \pgfpathrectangle{\pgfpoint{1.12889cm}{0.846667cm}}{\pgfpoint{3.10429cm}{2.59279cm}}
    \pgfusepath{clip}
    \begin{pgfscope}
      \pgfsetlinewidth{0.10pt}
      \definecolor{matfig2pgf_linecolor}{rgb}{1.000,0.000,0.000}
      \pgfsetstrokecolor{matfig2pgf_linecolor}
      \pgfsetdash{{2.00pt}{2.00pt}}{0pt}
      \pgfsetroundjoin
      \pgfplothandlerlineto
\pgfplotstreamstart
\foreach \x/\y in {1.129/0.847,1.145/0.848,1.162/0.856,1.178/0.874,1.194/0.901,1.211/0.935,1.227/0.974,1.276/1.095,1.292/1.134,1.309/1.170,1.325/1.203,1.341/1.234,1.358/1.262,1.374/1.287,1.390/1.310,1.407/1.330,1.423/1.348,1.439/1.364,1.455/1.376,1.471/1.386,1.486/1.395,1.533/1.417,1.564/1.434,1.580/1.444,1.595/1.455,1.611/1.467,1.627/1.480,1.642/1.495,1.658/1.510,1.673/1.527,1.689/1.545,1.720/1.584,1.751/1.628,1.783/1.674,1.830/1.750,1.892/1.856,1.954/1.963,2.001/2.041,2.048/2.115,2.079/2.162,2.110/2.206,2.142/2.247,2.173/2.287,2.204/2.324,2.235/2.358,2.267/2.390,2.298/2.420,2.329/2.448,2.360/2.474,2.407/2.509,2.454/2.541,2.501/2.568,2.547/2.593,2.594/2.615,2.641/2.634,2.688/2.651,2.735/2.666,2.782/2.680,2.844/2.694,2.906/2.707,2.985/2.721,3.063/2.731,3.156/2.741,3.265/2.749,3.390/2.756,3.546/2.762,3.749/2.767,4.030/2.770,4.233/2.771}
{
\pgfplotstreampoint{\pgfpoint{\x cm}{\y cm}}
}
\pgfplotstreamend
      \pgfusepath{stroke}
    \end{pgfscope}
  \end{pgfscope}
  \fi
    \pgftext[top,x=2.66692cm,y=0.268925cm,rotate=0]{$t$}
    \pgftext[bottom,x=0.4657cm,y=2.10079cm,rotate=90]{$$C3$$}
  \makeatletter\ifpgf@draftmode\makeatother\pgftext[x=5cm,y=4.09091cm]{\Huge{DRAFT}}\fi

%% file: matlab_plots/traj_NFkB.pgf
  \begin{pgfscope}
    \definecolor{matfig2pgf_color}{rgb}{1,1,1}\pgfsetfillcolor{matfig2pgf_color}
    \pgfpathrectangle{\pgfpoint{1.12889cm}{0.846667cm}}{\pgfpoint{3.10429cm}{2.59279cm}}
    \pgfusepath{fill}
  \end{pgfscope}
  \begin{pgfscope}
    \pgfsetlinewidth{0.5pt}
    \foreach \x in {1.90496,2.68103,3.4571}
    {
      \pgfpathmoveto{\pgfpoint{\x cm}{0.846667cm}}\pgfpathlineto{\pgfpoint{\x cm}{0.872595cm}}
      \pgfpathmoveto{\pgfpoint{\x cm}{3.43946cm}}\pgfpathlineto{\pgfpoint{\x cm}{3.41353cm}}
    }
    \foreach \y in {1.36523,1.88378,2.40234,2.9209}
    {
      \pgfpathmoveto{\pgfpoint{1.12889cm}{\y cm}}\pgfpathlineto{\pgfpoint{1.15993cm}{\y cm}}
      \pgfpathmoveto{\pgfpoint{4.23318cm}{\y cm}}\pgfpathlineto{\pgfpoint{4.20213cm}{\y cm}}
    }
    \pgfusepath{stroke}
  \end{pgfscope}
  \begin{pgfscope}
    \pgfsetlinewidth{0.5pt}
    \pgfpathrectangle{\pgfpoint{1.12889cm}{0.846667cm}}{\pgfpoint{3.10429cm}{2.59279cm}}
    \pgfusepath{stroke}
  \end{pgfscope}
  {\small
    \pgftext[x=1.12889cm,y=0.746667cm,top]{$0$}
    \pgftext[x=1.90496cm,y=0.746667cm,top]{$5$}
    \pgftext[x=2.68103cm,y=0.746667cm,top]{$10$}
    \pgftext[x=3.4571cm,y=0.746667cm,top]{$15$}
    \pgftext[x=4.23318cm,y=0.746667cm,top]{$20$}
    \pgftext[x=1.02889cm,y=0.846667cm,right]{$0.1$}
    \pgftext[x=1.02889cm,y=1.36523cm,right]{$0.2$}
    \pgftext[x=1.02889cm,y=1.88378cm,right]{$0.3$}
    \pgftext[x=1.02889cm,y=2.40234cm,right]{$0.4$}
    \pgftext[x=1.02889cm,y=2.9209cm,right]{$0.5$}
    \pgftext[x=1.02889cm,y=3.43946cm,right]{$0.6$}
  }
  \makeatletter\ifpgf@draftmode\makeatother\else
  \begin{pgfscope}
    \pgfpathrectangle{\pgfpoint{1.12889cm}{0.846667cm}}{\pgfpoint{3.10429cm}{2.59279cm}}
    \pgfusepath{clip}
    \begin{pgfscope}
      \pgfsetlinewidth{0.10pt}
      \definecolor{matfig2pgf_linecolor}{rgb}{0.000,0.000,1.000}
      \pgfsetstrokecolor{matfig2pgf_linecolor}
      \pgfsetdash{}{0pt}
      \pgfsetroundjoin
      \pgfplothandlerlineto
\pgfplotstreamstart
\foreach \x/\y in {1.129/1.892,1.145/1.900,1.162/1.924,1.178/1.962,1.194/2.013,1.211/2.074,1.227/2.144,1.243/2.220,1.276/2.385,1.325/2.636,1.341/2.716,1.358/2.792,1.374/2.864,1.390/2.931,1.407/2.993,1.423/3.050,1.439/3.102,1.455/3.132,1.471/3.132,1.486/3.109,1.502/3.069,1.517/3.018,1.533/2.959,1.564/2.829,1.611/2.631,1.642/2.508,1.658/2.451,1.673/2.397,1.689/2.346,1.705/2.299,1.720/2.256,1.736/2.215,1.751/2.178,1.767/2.145,1.783/2.114,1.798/2.086,1.814/2.060,1.830/2.037,1.845/2.017,1.861/1.998,1.876/1.982,1.892/1.967,1.908/1.955,1.923/1.943,1.939/1.933,1.954/1.925,1.970/1.917,1.986/1.911,2.001/1.905,2.017/1.901,2.048/1.893,2.079/1.888,2.110/1.885,2.157/1.883,2.220/1.883,2.516/1.890,2.735/1.892,4.233/1.892}
{
\pgfplotstreampoint{\pgfpoint{\x cm}{\y cm}}
}
\pgfplotstreamend
      \pgfusepath{stroke}
    \end{pgfscope}
  \end{pgfscope}
  \fi
  \makeatletter\ifpgf@draftmode\makeatother\else
  \begin{pgfscope}
    \pgfpathrectangle{\pgfpoint{1.12889cm}{0.846667cm}}{\pgfpoint{3.10429cm}{2.59279cm}}
    \pgfusepath{clip}
    \begin{pgfscope}
      \pgfsetlinewidth{0.10pt}
      \definecolor{matfig2pgf_linecolor}{rgb}{0.000,0.000,1.000}
      \pgfsetstrokecolor{matfig2pgf_linecolor}
      \pgfsetdash{}{0pt}
      \pgfsetroundjoin
      \pgfplothandlerlineto
\pgfplotstreamstart
\foreach \x/\y in {1.129/1.770,1.145/1.778,1.162/1.799,1.178/1.834,1.194/1.879,1.211/1.934,1.227/1.997,1.243/2.065,1.276/2.213,1.325/2.437,1.341/2.507,1.358/2.575,1.374/2.638,1.390/2.697,1.407/2.752,1.423/2.802,1.439/2.847,1.455/2.872,1.471/2.870,1.486/2.847,1.502/2.810,1.517/2.763,1.533/2.709,1.564/2.591,1.611/2.411,1.642/2.301,1.658/2.249,1.673/2.201,1.689/2.156,1.705/2.114,1.720/2.075,1.736/2.039,1.751/2.007,1.767/1.977,1.783/1.949,1.798/1.925,1.814/1.903,1.830/1.883,1.845/1.865,1.861/1.849,1.876/1.835,1.892/1.823,1.908/1.812,1.923/1.803,1.939/1.795,1.954/1.788,1.970/1.782,1.986/1.777,2.001/1.773,2.032/1.766,2.064/1.762,2.110/1.759,2.157/1.759,2.251/1.761,2.423/1.767,2.579/1.769,2.891/1.770,4.233/1.770}
{
\pgfplotstreampoint{\pgfpoint{\x cm}{\y cm}}
}
\pgfplotstreamend
      \pgfusepath{stroke}
    \end{pgfscope}
  \end{pgfscope}
  \fi
  \makeatletter\ifpgf@draftmode\makeatother\else
  \begin{pgfscope}
    \pgfpathrectangle{\pgfpoint{1.12889cm}{0.846667cm}}{\pgfpoint{3.10429cm}{2.59279cm}}
    \pgfusepath{clip}
    \begin{pgfscope}
      \pgfsetlinewidth{0.10pt}
      \definecolor{matfig2pgf_linecolor}{rgb}{0.000,0.000,1.000}
      \pgfsetstrokecolor{matfig2pgf_linecolor}
      \pgfsetdash{}{0pt}
      \pgfsetroundjoin
      \pgfplothandlerlineto
\pgfplotstreamstart
\foreach \x/\y in {1.129/1.890,1.145/1.898,1.162/1.922,1.178/1.960,1.194/2.011,1.211/2.072,1.227/2.141,1.243/2.218,1.276/2.382,1.325/2.632,1.341/2.711,1.358/2.787,1.374/2.858,1.390/2.925,1.407/2.987,1.423/3.044,1.439/3.095,1.455/3.125,1.471/3.124,1.486/3.101,1.502/3.061,1.517/3.010,1.533/2.951,1.564/2.822,1.611/2.624,1.642/2.501,1.658/2.444,1.673/2.390,1.689/2.340,1.705/2.293,1.720/2.250,1.736/2.210,1.751/2.173,1.767/2.139,1.783/2.109,1.798/2.081,1.814/2.056,1.830/2.033,1.845/2.013,1.861/1.994,1.876/1.978,1.892/1.964,1.908/1.951,1.923/1.940,1.939/1.930,1.954/1.922,1.970/1.914,1.986/1.908,2.001/1.902,2.017/1.898,2.048/1.891,2.079/1.886,2.110/1.883,2.157/1.881,2.220/1.880,2.516/1.888,2.735/1.889,4.233/1.890}
{
\pgfplotstreampoint{\pgfpoint{\x cm}{\y cm}}
}
\pgfplotstreamend
      \pgfusepath{stroke}
    \end{pgfscope}
  \end{pgfscope}
  \fi
  \makeatletter\ifpgf@draftmode\makeatother\else
  \begin{pgfscope}
    \pgfpathrectangle{\pgfpoint{1.12889cm}{0.846667cm}}{\pgfpoint{3.10429cm}{2.59279cm}}
    \pgfusepath{clip}
    \begin{pgfscope}
      \pgfsetlinewidth{0.10pt}
      \definecolor{matfig2pgf_linecolor}{rgb}{0.000,0.000,1.000}
      \pgfsetstrokecolor{matfig2pgf_linecolor}
      \pgfsetdash{}{0pt}
      \pgfsetroundjoin
      \pgfplothandlerlineto
\pgfplotstreamstart
\foreach \x/\y in {1.129/1.835,1.145/1.843,1.162/1.866,1.178/1.902,1.194/1.950,1.211/2.009,1.227/2.075,1.243/2.148,1.276/2.304,1.325/2.542,1.341/2.618,1.358/2.689,1.374/2.757,1.390/2.820,1.407/2.879,1.423/2.932,1.439/2.981,1.455/3.008,1.471/3.007,1.486/2.984,1.502/2.945,1.517/2.896,1.533/2.839,1.564/2.714,1.611/2.525,1.642/2.408,1.658/2.353,1.673/2.302,1.689/2.254,1.705/2.210,1.720/2.169,1.736/2.131,1.751/2.096,1.767/2.064,1.783/2.035,1.798/2.008,1.814/1.985,1.830/1.963,1.845/1.944,1.861/1.927,1.876/1.912,1.892/1.899,1.908/1.887,1.923/1.876,1.939/1.868,1.954/1.860,1.970/1.853,1.986/1.847,2.001/1.842,2.032/1.835,2.064/1.830,2.095/1.827,2.142/1.825,2.204/1.825,2.469/1.832,2.657/1.834,3.297/1.835,4.233/1.835}
{
\pgfplotstreampoint{\pgfpoint{\x cm}{\y cm}}
}
\pgfplotstreamend
      \pgfusepath{stroke}
    \end{pgfscope}
  \end{pgfscope}
  \fi
  \makeatletter\ifpgf@draftmode\makeatother\else
  \begin{pgfscope}
    \pgfpathrectangle{\pgfpoint{1.12889cm}{0.846667cm}}{\pgfpoint{3.10429cm}{2.59279cm}}
    \pgfusepath{clip}
    \begin{pgfscope}
      \pgfsetlinewidth{0.10pt}
      \definecolor{matfig2pgf_linecolor}{rgb}{0.000,0.000,1.000}
      \pgfsetstrokecolor{matfig2pgf_linecolor}
      \pgfsetdash{}{0pt}
      \pgfsetroundjoin
      \pgfplothandlerlineto
\pgfplotstreamstart
\foreach \x/\y in {1.129/1.960,1.145/1.969,1.162/1.994,1.178/2.035,1.194/2.088,1.211/2.153,1.227/2.227,1.243/2.308,1.276/2.482,1.325/2.750,1.358/2.918,1.374/2.995,1.390/3.068,1.407/3.136,1.423/3.198,1.439/3.255,1.455/3.289,1.471/3.291,1.486/3.268,1.502/3.229,1.517/3.176,1.533/3.115,1.564/2.979,1.611/2.770,1.642/2.639,1.658/2.578,1.673/2.521,1.689/2.467,1.705/2.417,1.720/2.370,1.736/2.327,1.751/2.287,1.767/2.250,1.783/2.217,1.798/2.186,1.814/2.159,1.830/2.134,1.845/2.111,1.861/2.091,1.876/2.072,1.892/2.056,1.908/2.041,1.923/2.028,1.939/2.017,1.954/2.007,1.970/1.998,1.986/1.991,2.001/1.984,2.017/1.978,2.032/1.973,2.064/1.966,2.095/1.960,2.126/1.957,2.173/1.954,2.235/1.953,2.360/1.954,2.579/1.959,2.844/1.960,4.233/1.960}
{
\pgfplotstreampoint{\pgfpoint{\x cm}{\y cm}}
}
\pgfplotstreamend
      \pgfusepath{stroke}
    \end{pgfscope}
  \end{pgfscope}
  \fi
  \makeatletter\ifpgf@draftmode\makeatother\else
  \begin{pgfscope}
    \pgfpathrectangle{\pgfpoint{1.12889cm}{0.846667cm}}{\pgfpoint{3.10429cm}{2.59279cm}}
    \pgfusepath{clip}
    \begin{pgfscope}
      \pgfsetlinewidth{0.10pt}
      \definecolor{matfig2pgf_linecolor}{rgb}{0.000,0.000,1.000}
      \pgfsetstrokecolor{matfig2pgf_linecolor}
      \pgfsetdash{}{0pt}
      \pgfsetroundjoin
      \pgfplothandlerlineto
\pgfplotstreamstart
\foreach \x/\y in {1.129/1.833,1.145/1.840,1.162/1.863,1.178/1.900,1.194/1.948,1.211/2.006,1.227/2.072,1.243/2.145,1.276/2.301,1.325/2.539,1.341/2.615,1.358/2.687,1.374/2.755,1.390/2.818,1.407/2.877,1.423/2.931,1.439/2.980,1.455/3.007,1.471/3.006,1.486/2.983,1.502/2.944,1.517/2.895,1.533/2.838,1.564/2.714,1.611/2.524,1.642/2.407,1.658/2.353,1.673/2.302,1.689/2.254,1.705/2.209,1.720/2.168,1.736/2.130,1.751/2.095,1.767/2.063,1.783/2.034,1.798/2.007,1.814/1.984,1.830/1.962,1.845/1.943,1.861/1.926,1.876/1.911,1.892/1.897,1.908/1.885,1.923/1.875,1.939/1.866,1.954/1.858,1.970/1.851,1.986/1.845,2.001/1.841,2.032/1.833,2.064/1.828,2.095/1.825,2.142/1.823,2.204/1.822,2.485/1.830,2.672/1.832,3.796/1.833,4.233/1.833}
{
\pgfplotstreampoint{\pgfpoint{\x cm}{\y cm}}
}
\pgfplotstreamend
      \pgfusepath{stroke}
    \end{pgfscope}
  \end{pgfscope}
  \fi
  \makeatletter\ifpgf@draftmode\makeatother\else
  \begin{pgfscope}
    \pgfpathrectangle{\pgfpoint{1.12889cm}{0.846667cm}}{\pgfpoint{3.10429cm}{2.59279cm}}
    \pgfusepath{clip}
    \begin{pgfscope}
      \pgfsetlinewidth{0.10pt}
      \definecolor{matfig2pgf_linecolor}{rgb}{0.000,0.000,1.000}
      \pgfsetstrokecolor{matfig2pgf_linecolor}
      \pgfsetdash{}{0pt}
      \pgfsetroundjoin
      \pgfplothandlerlineto
\pgfplotstreamstart
\foreach \x/\y in {1.129/1.696,1.145/1.703,1.162/1.724,1.178/1.756,1.194/1.798,1.211/1.849,1.227/1.907,1.243/1.971,1.292/2.177,1.325/2.313,1.341/2.378,1.358/2.439,1.374/2.496,1.390/2.550,1.407/2.599,1.423/2.645,1.439/2.686,1.455/2.707,1.471/2.704,1.486/2.682,1.502/2.647,1.517/2.603,1.533/2.552,1.564/2.442,1.611/2.275,1.642/2.173,1.658/2.125,1.673/2.081,1.689/2.039,1.705/2.000,1.720/1.965,1.736/1.932,1.751/1.901,1.767/1.874,1.783/1.849,1.798/1.827,1.814/1.806,1.830/1.788,1.845/1.772,1.861/1.758,1.876/1.745,1.892/1.735,1.908/1.725,1.923/1.717,1.939/1.710,1.954/1.704,1.970/1.699,1.986/1.695,2.017/1.689,2.048/1.685,2.095/1.683,2.142/1.683,2.251/1.687,2.391/1.693,2.532/1.695,2.782/1.696,4.233/1.696}
{
\pgfplotstreampoint{\pgfpoint{\x cm}{\y cm}}
}
\pgfplotstreamend
      \pgfusepath{stroke}
    \end{pgfscope}
  \end{pgfscope}
  \fi
  \makeatletter\ifpgf@draftmode\makeatother\else
  \begin{pgfscope}
    \pgfpathrectangle{\pgfpoint{1.12889cm}{0.846667cm}}{\pgfpoint{3.10429cm}{2.59279cm}}
    \pgfusepath{clip}
    \begin{pgfscope}
      \pgfsetlinewidth{0.10pt}
      \definecolor{matfig2pgf_linecolor}{rgb}{0.000,0.000,1.000}
      \pgfsetstrokecolor{matfig2pgf_linecolor}
      \pgfsetdash{}{0pt}
      \pgfsetroundjoin
      \pgfplothandlerlineto
\pgfplotstreamstart
\foreach \x/\y in {1.129/1.863,1.145/1.871,1.162/1.895,1.178/1.932,1.194/1.982,1.211/2.041,1.227/2.109,1.243/2.184,1.276/2.344,1.325/2.587,1.341/2.664,1.358/2.738,1.374/2.807,1.390/2.871,1.407/2.931,1.423/2.986,1.439/3.035,1.455/3.063,1.471/3.062,1.486/3.039,1.502/2.999,1.517/2.949,1.533/2.891,1.564/2.764,1.611/2.570,1.642/2.451,1.658/2.395,1.673/2.343,1.689/2.294,1.705/2.248,1.720/2.206,1.736/2.167,1.751/2.131,1.767/2.099,1.783/2.069,1.798/2.042,1.814/2.018,1.830/1.996,1.845/1.976,1.861/1.959,1.876/1.943,1.892/1.929,1.908/1.917,1.923/1.907,1.939/1.898,1.954/1.890,1.970/1.883,1.986/1.877,2.001/1.872,2.032/1.864,2.064/1.859,2.095/1.856,2.142/1.854,2.204/1.853,2.485/1.861,2.672/1.863,3.656/1.863,4.233/1.863}
{
\pgfplotstreampoint{\pgfpoint{\x cm}{\y cm}}
}
\pgfplotstreamend
      \pgfusepath{stroke}
    \end{pgfscope}
  \end{pgfscope}
  \fi
  \makeatletter\ifpgf@draftmode\makeatother\else
  \begin{pgfscope}
    \pgfpathrectangle{\pgfpoint{1.12889cm}{0.846667cm}}{\pgfpoint{3.10429cm}{2.59279cm}}
    \pgfusepath{clip}
    \begin{pgfscope}
      \pgfsetlinewidth{0.10pt}
      \definecolor{matfig2pgf_linecolor}{rgb}{1.000,0.000,0.000}
      \pgfsetstrokecolor{matfig2pgf_linecolor}
      \pgfsetdash{{2.00pt}{2.00pt}}{0pt}
      \pgfsetroundjoin
      \pgfplothandlerlineto
\pgfplotstreamstart
\foreach \x/\y in {1.129/1.743,1.145/1.750,1.162/1.771,1.178/1.805,1.194/1.849,1.211/1.903,1.227/1.963,1.243/2.029,1.309/2.310,1.325/2.377,1.341/2.441,1.358/2.500,1.374/2.555,1.390/2.605,1.407/2.650,1.423/2.691,1.439/2.726,1.455/2.743,1.471/2.735,1.486/2.708,1.502/2.667,1.517/2.618,1.533/2.563,1.580/2.382,1.611/2.263,1.642/2.151,1.658/2.098,1.673/2.048,1.689/2.001,1.705/1.956,1.720/1.914,1.736/1.875,1.751/1.838,1.767/1.803,1.783/1.771,1.798/1.740,1.814/1.712,1.830/1.686,1.845/1.661,1.861/1.637,1.876/1.616,1.892/1.595,1.908/1.576,1.923/1.558,1.954/1.524,1.986/1.495,2.017/1.469,2.048/1.445,2.079/1.424,2.110/1.405,2.142/1.388,2.189/1.365,2.235/1.345,2.282/1.329,2.329/1.314,2.376/1.301,2.423/1.290,2.485/1.278,2.563/1.266,2.641/1.256,2.719/1.249,2.828/1.242,2.953/1.236,3.125/1.231,3.344/1.228,3.718/1.226,4.233/1.225}
{
\pgfplotstreampoint{\pgfpoint{\x cm}{\y cm}}
}
\pgfplotstreamend
      \pgfusepath{stroke}
    \end{pgfscope}
  \end{pgfscope}
  \fi
  \makeatletter\ifpgf@draftmode\makeatother\else
  \begin{pgfscope}
    \pgfpathrectangle{\pgfpoint{1.12889cm}{0.846667cm}}{\pgfpoint{3.10429cm}{2.59279cm}}
    \pgfusepath{clip}
    \begin{pgfscope}
      \pgfsetlinewidth{0.10pt}
      \definecolor{matfig2pgf_linecolor}{rgb}{1.000,0.000,0.000}
      \pgfsetstrokecolor{matfig2pgf_linecolor}
      \pgfsetdash{{2.00pt}{2.00pt}}{0pt}
      \pgfsetroundjoin
      \pgfplothandlerlineto
\pgfplotstreamstart
\foreach \x/\y in {1.129/1.676,1.145/1.683,1.162/1.703,1.178/1.734,1.194/1.776,1.211/1.826,1.227/1.883,1.243/1.944,1.309/2.207,1.325/2.270,1.341/2.330,1.358/2.385,1.374/2.437,1.390/2.484,1.407/2.526,1.423/2.564,1.439/2.598,1.455/2.614,1.471/2.606,1.486/2.580,1.502/2.542,1.517/2.496,1.533/2.444,1.611/2.163,1.642/2.058,1.658/2.009,1.673/1.962,1.689/1.918,1.705/1.876,1.720/1.837,1.736/1.800,1.751/1.765,1.767/1.733,1.783/1.702,1.798/1.674,1.814/1.647,1.830/1.622,1.845/1.598,1.861/1.576,1.876/1.555,1.908/1.517,1.939/1.483,1.970/1.452,2.001/1.424,2.032/1.399,2.064/1.376,2.095/1.355,2.142/1.326,2.189/1.301,2.235/1.279,2.282/1.260,2.329/1.243,2.376/1.228,2.423/1.214,2.485/1.200,2.547/1.187,2.610/1.177,2.688/1.166,2.782/1.157,2.875/1.149,2.985/1.143,3.141/1.138,3.344/1.134,3.640/1.131,4.233/1.130}
{
\pgfplotstreampoint{\pgfpoint{\x cm}{\y cm}}
}
\pgfplotstreamend
      \pgfusepath{stroke}
    \end{pgfscope}
  \end{pgfscope}
  \fi
  \makeatletter\ifpgf@draftmode\makeatother\else
  \begin{pgfscope}
    \pgfpathrectangle{\pgfpoint{1.12889cm}{0.846667cm}}{\pgfpoint{3.10429cm}{2.59279cm}}
    \pgfusepath{clip}
    \begin{pgfscope}
      \pgfsetlinewidth{0.10pt}
      \definecolor{matfig2pgf_linecolor}{rgb}{1.000,0.000,0.000}
      \pgfsetstrokecolor{matfig2pgf_linecolor}
      \pgfsetdash{{2.00pt}{2.00pt}}{0pt}
      \pgfsetroundjoin
      \pgfplothandlerlineto
\pgfplotstreamstart
\foreach \x/\y in {1.129/1.670,1.145/1.677,1.162/1.696,1.178/1.728,1.194/1.769,1.211/1.819,1.227/1.875,1.243/1.937,1.309/2.199,1.325/2.262,1.341/2.322,1.358/2.378,1.374/2.431,1.390/2.479,1.407/2.523,1.423/2.563,1.439/2.598,1.455/2.616,1.471/2.610,1.486/2.586,1.502/2.550,1.517/2.505,1.533/2.455,1.580/2.290,1.611/2.180,1.642/2.078,1.658/2.031,1.673/1.986,1.689/1.943,1.705/1.903,1.720/1.866,1.736/1.831,1.751/1.799,1.767/1.769,1.783/1.741,1.798/1.715,1.814/1.691,1.830/1.668,1.845/1.647,1.861/1.628,1.876/1.610,1.892/1.593,1.908/1.577,1.939/1.548,1.970/1.523,2.001/1.500,2.032/1.479,2.079/1.451,2.126/1.426,2.173/1.403,2.220/1.383,2.282/1.358,2.345/1.336,2.407/1.317,2.469/1.301,2.532/1.286,2.594/1.274,2.672/1.261,2.766/1.248,2.860/1.238,2.969/1.230,3.094/1.223,3.234/1.217,3.422/1.213,3.702/1.209,4.171/1.207,4.233/1.207}
{
\pgfplotstreampoint{\pgfpoint{\x cm}{\y cm}}
}
\pgfplotstreamend
      \pgfusepath{stroke}
    \end{pgfscope}
  \end{pgfscope}
  \fi
  \makeatletter\ifpgf@draftmode\makeatother\else
  \begin{pgfscope}
    \pgfpathrectangle{\pgfpoint{1.12889cm}{0.846667cm}}{\pgfpoint{3.10429cm}{2.59279cm}}
    \pgfusepath{clip}
    \begin{pgfscope}
      \pgfsetlinewidth{0.10pt}
      \definecolor{matfig2pgf_linecolor}{rgb}{1.000,0.000,0.000}
      \pgfsetstrokecolor{matfig2pgf_linecolor}
      \pgfsetdash{{2.00pt}{2.00pt}}{0pt}
      \pgfsetroundjoin
      \pgfplothandlerlineto
\pgfplotstreamstart
\foreach \x/\y in {1.129/1.835,1.145/1.843,1.162/1.866,1.178/1.902,1.194/1.950,1.211/2.008,1.227/2.074,1.243/2.146,1.309/2.452,1.325/2.524,1.341/2.592,1.358/2.656,1.374/2.714,1.390/2.766,1.407/2.813,1.423/2.853,1.439/2.888,1.455/2.903,1.471/2.890,1.486/2.858,1.502/2.812,1.517/2.756,1.533/2.693,1.580/2.489,1.611/2.354,1.642/2.226,1.673/2.108,1.689/2.053,1.705/2.001,1.720/1.952,1.736/1.905,1.751/1.861,1.767/1.819,1.783/1.779,1.798/1.741,1.814/1.706,1.830/1.673,1.845/1.641,1.861/1.611,1.876/1.583,1.892/1.556,1.908/1.531,1.923/1.508,1.939/1.485,1.970/1.444,2.001/1.407,2.032/1.373,2.064/1.344,2.095/1.317,2.126/1.294,2.157/1.272,2.189/1.253,2.220/1.236,2.251/1.221,2.282/1.208,2.329/1.190,2.376/1.175,2.423/1.163,2.469/1.152,2.532/1.141,2.594/1.132,2.672/1.123,2.750/1.117,2.860/1.111,3.000/1.106,3.187/1.102,3.484/1.100,4.233/1.099}
{
\pgfplotstreampoint{\pgfpoint{\x cm}{\y cm}}
}
\pgfplotstreamend
      \pgfusepath{stroke}
    \end{pgfscope}
  \end{pgfscope}
  \fi
  \makeatletter\ifpgf@draftmode\makeatother\else
  \begin{pgfscope}
    \pgfpathrectangle{\pgfpoint{1.12889cm}{0.846667cm}}{\pgfpoint{3.10429cm}{2.59279cm}}
    \pgfusepath{clip}
    \begin{pgfscope}
      \pgfsetlinewidth{0.10pt}
      \definecolor{matfig2pgf_linecolor}{rgb}{1.000,0.000,0.000}
      \pgfsetstrokecolor{matfig2pgf_linecolor}
      \pgfsetdash{{2.00pt}{2.00pt}}{0pt}
      \pgfsetroundjoin
      \pgfplothandlerlineto
\pgfplotstreamstart
\foreach \x/\y in {1.129/1.940,1.145/1.949,1.162/1.974,1.178/2.014,1.194/2.067,1.211/2.130,1.227/2.203,1.243/2.282,1.276/2.451,1.325/2.708,1.341/2.789,1.358/2.866,1.374/2.938,1.390/3.005,1.407/3.066,1.423/3.123,1.439/3.174,1.455/3.203,1.471/3.201,1.486/3.177,1.502/3.135,1.517/3.082,1.533/3.021,1.564/2.886,1.611/2.679,1.642/2.551,1.658/2.491,1.673/2.434,1.689/2.381,1.705/2.331,1.720/2.284,1.736/2.241,1.751/2.200,1.767/2.163,1.783/2.128,1.798/2.097,1.814/2.067,1.830/2.040,1.845/2.015,1.861/1.992,1.876/1.971,1.892/1.951,1.908/1.933,1.923/1.916,1.954/1.885,1.986/1.858,2.017/1.833,2.064/1.800,2.110/1.769,2.173/1.731,2.251/1.686,2.313/1.652,2.376/1.621,2.438/1.591,2.501/1.565,2.563/1.540,2.626/1.519,2.688/1.500,2.750/1.483,2.813/1.468,2.891/1.452,2.969/1.438,3.047/1.427,3.141/1.416,3.250/1.406,3.359/1.398,3.531/1.389,3.687/1.384,3.905/1.379,4.218/1.376,4.233/1.375}
{
\pgfplotstreampoint{\pgfpoint{\x cm}{\y cm}}
}
\pgfplotstreamend
      \pgfusepath{stroke}
    \end{pgfscope}
  \end{pgfscope}
  \fi
  \makeatletter\ifpgf@draftmode\makeatother\else
  \begin{pgfscope}
    \pgfpathrectangle{\pgfpoint{1.12889cm}{0.846667cm}}{\pgfpoint{3.10429cm}{2.59279cm}}
    \pgfusepath{clip}
    \begin{pgfscope}
      \pgfsetlinewidth{0.10pt}
      \definecolor{matfig2pgf_linecolor}{rgb}{1.000,0.000,0.000}
      \pgfsetstrokecolor{matfig2pgf_linecolor}
      \pgfsetdash{{2.00pt}{2.00pt}}{0pt}
      \pgfsetroundjoin
      \pgfplothandlerlineto
\pgfplotstreamstart
\foreach \x/\y in {1.129/1.830,1.145/1.838,1.162/1.861,1.178/1.897,1.194/1.945,1.211/2.003,1.227/2.069,1.243/2.140,1.309/2.441,1.325/2.512,1.341/2.578,1.358/2.640,1.374/2.696,1.390/2.746,1.407/2.791,1.423/2.830,1.439/2.863,1.455/2.877,1.471/2.865,1.486/2.833,1.502/2.787,1.517/2.731,1.533/2.669,1.611/2.335,1.642/2.209,1.673/2.093,1.689/2.039,1.705/1.988,1.720/1.940,1.736/1.894,1.751/1.851,1.767/1.809,1.783/1.771,1.798/1.734,1.814/1.699,1.830/1.666,1.845/1.636,1.861/1.606,1.876/1.579,1.892/1.553,1.908/1.528,1.923/1.505,1.954/1.462,1.986/1.423,2.017/1.388,2.048/1.357,2.079/1.329,2.110/1.304,2.142/1.282,2.173/1.261,2.204/1.243,2.235/1.226,2.267/1.212,2.313/1.193,2.360/1.176,2.407/1.162,2.454/1.150,2.501/1.140,2.563/1.130,2.641/1.119,2.719/1.111,2.813/1.104,2.938/1.097,3.078/1.093,3.297/1.089,3.656/1.087,4.233/1.087}
{
\pgfplotstreampoint{\pgfpoint{\x cm}{\y cm}}
}
\pgfplotstreamend
      \pgfusepath{stroke}
    \end{pgfscope}
  \end{pgfscope}
  \fi
  \makeatletter\ifpgf@draftmode\makeatother\else
  \begin{pgfscope}
    \pgfpathrectangle{\pgfpoint{1.12889cm}{0.846667cm}}{\pgfpoint{3.10429cm}{2.59279cm}}
    \pgfusepath{clip}
    \begin{pgfscope}
      \pgfsetlinewidth{0.10pt}
      \definecolor{matfig2pgf_linecolor}{rgb}{1.000,0.000,0.000}
      \pgfsetstrokecolor{matfig2pgf_linecolor}
      \pgfsetdash{{2.00pt}{2.00pt}}{0pt}
      \pgfsetroundjoin
      \pgfplothandlerlineto
\pgfplotstreamstart
\foreach \x/\y in {1.129/1.930,1.145/1.938,1.162/1.963,1.178/2.002,1.194/2.055,1.211/2.118,1.227/2.189,1.243/2.268,1.325/2.688,1.341/2.767,1.358/2.842,1.374/2.912,1.390/2.977,1.407/3.036,1.423/3.091,1.439/3.140,1.455/3.166,1.471/3.163,1.486/3.138,1.502/3.096,1.517/3.042,1.533/2.981,1.564/2.846,1.611/2.640,1.642/2.512,1.658/2.452,1.673/2.395,1.689/2.341,1.705/2.291,1.720/2.244,1.736/2.200,1.751/2.158,1.767/2.120,1.783/2.084,1.798/2.051,1.814/2.020,1.830/1.992,1.845/1.965,1.861/1.940,1.876/1.917,1.892/1.895,1.908/1.874,1.939/1.836,1.970/1.802,2.001/1.770,2.032/1.741,2.079/1.702,2.126/1.665,2.173/1.631,2.220/1.599,2.267/1.570,2.313/1.543,2.360/1.518,2.407/1.495,2.454/1.473,2.501/1.454,2.547/1.437,2.610/1.416,2.672/1.398,2.735/1.382,2.797/1.369,2.875/1.355,2.969/1.341,3.047/1.331,3.141/1.322,3.281/1.312,3.406/1.306,3.593/1.300,3.827/1.295,4.155/1.292,4.233/1.292}
{
\pgfplotstreampoint{\pgfpoint{\x cm}{\y cm}}
}
\pgfplotstreamend
      \pgfusepath{stroke}
    \end{pgfscope}
  \end{pgfscope}
  \fi
  \makeatletter\ifpgf@draftmode\makeatother\else
  \begin{pgfscope}
    \pgfpathrectangle{\pgfpoint{1.12889cm}{0.846667cm}}{\pgfpoint{3.10429cm}{2.59279cm}}
    \pgfusepath{clip}
    \begin{pgfscope}
      \pgfsetlinewidth{0.10pt}
      \definecolor{matfig2pgf_linecolor}{rgb}{1.000,0.000,0.000}
      \pgfsetstrokecolor{matfig2pgf_linecolor}
      \pgfsetdash{{2.00pt}{2.00pt}}{0pt}
      \pgfsetroundjoin
      \pgfplothandlerlineto
\pgfplotstreamstart
\foreach \x/\y in {1.129/1.935,1.145/1.943,1.162/1.968,1.178/2.008,1.194/2.060,1.211/2.123,1.227/2.196,1.243/2.275,1.276/2.445,1.325/2.703,1.341/2.785,1.358/2.863,1.374/2.937,1.390/3.005,1.407/3.069,1.423/3.127,1.439/3.179,1.455/3.209,1.471/3.209,1.486/3.185,1.502/3.144,1.517/3.091,1.533/3.031,1.564/2.896,1.611/2.690,1.642/2.562,1.658/2.502,1.673/2.446,1.689/2.393,1.705/2.343,1.720/2.297,1.736/2.254,1.751/2.215,1.767/2.178,1.783/2.145,1.798/2.114,1.814/2.086,1.830/2.060,1.845/2.036,1.861/2.015,1.876/1.995,1.892/1.977,1.908/1.961,1.923/1.946,1.939/1.932,1.954/1.919,1.986/1.897,2.017/1.878,2.048/1.860,2.095/1.838,2.157/1.810,2.251/1.774,2.329/1.745,2.407/1.719,2.485/1.695,2.563/1.674,2.641/1.656,2.719/1.641,2.797/1.629,2.875/1.618,2.969/1.608,3.078/1.599,3.203/1.592,3.344/1.586,3.546/1.581,3.812/1.577,4.233/1.575}
{
\pgfplotstreampoint{\pgfpoint{\x cm}{\y cm}}
}
\pgfplotstreamend
      \pgfusepath{stroke}
    \end{pgfscope}
  \end{pgfscope}
  \fi
  \makeatletter\ifpgf@draftmode\makeatother\else
  \begin{pgfscope}
    \pgfpathrectangle{\pgfpoint{1.12889cm}{0.846667cm}}{\pgfpoint{3.10429cm}{2.59279cm}}
    \pgfusepath{clip}
    \begin{pgfscope}
      \pgfsetlinewidth{0.10pt}
      \definecolor{matfig2pgf_linecolor}{rgb}{1.000,0.000,0.000}
      \pgfsetstrokecolor{matfig2pgf_linecolor}
      \pgfsetdash{{2.00pt}{2.00pt}}{0pt}
      \pgfsetroundjoin
      \pgfplothandlerlineto
\pgfplotstreamstart
\foreach \x/\y in {1.129/1.905,1.145/1.914,1.162/1.938,1.178/1.977,1.194/2.028,1.211/2.090,1.227/2.160,1.243/2.236,1.309/2.564,1.325/2.642,1.341/2.717,1.358/2.787,1.374/2.852,1.390/2.911,1.407/2.965,1.423/3.013,1.439/3.055,1.455/3.075,1.471/3.067,1.486/3.038,1.502/2.992,1.517/2.936,1.533/2.873,1.564/2.735,1.611/2.525,1.642/2.394,1.658/2.332,1.673/2.273,1.689/2.217,1.705/2.164,1.720/2.114,1.736/2.066,1.751/2.022,1.767/1.979,1.783/1.940,1.798/1.902,1.814/1.867,1.830/1.834,1.845/1.803,1.861/1.773,1.876/1.745,1.892/1.719,1.908/1.695,1.923/1.671,1.954/1.628,1.986/1.589,2.017/1.555,2.048/1.523,2.079/1.495,2.110/1.470,2.142/1.447,2.173/1.426,2.204/1.407,2.235/1.390,2.282/1.368,2.329/1.349,2.376/1.332,2.423/1.317,2.469/1.305,2.532/1.292,2.594/1.281,2.657/1.272,2.735/1.263,2.828/1.255,2.938/1.248,3.063/1.244,3.250/1.239,3.531/1.236,4.093/1.235,4.233/1.235}
{
\pgfplotstreampoint{\pgfpoint{\x cm}{\y cm}}
}
\pgfplotstreamend
      \pgfusepath{stroke}
    \end{pgfscope}
  \end{pgfscope}
  \fi
  \makeatletter\ifpgf@draftmode\makeatother\else
  \begin{pgfscope}
    \pgfpathrectangle{\pgfpoint{1.12889cm}{0.846667cm}}{\pgfpoint{3.10429cm}{2.59279cm}}
    \pgfusepath{clip}
    \begin{pgfscope}
      \pgfsetlinewidth{0.10pt}
      \definecolor{matfig2pgf_linecolor}{rgb}{1.000,0.000,0.000}
      \pgfsetstrokecolor{matfig2pgf_linecolor}
      \pgfsetdash{{2.00pt}{2.00pt}}{0pt}
      \pgfsetroundjoin
      \pgfplothandlerlineto
\pgfplotstreamstart
\foreach \x/\y in {1.129/1.848,1.145/1.856,1.162/1.879,1.178/1.916,1.194/1.964,1.211/2.023,1.227/2.090,1.243/2.163,1.292/2.401,1.325/2.557,1.341/2.631,1.358/2.702,1.374/2.768,1.390/2.829,1.407/2.885,1.423/2.937,1.439/2.983,1.455/3.009,1.471/3.005,1.486/2.981,1.502/2.941,1.517/2.890,1.533/2.833,1.564/2.706,1.611/2.514,1.642/2.395,1.658/2.340,1.673/2.288,1.689/2.239,1.705/2.193,1.720/2.151,1.736/2.112,1.751/2.075,1.767/2.042,1.783/2.011,1.798/1.983,1.814/1.957,1.830/1.934,1.845/1.912,1.861/1.893,1.876/1.875,1.892/1.858,1.908/1.843,1.923/1.829,1.939/1.817,1.970/1.795,2.001/1.775,2.032/1.758,2.079/1.735,2.142/1.708,2.267/1.657,2.360/1.621,2.438/1.593,2.516/1.568,2.594/1.545,2.672/1.526,2.750/1.509,2.828/1.495,2.906/1.483,3.000/1.471,3.094/1.461,3.219/1.452,3.344/1.445,3.515/1.438,3.702/1.433,3.968/1.430,4.233/1.428}
{
\pgfplotstreampoint{\pgfpoint{\x cm}{\y cm}}
}
\pgfplotstreamend
      \pgfusepath{stroke}
    \end{pgfscope}
  \end{pgfscope}
  \fi
  \makeatletter\ifpgf@draftmode\makeatother\else
  \begin{pgfscope}
    \pgfpathrectangle{\pgfpoint{1.12889cm}{0.846667cm}}{\pgfpoint{3.10429cm}{2.59279cm}}
    \pgfusepath{clip}
    \begin{pgfscope}
      \pgfsetlinewidth{0.10pt}
      \definecolor{matfig2pgf_linecolor}{rgb}{1.000,0.000,0.000}
      \pgfsetstrokecolor{matfig2pgf_linecolor}
      \pgfsetdash{{2.00pt}{2.00pt}}{0pt}
      \pgfsetroundjoin
      \pgfplothandlerlineto
\pgfplotstreamstart
\foreach \x/\y in {1.129/1.802,1.145/1.810,1.162/1.832,1.178/1.867,1.194/1.914,1.211/1.971,1.227/2.035,1.243/2.105,1.309/2.405,1.341/2.548,1.358/2.614,1.374/2.675,1.390/2.733,1.407/2.785,1.423/2.833,1.439/2.876,1.455/2.899,1.471/2.895,1.486/2.871,1.502/2.832,1.517/2.783,1.533/2.727,1.564/2.606,1.611/2.422,1.642/2.309,1.658/2.257,1.673/2.207,1.689/2.161,1.705/2.118,1.720/2.077,1.736/2.040,1.751/2.006,1.767/1.975,1.783/1.946,1.798/1.920,1.814/1.896,1.830/1.874,1.845/1.854,1.861/1.837,1.876/1.820,1.892/1.806,1.908/1.793,1.923/1.781,1.939/1.770,1.954/1.760,1.986/1.743,2.017/1.729,2.064/1.710,2.142/1.684,2.235/1.652,2.313/1.623,2.423/1.580,2.563/1.523,2.657/1.488,2.735/1.461,2.813/1.437,2.891/1.415,2.969/1.396,3.047/1.379,3.141/1.362,3.234/1.347,3.328/1.336,3.468/1.321,3.578/1.312,3.765/1.301,3.921/1.295,4.124/1.289,4.233/1.287}
{
\pgfplotstreampoint{\pgfpoint{\x cm}{\y cm}}
}
\pgfplotstreamend
      \pgfusepath{stroke}
    \end{pgfscope}
  \end{pgfscope}
  \fi
  \makeatletter\ifpgf@draftmode\makeatother\else
  \begin{pgfscope}
    \pgfpathrectangle{\pgfpoint{1.12889cm}{0.846667cm}}{\pgfpoint{3.10429cm}{2.59279cm}}
    \pgfusepath{clip}
    \begin{pgfscope}
      \pgfsetlinewidth{0.10pt}
      \definecolor{matfig2pgf_linecolor}{rgb}{1.000,0.000,0.000}
      \pgfsetstrokecolor{matfig2pgf_linecolor}
      \pgfsetdash{{2.00pt}{2.00pt}}{0pt}
      \pgfsetroundjoin
      \pgfplothandlerlineto
\pgfplotstreamstart
\foreach \x/\y in {1.129/1.798,1.145/1.806,1.162/1.828,1.178/1.864,1.194/1.910,1.211/1.966,1.227/2.030,1.243/2.099,1.309/2.391,1.325/2.459,1.341/2.524,1.358/2.584,1.374/2.638,1.390/2.687,1.407/2.731,1.423/2.769,1.439/2.802,1.455/2.816,1.471/2.805,1.486/2.774,1.502/2.730,1.517/2.676,1.533/2.616,1.611/2.295,1.642/2.174,1.658/2.118,1.673/2.064,1.689/2.012,1.705/1.964,1.720/1.918,1.736/1.875,1.751/1.834,1.767/1.795,1.783/1.759,1.798/1.725,1.814/1.693,1.830/1.663,1.845/1.635,1.861/1.608,1.876/1.583,1.892/1.560,1.908/1.537,1.923/1.517,1.939/1.497,1.970/1.461,2.001/1.429,2.032/1.401,2.064/1.375,2.095/1.353,2.126/1.332,2.157/1.314,2.189/1.298,2.220/1.284,2.267/1.265,2.313/1.249,2.360/1.236,2.407/1.224,2.454/1.214,2.516/1.204,2.594/1.194,2.672/1.186,2.766/1.179,2.891/1.173,3.047/1.168,3.281/1.165,3.687/1.163,4.233/1.162}
{
\pgfplotstreampoint{\pgfpoint{\x cm}{\y cm}}
}
\pgfplotstreamend
      \pgfusepath{stroke}
    \end{pgfscope}
  \end{pgfscope}
  \fi
  \makeatletter\ifpgf@draftmode\makeatother\else
  \begin{pgfscope}
    \pgfpathrectangle{\pgfpoint{1.12889cm}{0.846667cm}}{\pgfpoint{3.10429cm}{2.59279cm}}
    \pgfusepath{clip}
    \begin{pgfscope}
      \pgfsetlinewidth{0.10pt}
      \definecolor{matfig2pgf_linecolor}{rgb}{1.000,0.000,0.000}
      \pgfsetstrokecolor{matfig2pgf_linecolor}
      \pgfsetdash{{2.00pt}{2.00pt}}{0pt}
      \pgfsetroundjoin
      \pgfplothandlerlineto
\pgfplotstreamstart
\foreach \x/\y in {1.129/1.815,1.145/1.823,1.162/1.845,1.178/1.881,1.194/1.928,1.211/1.986,1.227/2.051,1.243/2.122,1.276/2.274,1.325/2.505,1.341/2.577,1.358/2.645,1.374/2.710,1.390/2.769,1.407/2.824,1.423/2.874,1.439/2.919,1.455/2.944,1.471/2.940,1.486/2.916,1.502/2.877,1.517/2.827,1.533/2.771,1.564/2.648,1.611/2.461,1.642/2.345,1.658/2.292,1.673/2.241,1.689/2.194,1.705/2.150,1.720/2.109,1.736/2.071,1.751/2.036,1.767/2.004,1.783/1.974,1.798/1.947,1.814/1.923,1.830/1.901,1.845/1.880,1.861/1.862,1.876/1.845,1.892/1.830,1.908/1.816,1.923/1.804,1.939/1.793,1.954/1.782,1.986/1.764,2.017/1.748,2.064/1.728,2.126/1.704,2.407/1.603,2.501/1.572,2.579/1.548,2.641/1.531,2.719/1.513,2.797/1.497,2.875/1.483,2.953/1.472,3.047/1.461,3.156/1.451,3.281/1.442,3.422/1.435,3.593/1.429,3.843/1.425,4.202/1.422,4.233/1.422}
{
\pgfplotstreampoint{\pgfpoint{\x cm}{\y cm}}
}
\pgfplotstreamend
      \pgfusepath{stroke}
    \end{pgfscope}
  \end{pgfscope}
  \fi
  \makeatletter\ifpgf@draftmode\makeatother\else
  \begin{pgfscope}
    \pgfpathrectangle{\pgfpoint{1.12889cm}{0.846667cm}}{\pgfpoint{3.10429cm}{2.59279cm}}
    \pgfusepath{clip}
    \begin{pgfscope}
      \pgfsetlinewidth{0.10pt}
      \definecolor{matfig2pgf_linecolor}{rgb}{1.000,0.000,0.000}
      \pgfsetstrokecolor{matfig2pgf_linecolor}
      \pgfsetdash{{2.00pt}{2.00pt}}{0pt}
      \pgfsetroundjoin
      \pgfplothandlerlineto
\pgfplotstreamstart
\foreach \x/\y in {1.129/1.929,1.145/1.938,1.162/1.963,1.178/2.002,1.194/2.054,1.211/2.117,1.227/2.189,1.243/2.267,1.292/2.521,1.325/2.688,1.341/2.768,1.358/2.843,1.374/2.913,1.390/2.978,1.407/3.038,1.423/3.093,1.439/3.143,1.455/3.170,1.471/3.167,1.486/3.142,1.502/3.100,1.517/3.047,1.533/2.986,1.564/2.852,1.611/2.648,1.642/2.520,1.658/2.461,1.673/2.405,1.689/2.352,1.705/2.303,1.720/2.257,1.736/2.214,1.751/2.174,1.767/2.137,1.783/2.102,1.798/2.071,1.814/2.042,1.830/2.015,1.845/1.990,1.861/1.967,1.876/1.945,1.892/1.925,1.908/1.907,1.939/1.873,1.970/1.843,2.001/1.815,2.048/1.777,2.126/1.716,2.267/1.611,2.345/1.556,2.407/1.514,2.469/1.474,2.516/1.447,2.563/1.422,2.626/1.391,2.672/1.370,2.719/1.350,2.766/1.333,2.828/1.312,2.891/1.294,2.953/1.277,3.016/1.263,3.094/1.249,3.172/1.236,3.250/1.225,3.344/1.215,3.468/1.205,3.578/1.198,3.765/1.190,3.952/1.184,4.186/1.180,4.233/1.180}
{
\pgfplotstreampoint{\pgfpoint{\x cm}{\y cm}}
}
\pgfplotstreamend
      \pgfusepath{stroke}
    \end{pgfscope}
  \end{pgfscope}
  \fi
  \makeatletter\ifpgf@draftmode\makeatother\else
  \begin{pgfscope}
    \pgfpathrectangle{\pgfpoint{1.12889cm}{0.846667cm}}{\pgfpoint{3.10429cm}{2.59279cm}}
    \pgfusepath{clip}
    \begin{pgfscope}
      \pgfsetlinewidth{0.10pt}
      \definecolor{matfig2pgf_linecolor}{rgb}{1.000,0.000,0.000}
      \pgfsetstrokecolor{matfig2pgf_linecolor}
      \pgfsetdash{{2.00pt}{2.00pt}}{0pt}
      \pgfsetroundjoin
      \pgfplothandlerlineto
\pgfplotstreamstart
\foreach \x/\y in {1.129/1.771,1.145/1.778,1.162/1.800,1.178/1.835,1.194/1.880,1.211/1.935,1.227/1.997,1.243/2.065,1.309/2.356,1.325/2.426,1.341/2.492,1.358/2.555,1.374/2.614,1.390/2.668,1.407/2.717,1.423/2.762,1.439/2.802,1.455/2.822,1.471/2.816,1.486/2.791,1.502/2.752,1.517/2.703,1.533/2.648,1.564/2.528,1.611/2.347,1.642/2.234,1.658/2.182,1.673/2.133,1.689/2.086,1.705/2.042,1.720/2.001,1.736/1.963,1.751/1.927,1.767/1.894,1.783/1.864,1.798/1.835,1.814/1.809,1.830/1.784,1.845/1.762,1.861/1.741,1.876/1.721,1.892/1.703,1.908/1.686,1.923/1.670,1.954/1.641,1.986/1.615,2.017/1.592,2.048/1.572,2.095/1.544,2.142/1.520,2.189/1.499,2.235/1.479,2.282/1.462,2.329/1.447,2.391/1.429,2.454/1.413,2.516/1.399,2.594/1.385,2.672/1.374,2.766/1.363,2.860/1.354,2.969/1.347,3.109/1.340,3.281/1.334,3.500/1.331,3.843/1.328,4.233/1.327}
{
\pgfplotstreampoint{\pgfpoint{\x cm}{\y cm}}
}
\pgfplotstreamend
      \pgfusepath{stroke}
    \end{pgfscope}
  \end{pgfscope}
  \fi
    \pgftext[top,x=2.66692cm,y=0.268925cm,rotate=0]{$t$}
    \pgftext[bottom,x=0.4657cm,y=2.10079cm,rotate=90]{$${NF-$\kappa$B}$$}
  \makeatletter\ifpgf@draftmode\makeatother\pgftext[x=5cm,y=4.09091cm]{\Huge{DRAFT}}\fi